%% file: ms.tex
\documentclass[12pt,preprint]{aastex}

\slugcomment{To appear in ApJS, Draft version: \today}

\shorttitle{Young Cluster Survey}
\shortauthors{Gutermuth et al.}

\begin{document}

\title{A {\it Spitzer} Survey of Young Stellar Clusters within One Kiloparsec of the Sun: Cluster Core Extraction and Basic Structural Analysis}

\author{R. A. Gutermuth\altaffilmark{1}, S. T. Megeath\altaffilmark{2}, P. C. Myers\altaffilmark{1}, L. E. Allen\altaffilmark{1}, J. L. Pipher\altaffilmark{3}, G. G. Fazio\altaffilmark{1}}

\altaffiltext{1}{Harvard-Smithsonian Center for Astrophysics, 60 Garden Street, MS-42, Cambridge, MA 02138}
\altaffiltext{2}{University of Toledo, Toledo, OH}
\altaffiltext{3}{University of Rochester, Rochester, NY}

\begin{abstract}

We present a uniform mid-infrared imaging and photometric survey of 36 young,
nearby, star-forming clusters and groups using {\it Spitzer} IRAC and MIPS.  We
have confidently identified and classified 2548 young stellar objects using
recently established mid-infrared color-based methods.  We have devised and
applied a new algorithm for the isolation of local surface density enhancements
from point source distributions, enabling us to extract the overdense cores of
the observed star forming regions for further analysis.  We have compiled
several basic structural measurements of these cluster cores from the data,
such as mean surface densities of sources, cluster core radii, and aspect
ratios, in order to characterize the ranges for these quantities.  We
find that a typical cluster core is 0.39~pc in radius, has 26 members with
infrared excess in a ratio of Class~II to Class~I sources of 3.7, is embedded
in a $A_K$=0.8 mag cloud clump, and has a surface density of 60~pc$^{-2}$.  We
examine the nearest neighbor distances among the YSOs in several ways,
demonstrating similarity in the spacings between Class~II and Class~I sources
but large member clusters appear more dense than smaller clusters.  We
demonstrate that near-uniform source spacings in cluster cores are common,
suggesting that simple Jeans fragmentation of parsec-scale cloud clumps may be
the dominant process governing star formation in nearby clusters and groups.
Finally, we compare our results to other similar surveys in the literature and
discuss potential biases in the data to guide further interpretation.

\end{abstract}


\section{Introduction\label{intro}}

The formation of stars in clusters is a central problem in the study of star
formation.  Surveys of molecular clouds in the nearest 1 kpc show that
approximately 75\% of embedded young stars are in groups and clusters with 10
or more members \citep{carp2000,alle07}.  Although the smaller groups are more
numerous, about 80\% of all young stars located in embedded clusters and groups
are found in the largest clusters with more than 100 stars
\citep{ll03,porr03,alle07}.  Embedded clusters differ from open and globular
clusters in that the mass in the cluster is dominated by the mass of the
molecular cloud; thus they are bound primarily by the molecular cloud from
which they form.  As a consequence, only 5\% of embedded clusters are thought
to evolve into open clusters \citep{ll03}; the remainder likely disperse when
the molecular gas is dispersed by feedback from the young stars.

The formation and early evolution of clusters involves the interplay of
fragmentation from turbulent motions in the parental cloud and/or gravity,
dynamical motions of young stars, and the feedback from the youngs stars.
Increasingly sophisticated numerical simulations are now being used to examine
these processes in detail \citep[e.g.][]{kb00,kles00,bbb03,bbv03,gamm03,li04};
however, the predictions of these simulations are difficult to verify by
observations.  Perhaps the most important observational constraints are the
initial mass function (IMF), the spatial structure of the clusters and the
molecular gas, the kinematics of the cluster and gas, and the star formation
history.  The detailed form of the IMF has not emerged as as definitive test of
theory; many different theoretical scenarios have been able to reproduce the
IMF.  That the initial mass function appears universal and does not seem to
strongly vary with the ``initial'' conditions of the molecular gas is an
important constraint on theory, but gives little insight into the physics that
controls the IMF.  The star formation history remains difficult to constrain
due to uncertainties in establishing the ages of young stars \citep{hillen08}.
Finally, although the kinematics of the gas are routinely measured,
complementary kinematic surveys of young stars are only being obtained for
clusters with large numbers of unobscured members detectable by visible light
spectrometers \citep{fure08}.

The spatial distribution of the young stellar objects (YSOs), in contrast, is
emerging as an important means for studying embedded clusters and constraining
theoretical models.  Recent measurements have shown that the YSOs in embedded
clusters follow elongated and clumpy spatial distributions that often parallel
structures in the parental molecular cloud
\citep[e.g.][]{gome93,lal96,man98,alle02,gute05,teix06,wins07,gute08}.
Obtaining relatively unbiased maps of the structure requires a means for both
detecting embedded sources and identifying members against a field of
background stars.  Due to the the high dust column densities typically found
toward young clusters, near-infrared surveys offered \citep[e.g.][]{hoda94} the
first near-complete census of the cluster members, penetrating regions entirely
obscured at visible wavelengths.  However, distinguishing members from
unrelated field stars has proved difficult and must be done by using
statistical approaches \citep[e.g.][]{gute05}.

The {\it Spitzer Space Telescope} \citep{wern04} is a huge step forward in this
regard, bringing sensitivity and angular resolution comparable to those of
near-infrared surveys at mid-infrared wavelengths with onboard imaging
instruments the Infrared Array Camera \citep[IRAC;][]{fazi04} at 3-8~$\mu$m and
the Mid-Infrared Photometer for {\it Spitzer} \citep[MIPS;][]{riek04} at
24-160~$\mu$m.  Such capability impacts embedded cluster research in two ways.
First, mid-infrared wavelength light penetrates dusty environments more
effectively than light at near-infrared wavelengths \citep{flah07}.  Second,
for embedded ($ < 3$~Myr old) clusters, the majority of embedded members have
dusty circumstellar material \citep{hll01,hern07} that emits excess infrared
emission beyond what would be predicted for a typical photosphere, enabling
unique identification of these sources.  As shown by \citet{alle07}, {\it
Spitzer}-identified YSO surveys yield a much more complete view of the source
spatial distribution than near-infrared surveys, which can only infer clusters
by their elevated surface density and the relatively higher flux of cluster
members relative to field stars.  With the mid-infrared data, even low surface
density star formation is detectable because we can uniquely identify the disk
and envelope bearing members, only losing the diskless pre-main sequence
members which are expected to be relatively small in number in the youngest
($<$1~Myr old) clusters.

{\it Spitzer} also offers a new means for establishing the evolutionary class
of detected cluster members. Analysis of the mid-IR colors and/or spectral
energy distribution can be used to distinguish between protostellar objects
with infalling envelopes and more evolved pre-main sequence stars with disks
\citep{alle04,whitney04}.  Using this new capability for YSO classification at
the angular resolution and sensitivity of {\it Spitzer}, we can distinguish
with reasonable reliability the youngest sources, which are closest to their
star formation sites, and the more evolved pre-main sequence stars which may
have moved significantly from their birth sites.

We present here the {\it Spitzer} Young Cluster Survey, a systematic IRAC and
MIPS 24~$\mu$m imaging and photometric survey of 36 star-forming clusters and
groups selected from the \citet{porr03} catalog.  The primary goal of this
survey is to study the spatial distribution of YSOs in clusters as a function
of their evolutionary class.  The clusters in the sample range in size from
small groups to rich clusters enabling a study of how the cluster properties
(density, size) vary as a function of the number of members.  In this paper,
we use ``cluster'' to denote both the groups with tens of YSOs and clusters
with many tens to hundreds of YSOs. We exploit the {\it Spitzer} multiple
bandpass photometry for YSO identification and classification by adapting
methodologies developed for work on both individual clusters
\citep[e.g.][]{mege04,teix06,wins07,muen07,gute08,alle08} and entire nearby
molecular clouds \citep[e.g.][]{jorg06,alle07,harv07,jorg08}.  An important
limitation of this work is extracting cluster members from more extended
distributions of YSOs given a limited field of view \citep{alle08}.  To address
this limitation, we develop an objective criterion for defining a cluster core
and establishing the membership of this core while minimizing biases due to
distance and completeness.  After the systematic application of this criterion
to our sample, we compare the number of members, physical size, and shape of a
representative sample of young clusters.  

The paper is presented as follows.  In Section~\ref{obs}, we describe the
target selection, data treatment, and ancillary data products.  In
Section~\ref{distros}, we present the resulting images, YSO distributions, and
surface density maps of both the YSOs and their natal clouds, as well as the
photometry tables of the YSOs identified.  Section~\ref{demmem} contains an
examination of the YSO membership counts among the clusters, an examination of
their nearest neighbor distance distributions, and an objective cluster core
isolation method.  In Section~\ref{demcore} we present an analysis of the
structural measurements of the isolated cluster cores.  We discuss the results
in the context of simple thermal fragmentation and perform some analysis of the
aggregated nearest neighbor distance data for the whole survey in
Section~\ref{discuss}.  Section~\ref{bias} outlines several known issues with
the survey data and the biases they may inflict on the measurements presented.
We summarize the paper in Section~\ref{summ}.  Finally, we present a revised
version of the YSO classification scheme of \citet{gute08} in
Appendix~\ref{app1}, and some additional detail regarding our chosen cluster
area geometries and their derived quantities in Appendix~\ref{app2}.

\section{Observations\label{obs}}

Here we summarize the survey target selection and observations presented in
this work, including 2MASS, IRAC, and MIPS 24~$\mu$m photometry, and the
derived data products, including the near-infrared extinction maps and nearest
neighbor surface density maps.

\subsection{Survey Target Selection\label{samp}}

The target sample of the {\it Spitzer} survey was primarily selected from the
literature compilation of embedded clusters within 1~kpc by \citet{porr03}.
This compilation focused on clusters identified in near-IR imaging of star
forming regions and molecular clouds.  It incorporated the results of several
systematic near-infrared surveys, including the giant molecular clouds surveys
of \citet{lada1991} and \citet{carp2000}, the outflow survey of \citet{hoda94}
and the Herbig Ae/Be star survey of \citet{testi1999}.  The initial selection
included all clusters and groups in \citet{porr03} with estimated memberships
of $\ge$10 YSOs.  The clusters in the Orion molecular cloud complex and the
NGC~2264 region were included in other planned GTO surveys and were excluded
from the sample.  The clusters in Vela were also excluded because of a lack of
available spacecraft time and uncertainties in the distances to these regions.
The HD~216629, HD~37490 and VV~Serpens clusters all had 10 or more members in
\citet{porr03} but were eliminated because the molecular gas survey of
\citet{ridge2003} failed to show evidence for associated dense gas. Finally,
the RNO~1/b cluster was determined to be smaller than the required 10 members
from a re-examination of the results of \citet{testi1999}, so it was also excluded
from our sample.

Several targets which are not in \citet{porr03} were added to the sample.  The
S131 cluster was added on the basis of near-IR imaging of bright rimmed clouds
by \citet{sugi1995}.  Fields in two of the nearest molecular clouds, the Taurus
and Ophiuchus clouds, were included in order to obtain data on areas of
distributed star formation without the presence of dense clusters or groups.
In Taurus, this included fields centered on sparse groups of young stars in the
L1551 and L1495 dark clouds.  In addition, two sections of the Ophiuchus cloud
which were known not to contain clusters were also added to the survey: the
L1709 filament and the L1689 cloud core.  In the course of our survey, three
regions were found not to meet our initial criteria.  XY~Perseus does not have
a cluster and the estimate of 11 members by \citet{testi1999} is either the
result of chance statistical fluctuations in the density of background stars or
a low fraction of members with excess infrared emission.  Although the S106 and
CB~34 clusters were included on the basis of 600~pc and 900~pc distances,
respectively, these are now thought to be at distances greater than 1~kpc
\citep[][respectively]{schneider2007,lh97}.  

The final sample includes all of the embedded clusters in the \citet{ll03}
tabulation within 1~kpc, with the exception of Orion and NGC~2264.  Together
with the {\it Spitzer} surveys of Orion and NGC~2264, the {\it Spitzer} cluster
survey provides a representative sample of clusters within one 1~kpc; however,
the recent discovery of the Serpens South cluster \citep{gute08a} and the
CepOB3b cluster \citep{getman2006,alleninprep} show that this combined sample
is not complete.

The field sizes were 30\arcmin~$\times$~30\arcmin\ for regions at distances of
160 to 400~pc (1.4~pc~$\times$~1.4~pc to 3.6~pc~$\times$~3.6~pc) and
15\arcmin~$\times$~15\arcmin\ for regions from 500~pc to~1~kpc
(2.2~pc~$\times$~2.2~pc to 4.5~pc~$\times$~4.5~pc. These sizes were motivated
by the cluster diameters reported in the literature compilation of \citet{ll03}
and the survey of \citet{carp2000}; these showed typical diameters of 1~pc.
Three exceptions to our guidelines were: XY~Perseus, which was surveyed with a
15\arcmin~$\times$~15\arcmin\ field of view despite its distance of 160~pc;
IC~5146, which was covered in a 30\arcmin~$\times$~30\arcmin\ field assuming it
was only $\sim$460~pc distant \citep{lal99}, though recent examination
suggests 950~pc \citep{harv08}; and L1709, which was mapped with an elongated
field of view aligned with the molecular filament.

Table~\ref{littable} lists some fundamental characteristics for the final list
of targeted regions, including far-infrared luminosity of the brightest source
as measured by the {\it Infrared Astronomy Satellite} ({\it IRAS}), local cloud
mass as derived from the near-infrared extinction maps (described in
Section~\ref{avmaps} below), adopted heliocentric distance and the reference
for each distance estimate.

\subsection{Infrared Imaging and Photometry}

We have provided Table~\ref{ysotable1} containing detailed information on the
{\it Spitzer} data used for each region, including the Astronomical Observation
Request (AOR) identification numbers of the datasets included for each region,
the dates that each were observed, and the SSC BCD calibration pipeline version
used to reduce each.  Table~\ref{ysotable1b} includes overall mosaic
information: pointing center coordinates (Right Ascension and Declination,
J2000), angular areas of the mosaics, total exposure time per pixel, and
field-averaged differential 90\% completeness limits \citep{gute05} for the
point source photometry for each region and bandpass.  All data reduction was
performed via Gutermuth's Cluster Grinder, an IDL software package that unifies
and increases automation for the many modules discussed in \citet{gute08} and
summarized below.  These modules rely heavily on the IDL Astronomy User's
Library \citep{land93}.

\subsubsection{IRAC}

Typical IRAC observations were $3\times4$ raster maps of four dithered images
at each position in the raster, though several of the nearby regions were
observed with $6\times7$ raster maps of only two dithered images per position.
All observations were obtained in High Dynamic Range (HDR) mode, whereby two
images are taken in succession at 0.4s and 10.4s integration times.  We
utilized standard Basic Calibrated Data (BCD) products from the {\it Spitzer}
Science Center's standard data pipeline.
The BCD images were first improved with custom treatments for bright source
artifacts \citep[``pulldown'', ``muxbleed'', and
``banding'';][]{hora04,piph04}.  Then all frames are combined into mosaics with
WCSmosaic \citep{gute08} such that rotation, spatial scale distortion, and
subpixel offset resampling are all performed in one transformation, to minimize
smoothing.  IRAC mosaics were constructed at their native pixel scale of
~1\farcs2 per pixel, and angular resolutions range from 2\farcs0 to 2\farcs5
full width at half maximum, varying by bandpass.  
Redundancy-based outlier detection and removal is also performed during mosaic
construction, eliminating transient signals such as cosmic ray hits.  

Automated source detection and aperture photometry were performed using PhotVis
version 1.10 \citep{gute08}.  PhotVis utilizes a modified DAOphot
\citep{stet87} source-finding algorithm which calculates local noise across the
mosaics and uses this ``noise map'' to set a position dependent detection
threshold. 
This technique provides robust source detection in fields with bright,
spatially varying nebulosity.  Aperture photometry is performed using synthetic
apertures of 2\farcs4 radius, with background annuli of inner and outer radii
of 2\farcs4 and 7\farcs2, respectively.  These aperture and annulus sizes have
been adopted as a practical way of adequately sampling the core of the point
spread function (PSF) while minimizing contamination from resolved background
emission and other nearby sources \citep{mege04}.  Photometric calibration
values adopted (Vega-standard magnitudes for 1~DN~s$^{-1}$) are identical to
those used by \citet{gute08}: 19.455, 18.699, 16.498, and 16.892 for 3.6, 4.5,
5.8 and 8.0~$\mu$m bands respectively.  These values are derived from the
calibration effort presented in \citet{reac05}, and they include standard
corrections for the aperture sizes chosen.  The calibration uncertainties are
approximately 5\% across all IRAC bands. 

Initial reduction efforts revealed that up to 50\% of the radial astrometric
residuals to 2MASS PSC positions are caused by simple full field offsets that
seem to vary from AOR to AOR.  Before building the final mosaics that are
sometimes composed of more than one AOR of BCD images, the data from each AOR
were separately run through the above mosaicking, point source extraction, and
catalog merger processes (the last item is described in Section~\ref{bandmerge}
below) in order to measure this offset and improve the astrometric
calibrations.  The median offset applied is $\sim$0\farcs3.  The final spread
of radial astrometric residuals relative to 2MASS are typically 160
milliarseconds in the 3.6, 4.5, and 5.8~$\mu$m channel mosaics, and 180
milliarcseconds for the 8.0~$\mu$m channel mosaic.

\subsubsection{MIPS}

Most MIPS mosaics were obtained at the medium scan rate, with full scan width
stepping.  The exceptions are: IC~5146, where data were used from the Gould
Belt Legacy Survey (Allen et al. in prep.), which adopted fast scanning and two
epochs of observations with a half scan width offset to enable full 70~$\mu$m
coverage; IC~348, NGC 1333, the Ophiuchus fields, and the Taurus fields were done to a
significantly higher depth as part of the MIPS GTO Disk Evolution Program.
Typical resolution is 5'' FWHM in the 24~$\mu$m band.  The 70 and 160~$\mu$m
data are inadequate for our purposes because of their incomplete coverage 
and low angular resolution.  The 24~$\mu$m data were reduced, mosaicked, and
point sources were extracted, with Cluster Grinder and PhotVis, as was
described above for the IRAC data.  We used a synthetic aperture radius of
7\farcs6, and background annulus inner and outer radii of 7\farcs6 and
17\farcs8 for the 24~$\mu$m data to match the core of the PSF and minimize the
effects of contamination, as described above for the IRAC data.  We adopt the
24~$\mu$m photometric calibration value derived by \citet{gute08}, 14.6 mag for
1~DN~s$^{-1}$ within the chosen apertures.  The astrometric calibration of MIPS
AORs is adjusted similarly to the method described above for IRAC.  The median
offset applied is $\sim$0\farcs5.  The final spread of radial astrometric
residuals relative to 2MASS is typically 350 milliarcseconds.

\subsection{Final Source Photometry Catalogs\label{bandmerge}}


Source list matching is performed in stages to minimize mismatches.  First, the
IRAC data of the four bandpasses are merged, using a maximum radial matching
tolerance of 1\arcsec.  Then the mean positions of each merged entry from that
catalog are merged with the 2MASS PSC, again using a maximum 1\arcsec\ radial
matching tolerance.  Finally, the MIPS 24~$\mu$m source list is matched to the
mean positions of each entry in this secondary catalog, using a 1.3\arcsec\
radial matching tolerance to account for the lower resolution and less precise
astrometry of these longer wavelength data.  Note that the latter matching
tolerance is significantly smaller than the 3\arcsec\ tolerance used by
\citet{gute08}.  This is largely a product of the astrometric recalibration
efforts applied to the IRAC and MIPS data, and results in more robust matching
of the lower resolution MIPS data, with considerably reduced mismatching rates.

Classification of sources is performed using the multi-phase mid-IR colors
method explained and justified in detail in \citet{gute08}, but updated to
account for complications found in the more active and typically more distant
regions of the survey (see Appendix~\ref{app1} for an updated description of
the method).  In summary, the method uses the numerous available flux ratios,
or colors, to identify and classify YSOs as robustly as possible while
mitigating the effects of contamination and reddening.  Contamination arises
from non-YSO sources with excess infrared emission, including star-forming
galaxies, broad-line active galactic nuclei (AGN), and unresolved knots of
shock emission from outflows colliding with cold cloud material.  Once
separated from the diskless photospheres (Class~III pre-main sequence stars and
unrelated field stars) and the aforementioned contaminants, the YSOs are
classified into the canonical categories of Class~I (protostars with infalling
envelopes, including flat spectrum objects) and Class~II (pre-main sequence
stars with optically thick disks) YSOs \citep{alle04}, with supplemental
categories of ``deeply embedded sources'' (added to the Class~I tallies) and
the ``transitional disks'' (added to the Class~II tallies).  

In total, we identified 2548 YSOs with infrared excess emission, with 472
having mid-infrared colors consistent with Class~I YSOs and 2076 having colors
consistent with Class~II YSOs.  Of these, we expect $33 \pm 4$ to be residual
extragalactic contaminants, with $20 \pm 3$ in the Class~I group, and $13 \pm
3$ in the Class~II group.  Note that the mean residual contamination is 2 per
target in the nearby regions surveyed over 30\arcmin\ $\times$ 30\arcmin\ (5 in
the large Tau L1495 survey region) and 0.5 per target in the regions beyond
500~pc surveyed over 15\arcmin\ $\times$ 15\arcmin .  These estimates are upper
limits to the mean contamination, as extinction from a dusty molecular cloud
will bias the AGN contamination downward somewhat.  Regardless, such small
amounts of contamination are highly unlikely to significantly affect the
subsequent physical measurements of the clustered structures as reported.
Tables~\ref{tab4}-\ref{tab39} hold the positions and merged photometry for all
high confidence YSO candidates for each region surveyed.  The entries are
sorted first by YSO evolutionary class (``I*'' are "deeply embedded sources",
``II*'' are transition disk candidates) and then by ascending R.A.


\subsection{YSO Nearest Neighbor Distances and Surface Density Maps\label{nnd}}

Once we have isolated the population of likely YSOs in a given region, we make
several measurements related to source spacing and surface density.  First, we
measure the projected distance from each YSO to its nearest YSO neighbor, often
noted as the simple nearest neighbor distance in the literature, and notated as
NN2 throughout this text.  Similarly, we measure the NN6 nearest neighbor
distance, the radial distance from each source such that a circular area of
that radius centered on the source contains the nearest five neighbors, for a
total of six YSOs.  

We also map the surface density structure of each region using the ``nearest
neighbors'' method \citep{gute05}.  At each position in a uniform grid, we
measure the radial distance necessary to encompass the $n$ nearest YSOs and
compute the local surface density $\sigma\{i,j\} = \frac{n-1}{\pi
r_n^2\{i,j\}}$ with fractional uncertainty of $(n-2)^{-0.5}$ \citep{ch85}.
This method is effectively an adaptive smoothing method, and the typical
resolution and the statistical significance of a given measurement are set by
adopting some value of $n$.  
For the purposes of this work, we have adopted $n=6$ for all YSO surface
density maps as a good compromise between resolution and sensitivity.  This
choice also makes the maps similar to the NN6 nearest neighbor distances
mentioned above.

\subsection{Near-infrared Extinction Maps\label{avmaps}}

For each surveyed region, we have mapped the surface density structure of dust,
a proxy for the natal molecular material using the measured $H-K_S$ colors of
field stars \citep{lal07}.  Intrinsic $H-K_S$ colors are assumed to be 0.2~mag, and
each source's extinction is measured using the reddening law of \citet{rl85}.
The mapping technique is an adaptive one, described in detail in
\citet{gute05}, and is similar to our surface density mapping technique above.
Most maps have a dynamic range of $A_V$ = 2 to 12 mag \citep[$1 A_V = 0.112
A_K$][]{rl85}, and the typical resolution is 90\arcsec\, though both vary based
on the surface density of field stars and their unextinguished
distribution in brightness.

\section{Mid-infrared Images and YSO Spatial Distributions\label{distros}}

Figures~\ref{first}-\ref{last} show the observations and fundamental analysis
of the entire survey, with the objects sorted by ascending Right Ascension.  In
all figures, the upper left panel is a color composite image of the 3.6, 5.8,
and 24~$\mu$m mosaics in blue, green, and red respectively.  The gray outline
tracing the limits of the simultaneous field of view of all four IRAC channels
(reproduced in all subsequent panels), and MIPS 24~$\mu$m coverage is typically
$>$95\% of this area and extends considerably beyond the fields of view shown.
The upper right panel shows the column density distribution of the molecular
cloud in the same field of view, as represented by our maps of near-infrared
reddening of stars in the field.  Contours are placed at intervals of 2~mag in
$A_V$, and start at $A_V = 5$.  The grayscale is uniformly chosen for all
clouds such that white is $A_V = -1$ and black is $A_V = 15$.  The left middle
panel overlays the distribution of YSOs on the $A_V$ contours from the upper
right panel in order to facilitate direct comparison of the YSO positions and
the molecular cloud morphology.  Class~I objects are marked as red circles and
Class~II objects are marked as black squares here.  The right middle panel
displays the surface density map of the {\it Spitzer}-identified YSOs, as
described in Section~\ref{nnd}.  The grayscale is chosen such that white is
0~pc$^{-2}$ and black is 1000~pc$^{-2}$ in all maps, and the contour intervals
are 1$\sigma$ (50\%) from the next higher level.  The lower left panel shows
the Minimal Spanning Tree (MST) of the YSO positions.  Black points and MST
connections are for those objects that are more closely spaced than the
critical length measured in the cumulative distribution of MST branch lengths
(lower right panel), while gray points and MST branches mark the more isolated
objects.  Colored outlines of the cluster cores identified (green, blue, and
red in order of descending membership) are also overlaid.  The MST-based cluster core isolation method is described in detail in Section~\ref{mstchop} below.

\section{Characterization of Cluster Membership\label{demmem}}

In Table~\ref{ysotable2}, we summarize the YSOs detected by class as well as in
total for each region that we observed.  The number of YSOs in each class is
useful for estimating the relative age of a star-forming region, assuming a
constant birthrate \citep{hatc07}.
In this work, we present the ratio of the number of Class~II sources to the
number of Class~I sources.  In addition to assuming a constant star-formation
rate, relative age interpretation using this ratio hinges on the assumption
that no disks are lost in the Class~II population; we will discuss biases
related to these issues in Section~\ref{bias}.  We also have included the YSO
tallies and ratios separately for the dense ``cores'' of each cluster region
and their lower density distributed star-forming areas.  This density
distinction is a common feature of the YSO spatial distributions for most of the
regions surveyed; we will describe a method to isolate these regions below.


\subsection{Nearest Neighbor Distance Distributions}

The spatial distribution of YSOs in a region is often analyzed in terms of a
typical spacing between sources in order to compare this spacing to the Jeans
fragmentation scale for a self-gravitating medium with thermal pressure.
For example, observational studies by \citet{gome93} and \citet{man98}
presented median NN2 nearest neighbor distances of YSOs and pre-stellar
molecular cloud cores, respectively.  The former work made a qualitative
comparison to the fragmentation process by referencing the typical size of
star-forming clumps in the natal molecular cloud, while the latter performed a
more direct comparison to the Jeans process by comparing the submillimeter core
separation lengths and the inferred Jeans length from independently measured
cloud conditions.  More recently, fragmentation in gas with
turbulence \citep[e.g.][]{ball07} and magnetic fields \citep[e.g.][]{ward07}
have been discussed, leading to more detailed predictions of the distributions
of fragment spacings.  Similarly, some
recent observations of star forming regions have been analyzed in terms of the
distribution of NN2 distances.
\citet{teix06} found a strong peak in their histogram of NN2 spacings for the
protostars in the ``Spokes'' embedded cluster in NGC~2264.  These authors
suggested that this peak indicated a significant degree of Jeans fragmentation,
since this most frequent spacing agreed with an 
estimate of the Jeans length for the dense gas within which the YSOs are
embedded.  \citet{gute08} also found a clear peak in the NN2 spacings in both
protostars and more evolved pre-main sequence stars with disks in the NGC~1333
embedded cluster.  This result suggests that the tendency for a narrow range of
spacings among YSOs in a cluster can last into the Class~II phase of YSO
evolution.  

\begin{figure}
\epsscale{.80}
\plotone{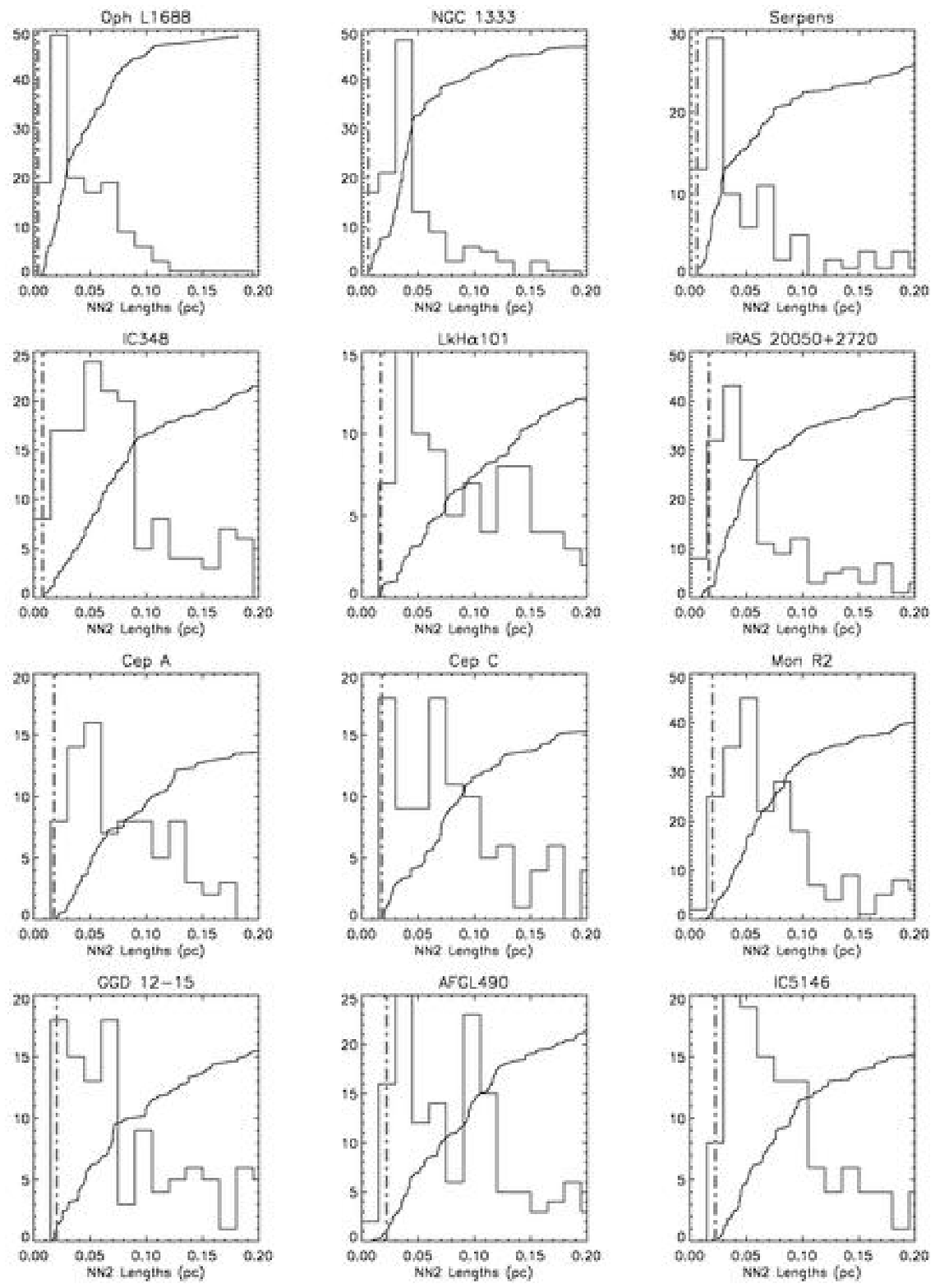}
\caption{Histograms of NN2 lengths for each of the 12 largest clusters, sorted by ascending heliocentric distance from left to right and top to bottom.  The vertical dot-dashed line marks the roughly 5\arcsec\ boundary below which source confusion becomes a significant problem.  The bin size is 0.015~pc for all plots.  The cumulative distribution function of the data is also overplotted.\label{nnhists}} 
\end{figure}

A statistically larger sample of star-forming clusters is needed to see how
often such peaks are observed in order to estimate what is typical in clustered
star formation.  We have plotted histograms of the NN2 distances for each YSO
for the twelve clusters with the most members in Figure~\ref{nnhists}, sorted
by heliocentric distance.  From these plots, it is clear that many clustered
star forming regions have a peak at small spacings that is not caused by
resolution limitations, and that they also have a relatively long tail of large
spacings.  Although the peak may be sharp, or broad, or one of several near
equivalent peaks, such peaked character is often observed, regardless of the
two dimensional distribution of sources.

Of the twelve regions, all have more short spacings than long.  Eight have
well-defined single peaks at least one bin greater than the confusion limit,
though the limit shown is a typical value and may vary considerable from region
to region.  Six regions (NGC~1333, IC~348, Serpens, LkH$\alpha$~101,
IRAS~20050+2720, and Mon~R2) have both a well-defined peak at 0.02-0.05~pc and
a tail extending to spacings of 0.2~pc and greater.  With coarser binning,
GGD~12-15 and IC~5146 would join this latter group.  Similarly, all the
cumulative distributions in Fig.~\ref{nnhists} have slopes that decrease with
increasing spacing.  When the histograms show a pronounced peak and tail, the
cumulative distributions have a steep inner slope and a shallow outer slope.

Peaked NN2 distance distributions typically suggest a significant sub-region
(or sub-regions) of relatively uniform, elevated surface density, but the
specific two dimensional geometry is unspecified in such a characterization.
By adopting an NN2 length threshold, we can isolate those sources which are
closer together than this threshold, yielding populations of sources that make
up a local surface density enhancement, as well as some isolated, resolved
binaries from nearby, lower density star formation.  However, this treatment
isolates a locally dense population without grouping more than a few of the
sources together based on the linkages shorter than the threshold length.
Another algorithm must be performed to group together the close neighbors into
larger structures, typically involving chosen surface density thresholds to
subdivide the population into structures via contour delimitation on a smoothed
surface density map \citep[e.g.][]{chav08,jorg08,schm08,roma08}. Alternatively,
a similar yet more complete algorithm could be adopted to isolate the more
closely spaced population and immediately group them together.  


\subsection{Extracting Clusters From Clouds}

The review of structure in young stellar clusters presented by \citet{alle07}
demonstrated that most embedded clusters appear to be the densest regions of
much larger distributions of YSOs that span entire molecular clouds.  Thus
measurements of cluster characteristics hinge greatly on what one chooses to
include as part of a cluster.  The cluster isolation method adopted by
\citet{alle07} is the Path Linkage Criterion or PLC \citep{batt91}, effectively
a single surface density cutoff in physical units for the assumed distance to
each cloud.  This method, and other surface density threshold methods like it
\citep[e.g.][]{ll03,roma08,jorg08,schm08}, have the advantage of uniform
treatment of all regions, but the main drawback is that the thresholds adopted
are arbitrary.  Since surface density profiles, fractions of members with
infrared excess, mass sensitivities, and heliocentric distance uncertainties
vary from region to region, no single surface density threshold can be
appropriate for all clusters.  Similarly, it is often difficult to identify a
simple surface density background in these regions, as low density distributed
star formation is also potentially non-uniform.
These complications support the development of a new technique in which a
spacing or density threshold is not selected arbitrarily, but is instead
determined in a systematic manner for each region.


As mentioned in Section~\ref{obs}, the survey presented here was a targeted
one, where each set of observations was aimed at a previously known young
cluster.  Some clusters were obviously surveyed completely as evidenced by the
lack of any YSOs or dense gas near the field edge, such as CB~34.  However,
other clusters are much larger than was previously known, such as AFGL~490, and
many of these clearly have been truncated by the field of view of the
observations.  The variable degree of coverage of the distributed population of
YSOs around many of the clusters severely hinders our ability to distinguish
clustered YSOs from their more diffuse siblings in an objective way.  To
counter this difficulty, we have constructed a new empirical method to identify
the local surface density threshold for extracting the overdense regions of
each star-forming region.  This method utilizes the spectrum of source spacings
in a given region to isolate dense regions from the less dense, regardless of
the two-dimensional projected geometries of the source distributions.


\subsubsection{Minimum Spanning Tree Branch Length Distributions}

A {\it more complete} and more convenient characterization of the spectrum of
source spacings can be obtained from the Minimum Spanning Tree (MST) of the
source positions, a mathematical device that can be used to perform a
near-identical analysis to the PLC mentioned above \citep{batt91}.  The MST is
defined as the network of lines, or {\it branches}, that connect a set of
points together such that the total length of the branches is minimized and
there are no closed loops.  

The MST yields a more complete characterization of the spacing spectrum than
the nearest neighbor distances because it connects all points together into a
single network, while the NN2 linkages for the same set of points only connect
each source to its nearest neighbor, creating numerous ``islands'' of
disconnected small-$N$ groups.  For a given source distribution, the set of NN2
lengths often studied in the literature is in fact a partially degenerate
subset of the MST branch lengths, specifically the shortest MST branch length
connected to each source (close pairs are each other's nearest neighbors, thus their connecting MST branch is counted twice).  The MST is more
convenient than the NN2 links because it allows immediate partitioning of
substructured distributions \citep[e.g.][]{cw04,sk06,bast07,bast08} upon
adoption of a surface density threshold as expressed by a critical branch
length.  Furthermore, since the MST has no inherent smoothing associated with
it, grouping objects in this way imposes no bias with regard to the shapes of
the distributions one can isolate, preserving the underlying geometry from
shapes that are round to filamentary.   

\begin{figure}
\epsscale{1.0}
\plotone{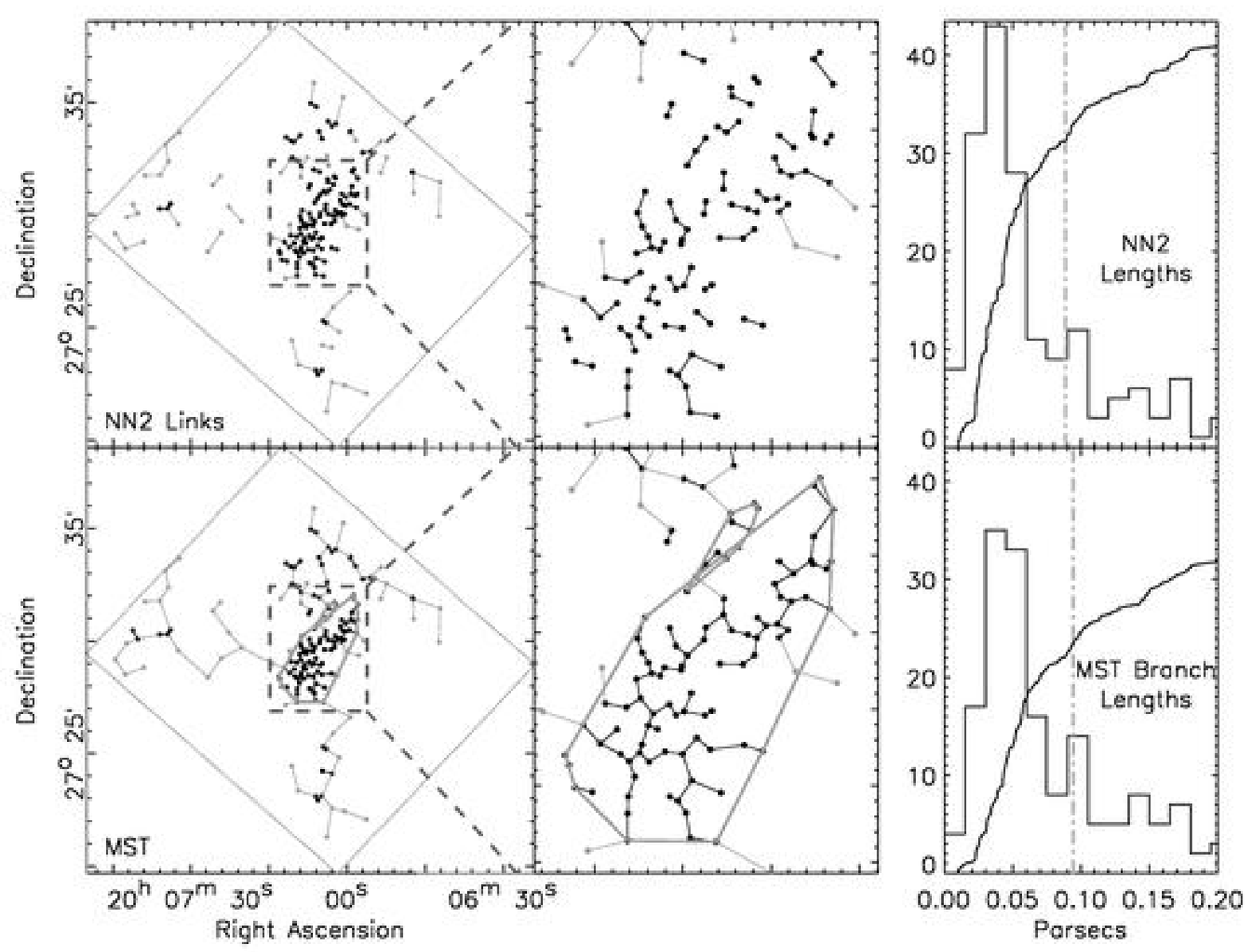}
\caption{Source linkage diagrams using both the NN2 links (top) and MST (bottom), using IRAS~20050+2720 as an example to demonstrate the ``full field'' distribution (all points) and the isolated high density population (black points).  The NN2 links and the MST itself are plotted as gray line segments if those lengths are longer than the critical length (as noted by the gray dot-dashed lines over the histograms in the right column) and black segments if those links are shorter than that length.  In the MST's case, after all branches longer then the local critical length are clipped, two groups of still-connected sources that have 10 or more members are found at the center.  Thus we identify these as the ``cores'' of IRAS~20050+2720 and mark their boundaries with the convex hull of the source distributions in thick gray.\label{mstex}}
\end{figure}

In Figure~\ref{mstex}, we demonstrate the similarity in the histograms of NN2
distances and the MST branch lengths.  Both distributions have clear peaks, but
by clipping the MST branch lengths that are longer than some critical distance,
cohesive structures are immediately isolated by their contiguous residual
networks of MST branches shorter than that distance.  As mentioned in
Section~\ref{nnd}, the NN2 links longer than a given threshold can be clipped
in a similar fashion, but this only isolates the populations in dense regions
and disparate close pairings, while having very little bearing on the grouping
of cluster-like structures.  We will now describe how the critical distances
shown here were derived.

\begin{figure}
\epsscale{.50}
\plotone{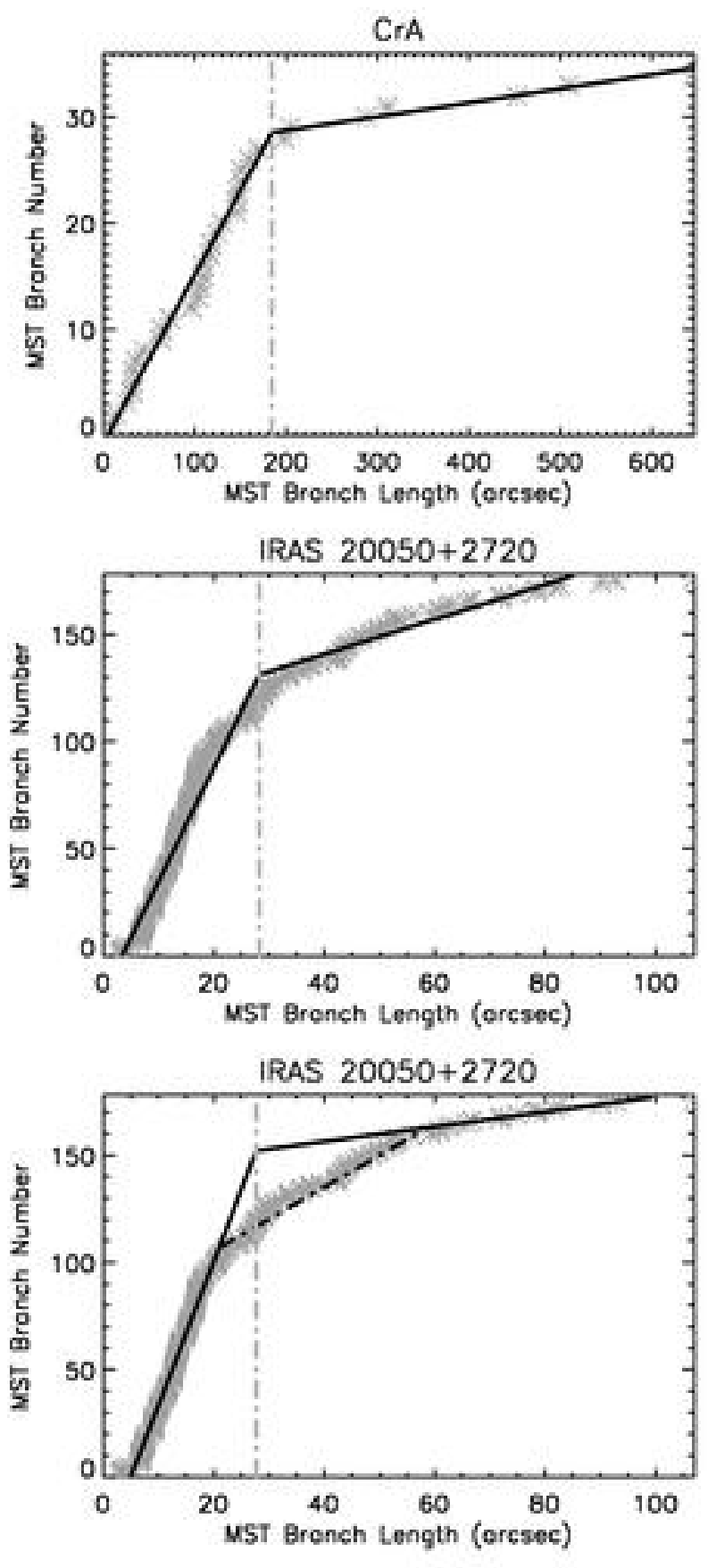}
\caption{Example MST critical length analyses for two survey clusters.  The data plotted are the cumulative distribution functions, with the sorted length values on the vertical axis and a rising integer counting index on the horizontal axis.  The top plot demonstrates a ``true'' two-line fit to CrA, where the fit is a reasonable characterization of the data because of the lack of a curved transition region.  The middle plot is also a ``true'' two-line fit, but for IRAS~20050+2720, demonstrating the poor fit quality because of the curved transition.  The bottom plot shows the three-line refinement for IRAS~20050+2720 and its condensed, two line counterpart for the purpose of finding a single MST critical length.\label{mstcdf}}
\end{figure}

We first plot the cumulative distribution function of the MST branch lengths
(the identical process can be performed with the NN2 link lengths).  In this
representation, the cumulative distribution can be approximated by two or three
line segments
(see Figure~\ref{mstcdf}): a steep-sloped segment at short spacings, a
transition segment that approximates the curved character of the intermediate
length spacings, and a shallow-sloped segment at long spacings.  By number, most
sources are typically found in the steep regime in our survey, where source
spacings are small, e.g. in a clustered subset of the population.     
%
%
%

To isolate the local clustered regime in each region, we fit two or three line
segments via a custom iterative fitting routine in IDL.
In the case where there is no evidence for a transition regime, e.g. CrA in
Fig.~\ref{mstcdf}, a true two line fit is used.  This is often a poor fit to
the data, however, e.g.  IRAS~20050+2720.  Here we see the curved transition
region between the shallow and steep slope linear portions.  Since the middle
regime is typically poorly characterized by a line segment, we ignore this
piece of the fit, and extend the shallow and steep line segments to connect
together\footnote{Improved fitting and characterization techniques for this
process are in development and beyond the scope of this paper.}.  With this
method, we effectively obtain a two line characterization of the sorted MST
branch lengths for every region observed.  We then adopt the MST branch length
of the intersection point between the two lines as the critical cutting length
for that region.    

Once the critical length is adopted, we perform two analyses for each region:
one ``full field'' analysis encompassing every YSO we have detected, and one
for the cluster ``core'' or ``cores'' we isolate from the lower density star
formation by clipping MST branches longer than the critical length found above.
As an example, see Figure~\ref{mstex} for the resulting core isolation process
applied to IRAS~20050+2720.  When applied across our entire sample of 36 observed regions, we isolate 39 cluster cores containing 1573 of 2548 (62\%) of the YSOs identified in our survey.

\subsubsection{Clustered/Distributed Population Separation Modeling\label{mstchop}} 

\begin{figure}
\epsscale{1}
\plotone{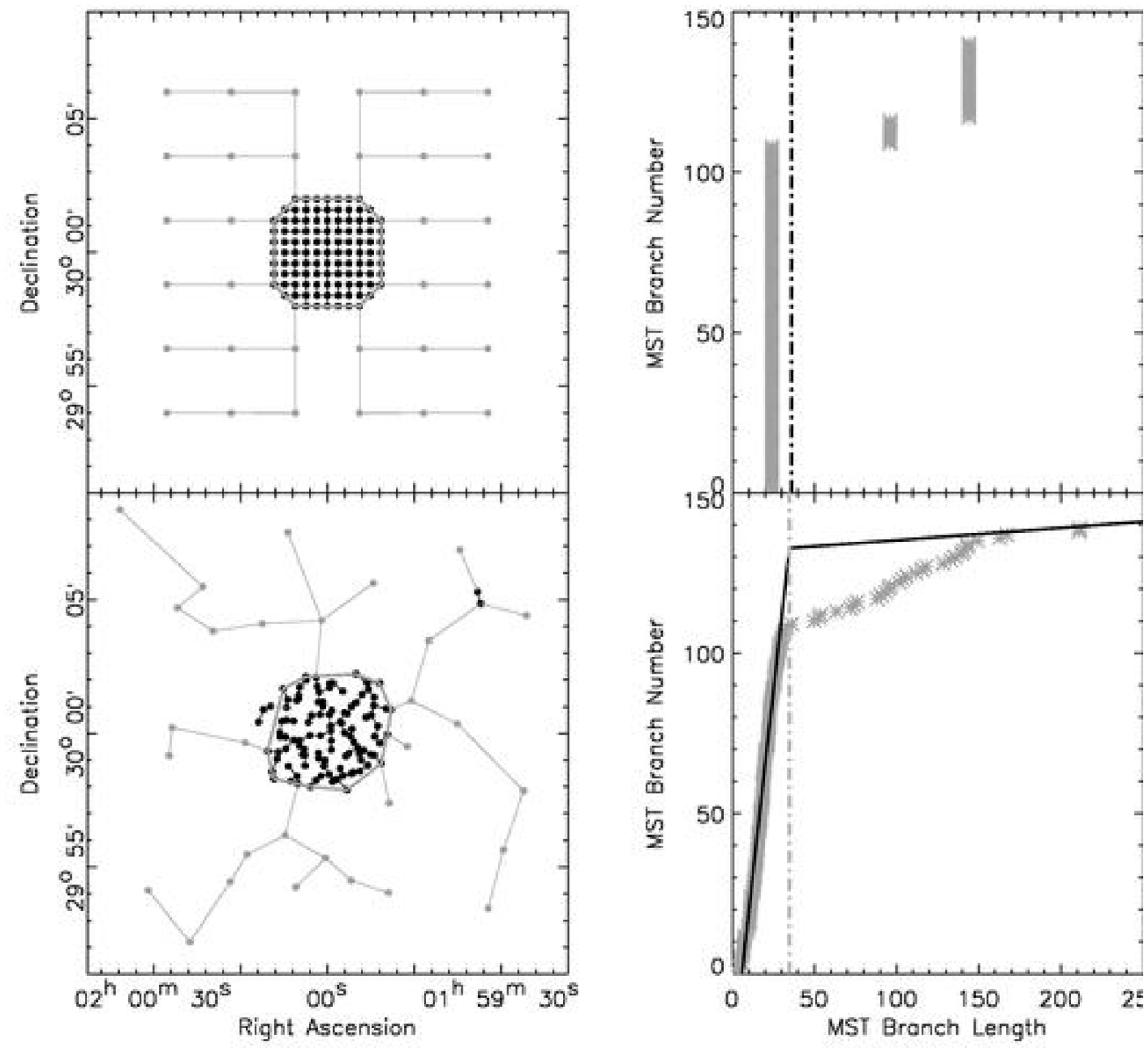}
\caption{Grid cluster/background models for demonstrating and testing the cluster extraction technique as applied to a region with a single round core of uniform surface density.  The top row is the noiseless grid model, and the bottom row is for the same grid model with Gaussian distributed offsets applied to each source away from its grid position.  Three-line to two-line fits are shown in the right column.  The computed MST critical lengths are applied to get the groupings plotted in green.\label{gridmodel}}
\end{figure}

\begin{figure}
\epsscale{1}
\plotone{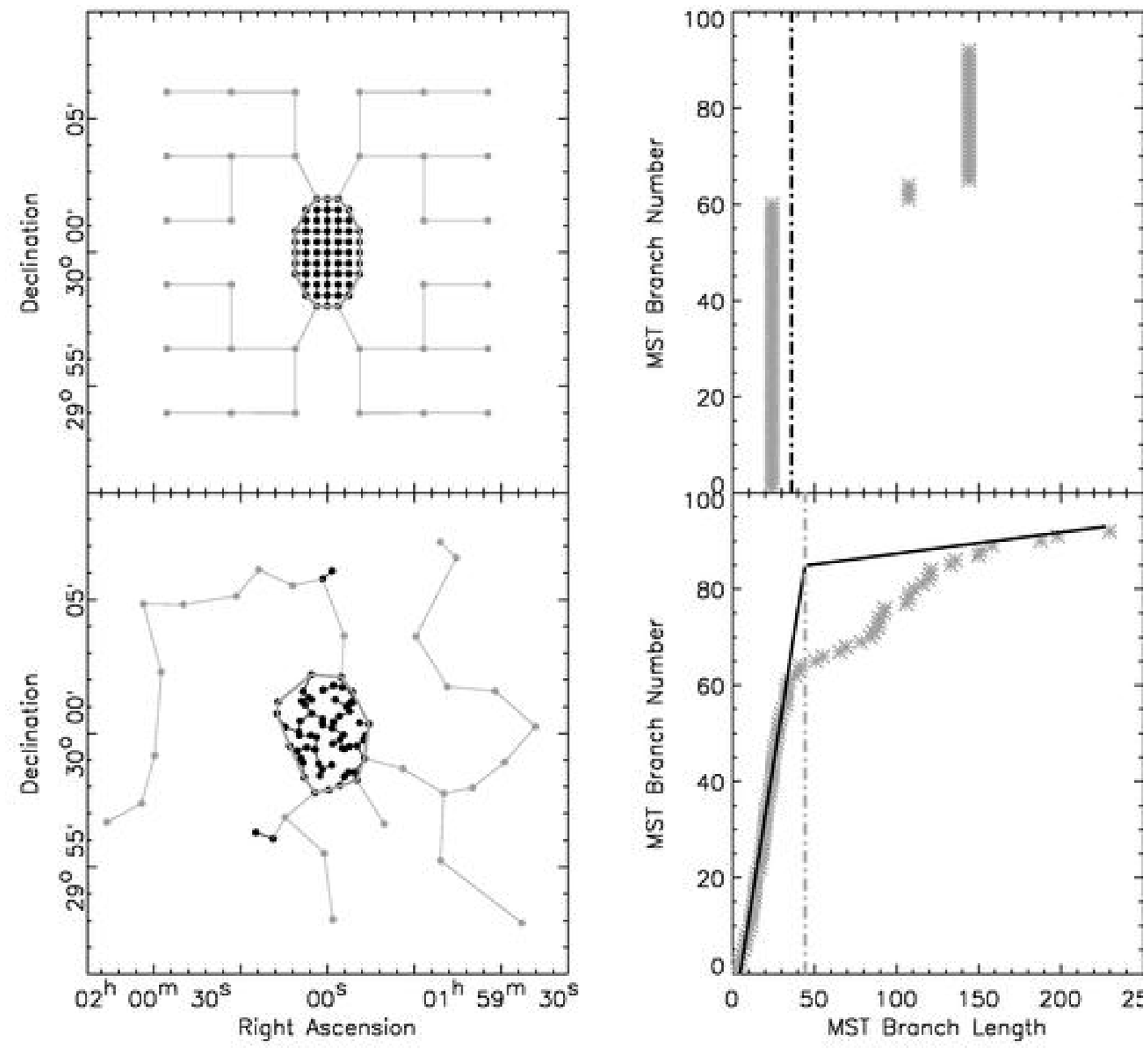}
\caption{Grid cluster/background models for demonstrating and testing the cluster extraction technique as applied to a region with a single elongated core of uniform surface density.  See Fig.~\ref{gridmodel} for a detailed description of the four panels.\label{gridmodel2}}
\end{figure}

\begin{figure}
\epsscale{1}
\plotone{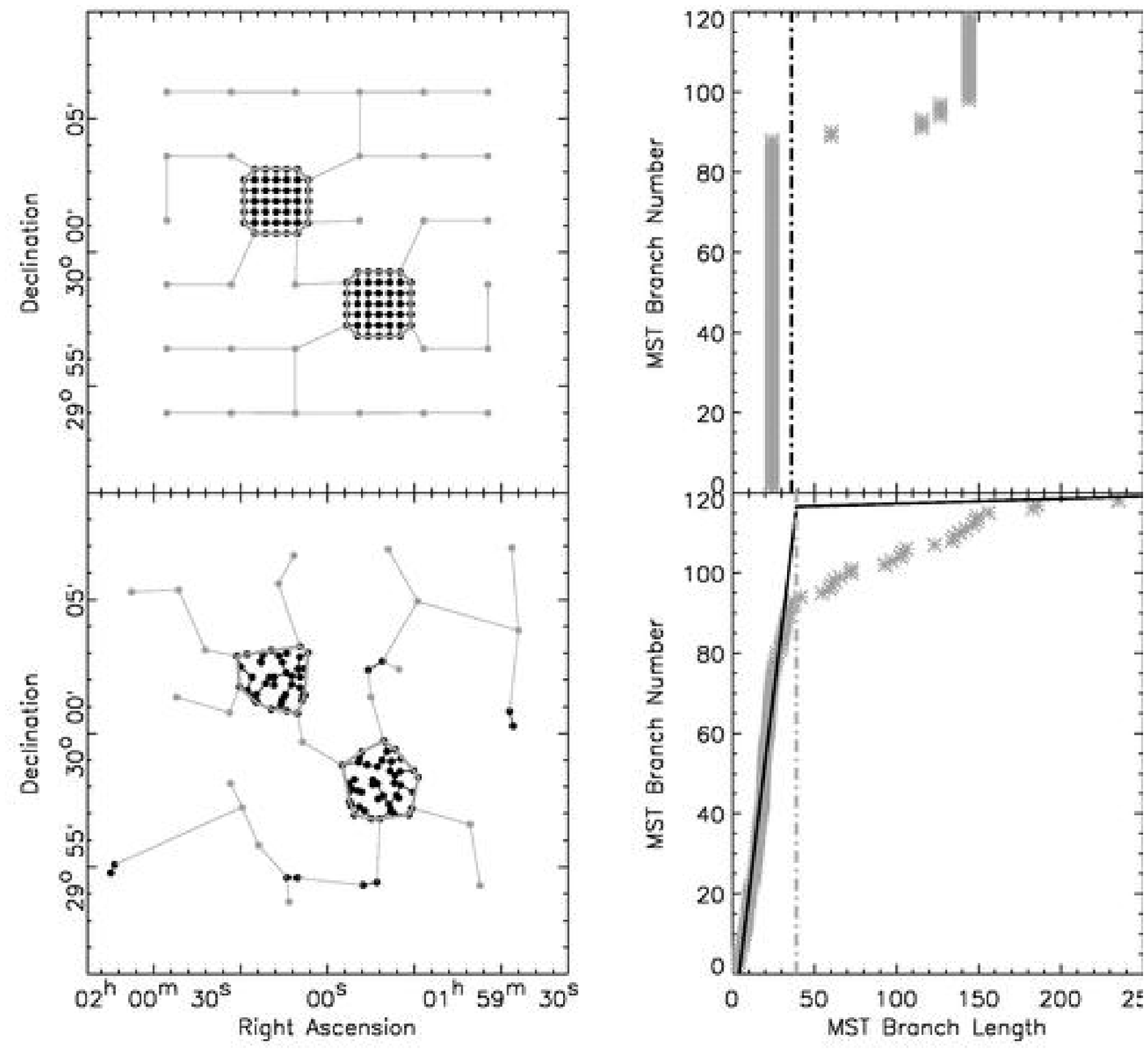}
\caption{Grid cluster/background models for demonstrating and testing the cluster extraction technique as applied to a region with a pair of round cores of uniform surface density.  See Fig.~\ref{gridmodel} for a detailed description of the four panels.\label{gridmodel3}}
\end{figure}

We apply the method to some idealized cases of cluster geometry (central round
core, central elongated core, two round cores) in
Figures~\ref{gridmodel},~\ref{gridmodel2},~\&~\ref{gridmodel3}.  The models are simply artificially
uniform clusters embedded in lower density, artificially uniform background distributions.
Regardless of geometry, the sorted MST branch lengths fall into three bins:
short spacings within the cluster, long spacings in the background, and a small
set of intermediate spacings where the background objects are near to the edge
of the cluster.  In order to isolate the cluster from the background, one would
adopt a critical length longer than the typical cluster member spacings and
shorter than the intermediate spacings between the cluster and background, such
as the dot-dashed vertical line shown in the upper right plots in
Figs.~\ref{gridmodel}-\ref{gridmodel3}.  Deleting the MST branch connections longer than this
value (those in gray in the upper left plot) leaves behind the cluster(s), still
connected with the remaining MST branches, and easily isolated.  We draw a
convex hull in thick gray around the resulting grouping(s).

Making these artificial cluster fields resemble our data is achieved by adding
some noise to each grid position that is scaled by the grid spacings of the
cluster and background members (see the bottom row of plots in
Figs.~\ref{gridmodel}-\ref{gridmodel3}).  This creates a more smoothly varying distribution of MST
branch lengths.  Given the sorting aspect of the method, adding any measurable
amount of noise produces non-infinite slopes in each of these three spacing
sets, and non-uniformity in the surface density structure of the cluster such
as clumpiness or a radial density fall-off would add some curvature to the
transition zone.  Fitting a multiple-line model to the data enables us to
algorithmically determine a similar cutting length to the grid-like example
above.  When that critical length is applied to the MST, we once again isolate
the overdense grouping(s) successfully.


\subsection{Selected YSO Membership Comparisons to Existing {\it Spitzer} Surveys}

With the YSOs identified and the cluster cores isolated and characterized in
each region, we can compare our results with those of some recent, similar {\it
Spitzer} surveys covering regions presented here.  This comparison illustrates
the range of uncertainty induced by using differing data treatments, source
classification schemes, and cluster isolation and analysis techniques.

The regions L1688 and L1689 in Ophiuchus and NGC~1333 and IC~348 in Perseus
were recently studied as part of a {\it Spitzer} and submillimeter comparative
survey \citep{jorg08}.  The {\it Spitzer} data and analysis come from the c2d
Legacy Survey pipeline data products and source classification scheme that is
primarily based on a series of color and magnitude cuts to minimize
extragalactic contaminants and a 2-24~$\mu$m multiband least squares fit to
obtain the SED slope for YSO classification.  For L1688, our full field YSO
count is similar to that of the 25~$M_\odot$~pc$^{-3}$ volume density limited
``cluster'' as defined in that work \citep[147 versus 154 in][; ``our value''
versus ``others' value'' order preserved throughout this section.]{jorg08},
although our largest MST branch length corresponds to a local volume density of
just over 40~$M_\odot$~pc$^{-3}$.  However, the number of Class~I sources
reported is considerably different (33 versus 20).  The small grouping in L1689
is quite different in both total number of YSOs (5 versus 11) and Class~I
sources (0 versus 3).  

The full field results for the Perseus clusters were expected to be somewhat
more comparable since the longest MST branch lengths in our analyses of these
clusters are nearly equivalent to the density threshold adopted by \citet{jorg08}, 1~$M_\odot$~pc$^{-3}$.  In the case of
NGC~1333, we have reported several more YSOs (133 versus 115) and similar
Class~I sources (36 versus 41).  (It is worth noting here that changes to the
classification scheme of \citet{gute08} resulted in slight redutions in the YSO
count (133 vs 137) and the Class I count (36 vs 39) reported, though the
photometry is the same.)  In IC~348, we report considerably fewer YSOs (160
versus 189) and a similar number of Class~I sources (16 versus 15).  Our
limited survey of this region likely truncates its true extent, reducing the
total number of YSOs detected.  Our survey of IC~348 is more comparable in area
to that of \citet{muen07}, who further examined {\it Spitzer}-identified YSOs
with spectroscopy.  Our results for IC~348 are similar to those reported in
\citet{muen07}, both in total YSOs (160 versus 156, though only 138 were
confirmed by spectroscopy) and in Class~I sources (16 versus 20).

The identical data to what we have presented here for Serpens were analyzed by
\citet{wins07}, revealing a somewhat larger number of YSOs with infrared excess
(97 versus 110) and a somewhat larger number of Class I and flat spectrum
sources (24 versus 38).  The discrepancy in the former is likely explained by
the more conservative transition disk criteria used here.  The latter
discrepancy may stem from some combination of disagreement in the two methods
regarding the flat spectrum and Class~II source classification threshold,
including differences in accounting for variable reddening, as well as the
inclusion by \citet{wins07} of two sources detected at only MIPS 24 and
70~$\mu$m with significant submillimeter emission that we have omitted here.
Also, the Serpens Core as isolated in Section~\ref{mstcdf} is a structure of
nearly identical area (0.2~pc$^2$) to ``Cluster A'' as described in
\citet{harv07}, another paper from the c2d Legacy Survey.  However, the
difference in YSO counts is significant (55 versus 44).  YSO class was not
distinguished for the Cluster~A subset in that paper, so we have not performed
a comparison of the Class~I source counts.  The discrepancy here is most likely
explained by a more conservative set of filtering criteria used to remove
extragalactic contaminants in \citet{harv07}.  

\citet{schm08} characterized the membership and structure of L1688, Serpens,
NGC 1333, and IC348 using publically available c2d delivery catalogs.  
Class II tallies are largely consistent between our work, their work, and c2d
papers such as \citet{jorg08}, but the numbers of Class I sources reported by
\citet{schm08} for three of the four clusters are highly discrepant (33 versus
194 in L1688; 16 versus 72 in IC348; 33 versus 62 in NGC 1333).  It appears
that the catalog flags used to identify these youngest sources were prone to
contain many contaminants (N. Evans, private communication).

A small number of the more distant clusters in our survey have also been
studied via {\it Spitzer}.  L988e was studied using very similar techniques to
those used here by \citet{alle08}, although they included deep near-infrared
photometry, improving their overall sensitivity relative to this work.  They
reported 92 YSOs total, with 19 Class~I sources, while we have found 80 YSOs,
of which 13 are Class~I.  With the overall similarity in applied method, this
comparison demonstrates the clear benefit of using deep near-infrared data to
supplement IRAC photometry.  \citet{harv08} studied IC~5146 as part of the
Gould Belt Legacy Survey.  They found 79 YSOs, and 18 of those were Class~I or
flat spectrum sources down to a density threshold of 1 $M_\odot$/pc$^{-3}$ in
their ``Cluster A'', and 29 YSOs, 5 of them Class~I or flat at 25
$M_\odot$/pc$^{-3}$ in their ``Cluster A East''.  These structures roughly
correspond to the IC~5146 Core as reported here, where we found 96 YSOs, 3 of
them Class~I, at a MST branch length threshold equivalent to 22
$M_\odot$/pc$^{-3}$.  As with the Serpens comparison above, the best
explanation for these discrepant numbers of sources is the more restrictive
filter for extragalactic contaminants used by \citet{harv08}.


%

\section{Physical Measurements of Cluster Cores\label{demcore}}

For this work, we have chosen to measure several fundamental properties of the
clusters and groups studied: radial size, mean and peak surface densities for
both stars and gas, and relative evolutionary state of the population.  While
these data are useful in their own right, the ranges over which these
quantities vary through the survey is even more useful, as they inform our
current understanding of what is ``typical'' in young stellar clusters within
the nearest kiloparsec.

The values compiled in Table~\ref{ysotable3} are measured as follows.  The
circular radial size, $R_{circ}$, 
is calculated as half of the largest distance between any two members, the
radius of the minimum area circle that encloses the entire grouping.  The
effective radial size, $R_{hull}$, is the square root of the area (divided by
$\pi$) of the convex hull\footnote{The convex hull of a set of points in two
dimensions is the minimum area polygon that contains those points such that all
internal angles between adjacent edges are less than 180 degrees.} that
contains each grouping, adjusted to account for the fraction of sources that
are vertices of the convex hull
\citep[$A_{adjusted}=A_{hull}/(1-n_{hull}/n_{total})$;][]{sk06}.  This is an
effective radius that accounts for the often elongated geometries of these
regions, and the ratio $R^2_{circ} / R^2_{hull}$, also recorded in the table,
corresponds to the aspect ratio of elongated distributions (see
Appendix~\ref{app2}).  The value of the adjusted convex hull area is also
divided into the number of {\it Spitzer}-identified IR-excess sources enclosed
by the hull to obtain a mean surface density for each grouping,
$\sigma_{mean}$.  The peak density, $\sigma_{peak}$ is simply the maximum value
of the portion of the NN6 surface density map that falls within the convex hull
of each grouping.  Similarly, the mean and peak $A_K$ values are measured over
the same area in the near-infrared extinction maps.  These extinction values
are background-subtracted to account for unassociated clouds along the line of
sight that would bias these measurements. 

\begin{figure}
\epsscale{1.0}
\plotone{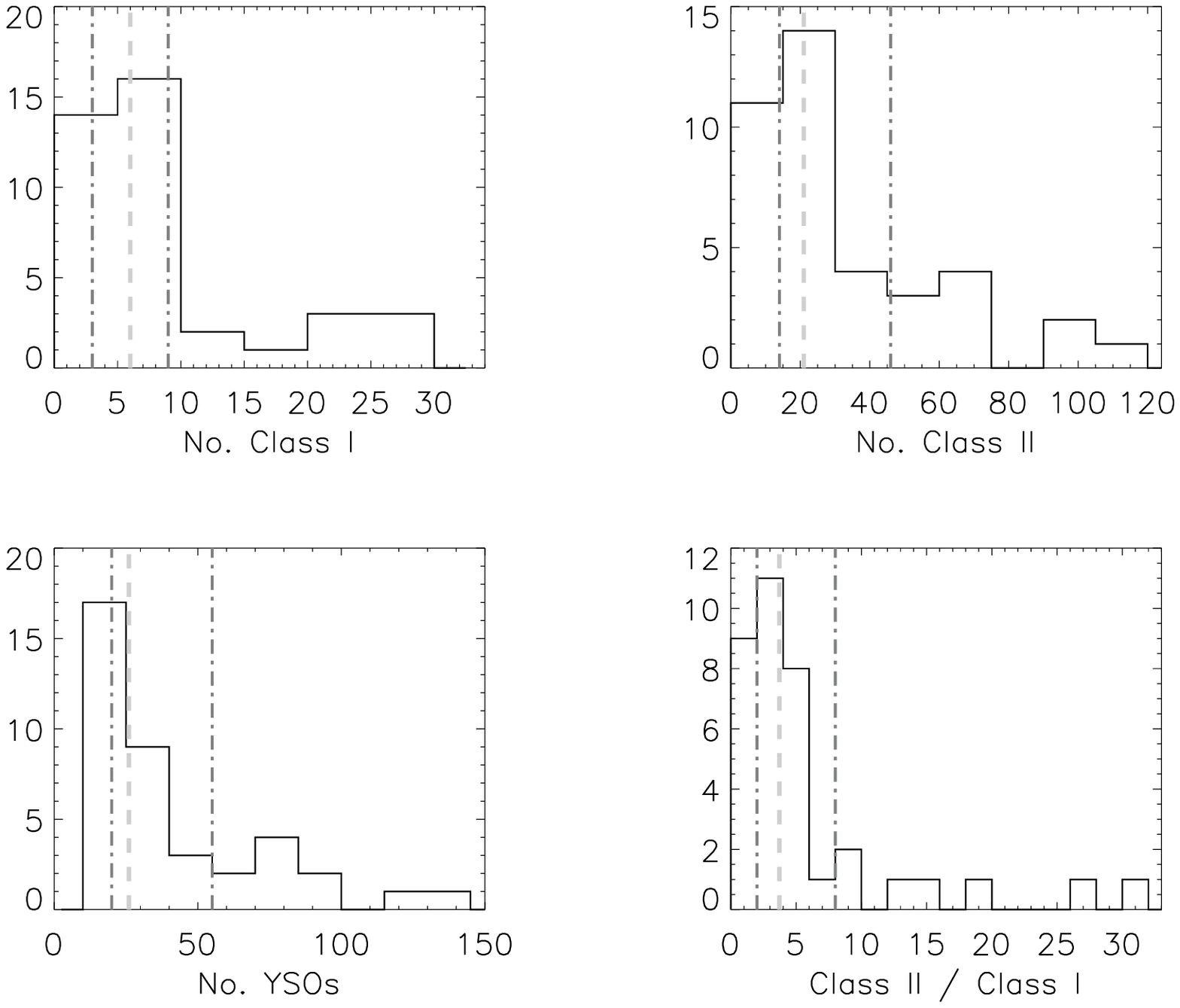}
\caption{Selected histograms of measured physical properties for the entire sample of successfully isolated cluster cores in all the studied cluster fields.  The median values are marked as light gray vertical dashed lines.  The 25th and 75th percentile values are marked on each side of the median by dark gray vertical dot-dashed lines.\label{global1}}
\end{figure}

\begin{figure}
\epsscale{1.0}
\plotone{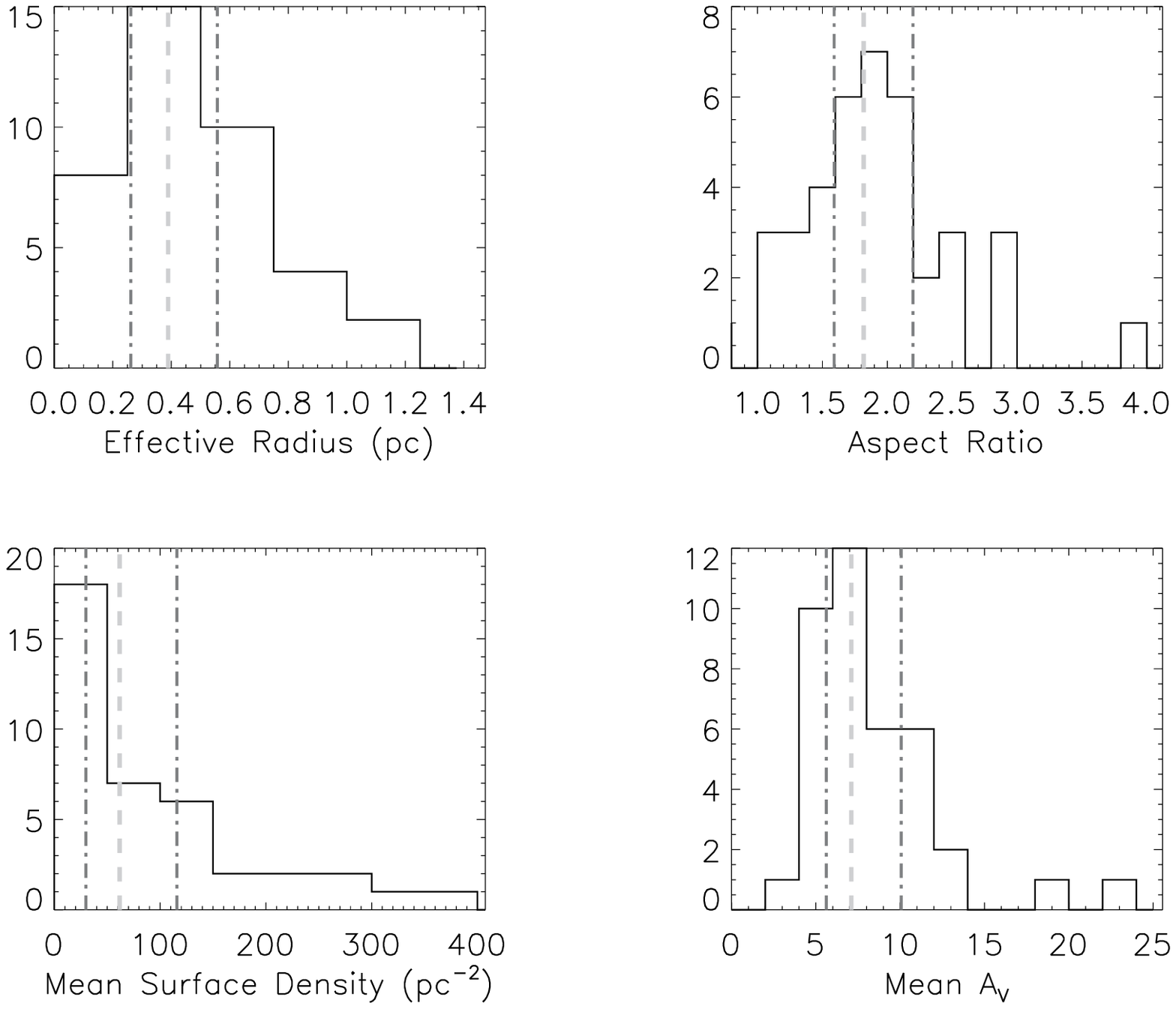}
\caption{Selected histograms of measured physical properties for the entire sample of successfully isolated cluster cores in all the studied cluster fields.  Overlays are the same as in Fig.~\ref{global1}.\label{global3}}
\end{figure}

We show the distributions for some of these fundamental quantities in
Figures~\ref{global1}~and~\ref{global3}.  Median values are marked by light
gray dashed lines and the 25th and 75th percentile values are marked by darker
gray dot-dashed lines to the left and right, respectively.  These values are
summarized in Table~\ref{restable}.

Class~I and Class~II counts are highly peaked, with medians of 6 and 21
respectively.  Both also have long tails to higher values.  The two cluster
cores with the largest number of protostars are the AFGL490~Core and
IRAS~20050+2720~Core-1, with 28 and 29, respectively.  The clump of three
cluster cores with greater than 80 Class~II sources are the cores of AFGL490,
MonR2, and IC~5146.  Most groups are small in our sample, with total YSO
membership counts peaked at groups of 10 to 25 members.  The median of the
distribution is only 26 members, but there is a long tail to larger clusters.
The two extreme outliers with over 100 members are the cores of Mon~R2 and
AFGL490.  This being a survey for young and active star-forming clusters, it
comes as no surprise that most of the cluster cores isolated have a high
fraction of protostellar sources, translating into a Class~II / Class~I ratio
peaked around the median of 3.7.  Three cluster cores with greater than 50
members stand out as having considerably higher values for this ratio:
IC348~Core-1, LkH$\alpha$~101~Core, and IC~5146~Core.

The median effective radius of the sample is 0.39~pc, with some asymmetry in
the distribution of radii to long values.  Two cluster cores, AFGL490 and
BD+40$^{\circ}$4124, are distinctly large relative to the rest of the sample,
with radii greater than 1~pc.  These regions are composed of large, relatively
uniform distributions of YSOs with no apparent cores of enhanced density.  Most
of the cores isolated in this survey are elongated, with the central 50th
percentile of aspect ratio in the range of 1.6 to 2.2.  Some small-$N$ clusters have unphysical
aspect ratios (below 1), and these are discussed briefly in
Appendix~\ref{app2}.  The two most elongated cluster cores with greater than 50
members are the S171~Core and IRAS~20050+2720~Core-1.  The latter is fairly
young (Class~II / Class~I of 3.71 and 1.89, respectively) and dense (mean
densities of 100 and 200~pc$^{-2}$) relative to the rest of the sample.  The
central 50th percentile of the survey have mean background subtracted $A_K$
values in the range of 0.6 to 1.0 mag in a relatively symmetric distribution.
The two high extinction outlying cores are the two Ophiuchus cores in L1688.
The Ophiuchus extinction maps are some of the highest quality maps presented in
this work because of the region's projection onto the galactic bulge and its
rich field star population.  Therefore we anticipate that many of the cluster
core regions of our extinction maps are saturated, leading to underestimates of
the true mean $A_K$ in these regions.  Like the number of YSOs, the mean
surface densities are skewed to low values.  With a median value of
60~pc$^{-2}$ and over 64$\%$ of the survey below 100~pc$^{-2}$, it is clear
that the mean densities of most YSO groups are rather low.  However, there is a
long tail toward large mean surface densities.  The densest two cluster cores,
Serpens~Core and IRAS~20050+2720~Core-2, range from $\sim$350~pc$^{-2}$.  These
two cores are elongated (aspect ratios of 2.16 and 2.91, respectively), small
(0.29~pc and 0.10~pc, respectively), and relatively young (Class~II / Class~I
of 1.39 and 1.75, respectively).

\begin{figure}
\epsscale{.50}
\plotone{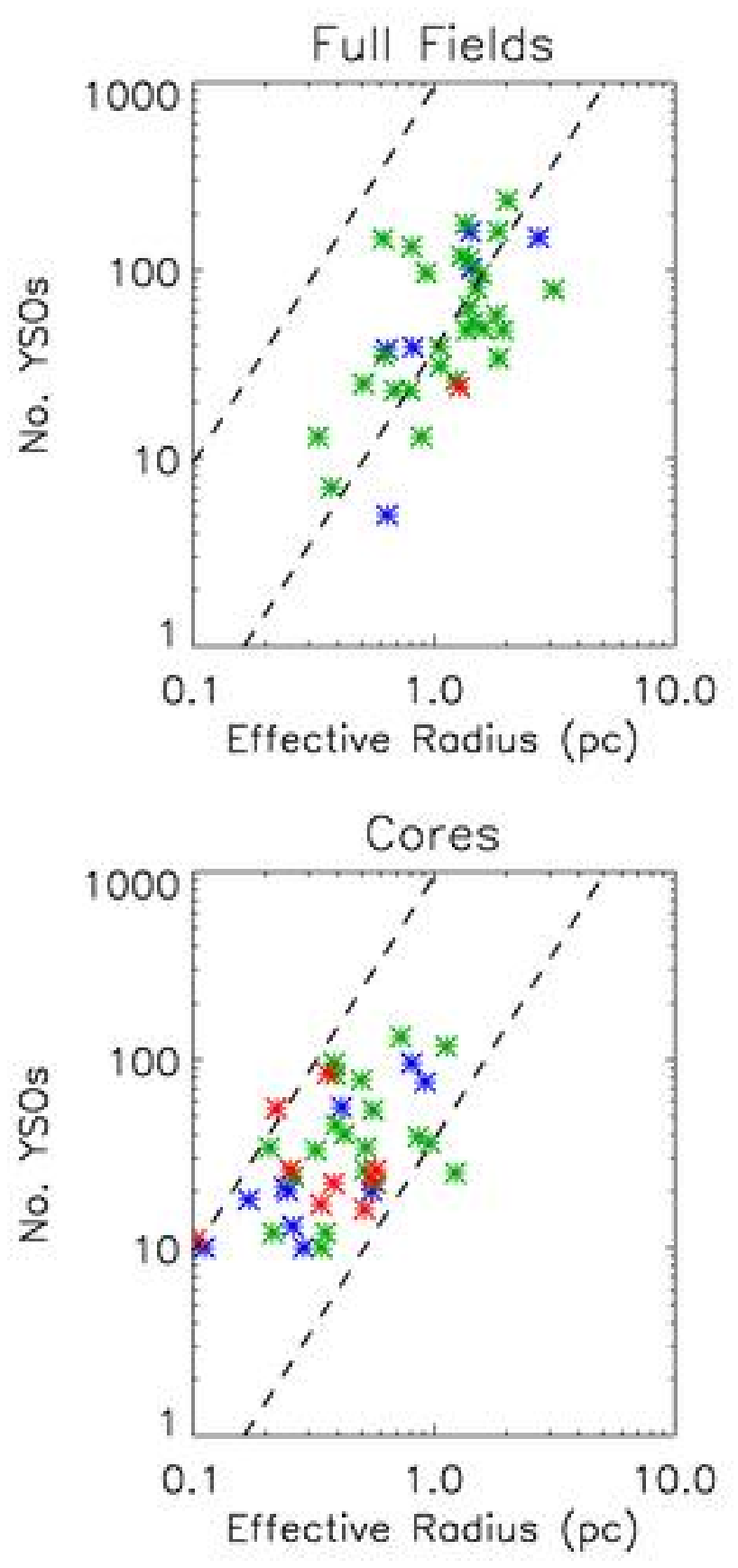}
\caption{Select correlation plots of measured physical quantities for the survey.  The color coding of both plots is indicative of each region's qualitative evolutionary state: red for young regions with Class II / Class I $<$ 2.0, green for the central 50th percentile with 2.0 $<$ Class II / Class I $<$ 8, and blue for the more evolved regions with Class II / Class I $>$ 8.  The black dashed lines mark lines of constant surface density at 12 and 300~pc$^{-2}$, the range spanned by the bulk of the cluster cores.\label{global2}}
\end{figure}

We have explored several potential correlations in the survey data, but found
little, suggesting quite wide variety in the phenomena governing a cluster
population's typical evolutionary state, spatial configuration, and natal gas
content.  We have provided an example in Figure~\ref{global2}, the number of
YSOs versus the effective radius, a common analysis published in some recent
survey papers \citep{adam06,alle07}.  The color coding of the points has been
added to show the lack of correlation between evolutionary state and mean
surface density.  Regions in the lower 25th percentile in Class~II / Class~I
ratio ($<2.0$) are plotted in red, the upper 25th percentile ($>8.0$) in blue,
and the central 50th percentile in green.  The top plot shows the full field
distribution from all the YSOs identified.  This distribution shows a similar
character to the others in the literature, e.g. a rough correspondence with a
line of constant column density.  However, when we plot the same data for the
extracted cluster cores, we see a much more uniform distribution spanning a
factor of $\sim$25 in column density.  


\section{Discussion\label{discuss}}

\subsection{Aggregated Nearest Neighbor Distance Analysis}

\begin{figure}
\epsscale{.80}
\plotone{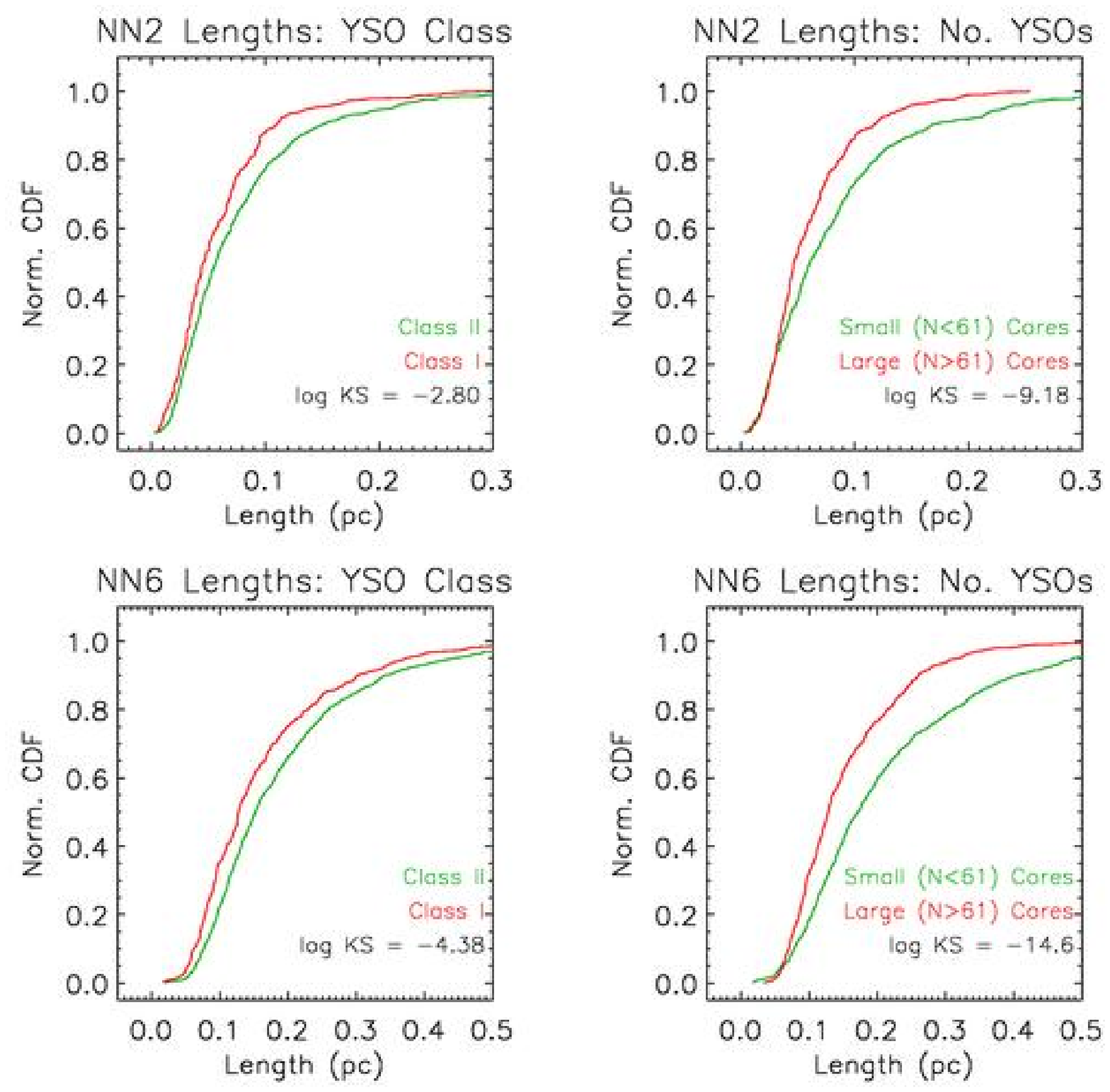}
\caption{Normalized cumulative distribution functions (CDF) of the nearest neighbor distances for all cluster core sources in the survey in aggregate.  The top row is NN2 lengths, while the bottom row is the NN6 lengths, a more robust surface density estimator.  In the right column, we have split the sample evenly by cluster size, while in the left column, we have split the sample by YSO classification.\label{nndist1}}
\end{figure}

In Figure~\ref{nndist1}, we have plotted the normalized cumulative distribution
functions for nearest neighbor distances of the YSOs in all of the cluster
cores in the survey in aggregate; the top row shows the NN2 distances, while
the bottom row is the NN6 distances, a more statistically robust surface
density estimator \citep{ch85}.  In the left column of plots, we have split the
dataset by YSO class to look for clustering differences among protostars versus
stars with disks.  Kolmogorov-Smirnov (KS) tests result in 0.1\% chance of
similarity, a $\sim$3$\sigma$ result; weak evidence that Class~I
sources are typically closer to their nearest neighbor(s) than Class~II
sources.  Such a difference was shown much more clearly for one member of the
survey, Serpens, in a recent paper \citep{wins07}.

We have also split the dataset at the 50$^{th}$ percentile by cluster core size
(right column of plots) to examine the similarity of the source spacing spectra
of large and small membership clusters.  The KS similarity tests for both the
NN2 and NN6 distances are conclusively different; the aggregated sources in
$N>60$ member cluster cores in the survey are typically much closer to their
neighbors, with the effect even more pronounced for NN6 than for NN2.  Since
these are projected spacings, we cannot rule out the possibility that large $N$
clusters are physically larger along the line of sight and that projection
effects are the cause for the shorter spacings.

\subsection{A Monte Carlo Evaluation of the Convex Hull Structural Characterization}

The properties we tabulate in this paper are those determined from the cluster
cores extracted via our MST analysis, i.e. the area and aspect ratio of each
core as described by the convex hull of the source positions, and the number of
Class I and II sources within the hull.  Thus, it is an important question
whether the mean surface density and the convex hull comprise a complete
description of a cluster core's structure, or whether there is significant
sub--structure which has not been characterized in our analysis.  The only
statistic tabulated in this paper which probes the structure {\it interior} to
the cluster core is the peak density.  The fact that the peak densities are
higher than the mean densities suggest that there may be sub--structure,
although this could also be a result of statistical fluctuations in a random
distribution of sources.  

To test for the presence of sub--structure in a statistically robust manner, we
have modeled the observed cumulative distribution functions for the aggregated
NN2 and NN6 distances of the YSOs in all of the large ($N>60$) and small
($N<60$) cluster cores in the survey, as described above.  The aggregate
nearest neighbor distance distribution models were constructed by
redistributing all the sources in the cluster cores uniformly and randomly
within their area.  Specifically, we approximated the boundary for each cluster
core as an ellipse with the area and aspect ratio of each cluster core's convex
hull.  As in the previous analysis, the clusters were segregated into large and
small clusters, and for each set of clusters, the model distributions of the
aggregated NN2 and NN6 distances were constructed.  This procedure was then
repeated using the NN2 and NN6 distances calculated from the model data for
each of the 1000 interations of the simulations.

\begin{figure}
\epsscale{.80}
\plotone{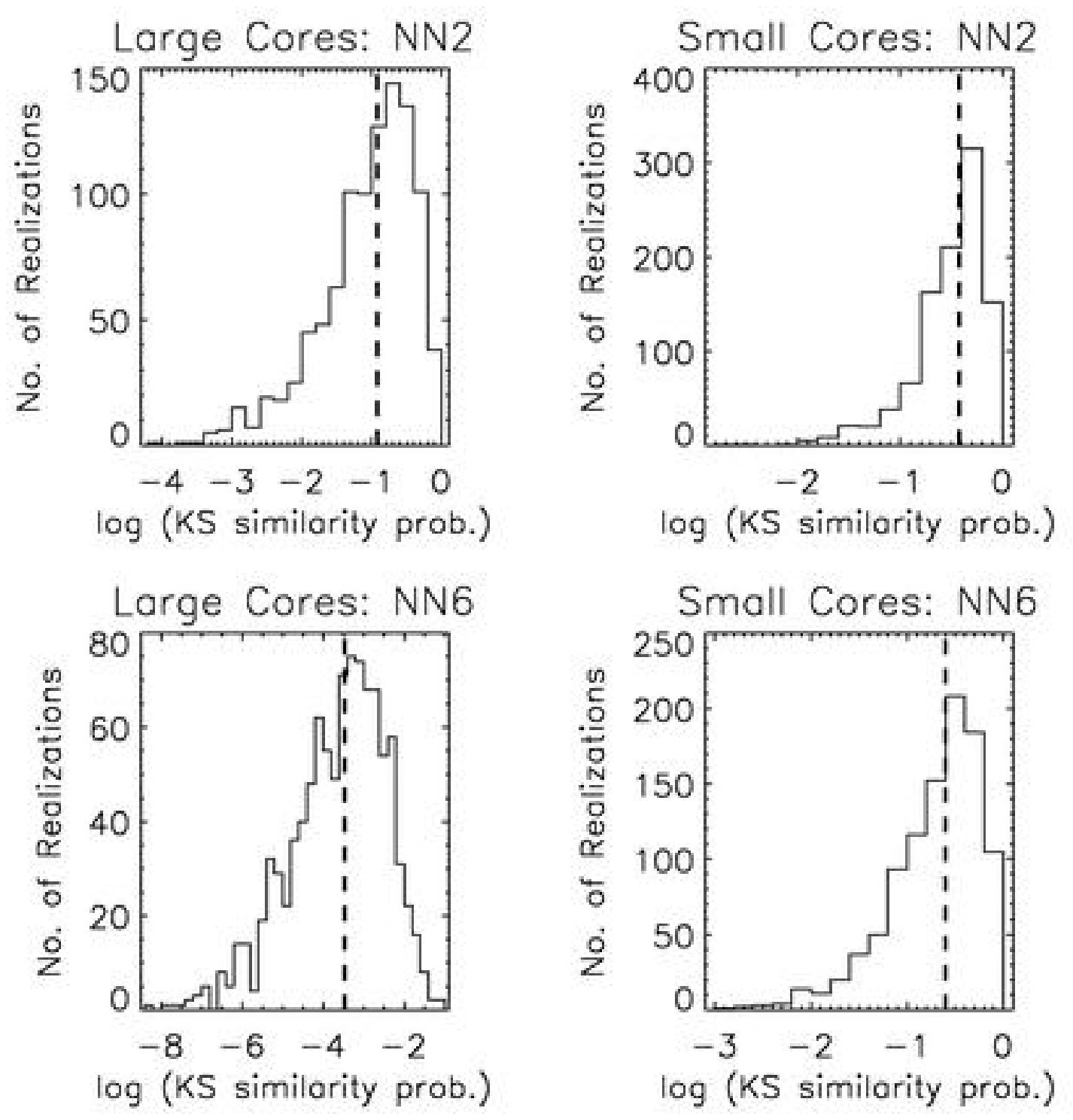}
\caption{Histograms of KS similarity probabilities for 1000 realizations of uniform surface density random source distribution models, compared to the aggregate data subdivided by cluster core size ($N=61$) as in Fig.~\ref{nndist1}.  The dashed vertical lines mark the median values of the distributions.\label{kshist}}
\end{figure}

A KS test was performed comparing the observed cumulative distribution to each
of the 1000 model realizations and a probability was calculated for whether the
observed and model data were derived from the same parent distribution.  The
histograms of these probabilities are shown in Figure~\ref{kshist}.  For the
large clusters, the distribution of sources are highly unlikely to be uniformly
distributed in the elliptical cluster boundaries.  This demonstrates the
presence of sub--structure within the large cores, although the nature of the
structure requires further study.  Interestingly, the smaller clusters do not
show a significant difference from a random distribution.  In aggregate, they
appear to be well described by three physical qualities: mean surface density,
area, and aspect ratio.

The differences demonstrated between the aggregate nearest neighbor distance
subsets do not imply that {\it all clusters} of any subset are consistent or
inconsistent with the simple three parameter model.  Further investigations of
the structure of individual clusters using this method as well as other cluster
structure indicators such as the AAP \citep{gute05} and Q \citep{cw04}
parameters, which are sensitive to various asymmetries in source distributions
but are also more sensitive to observational limitations and statistical biases
\citep{gt05,bast08}, will be performed in a future paper. 




\subsection{Similar Spacings of YSOs}

Many of the clusters in the present sample show a concentration of sources with nearest neighbor projected spacing of several times 0.01~pc, as shown in the histograms of Figure~\ref{nnhists}.  The median spacing of the YSOs in each cluster core, listed in Table~\ref{ysotable3f}, has a  mean and standard error  0.072~$\pm$~0.006 pc over 39~clusters.  This tendency toward similar spacings does not appear to be a consequence of resolution, since at a typical cluster distance  of 500 pc, the effective resolution is $\sim$0.01~pc.

The similarity of protostar spacings along linear filaments was interpreted as a signature of Jeans fragmentation in NGC~2264 \citep{teix06}, and the similarity of spacings of Class~I and Class~II YSOs in NGC~1333 was interpreted as indicating that Class~II YSOs have not moved enough since birth to erase their initial distribution of spacings \citep{gute08}.

The present sample is much larger than those in NGC~1333 and NGC~2264, but it still displays a similarity of YSO spacings, including both protostars and Class~II YSOs.  Since this tendency is now evident in a large sample, it will be important to investigate it as a well-defined property of young clusters.

Jeans fragmentation of a uniform isothermal medium \citep[][, equation 13-33]{spit78} is a starting point for discussing YSO spacing.  The spacing 0.072~pc is equal to the Jeans length for initial temperature 20~K and density $2 \times 10^5$~cm$^{-3}$.  These properties are typical for the regions of infrared dark clouds believed to be forming young clusters \citep{rjs06}.  It will be useful to see whether simulations of thermal and turbulent fragmentation in cluster forming regions can reproduce the distributions of YSO spacings reported here.

\section{Biases\label{bias}}

While this survey and its associated data products are some of the most
sensitive and uniform of their kind, they do suffer from several sources of
bias, and these biases should be considered carefully when interpreting
these data.  

\subsection{Distance}

The heliocentric distances to the regions in this survey vary by over an order
of magnitude.  Such variation has two effects on the results presented here.
First, for a given typical minimum separation of sources in angular units, the
corresponding physical separation limit scales directly with distance.
Therefore, we cannot resolve as closely spaced sources in the more distant
regions as we can in the closer regions.  Second, for a given observational
depth and a given mean age of cluster members, assuming a consistent initial
mass function from region to region, the mass sensitivity limit is higher in
our observations of the more distant regions.  

For example, assuming that a 1~Myr-old diskless pre-main sequence star
\citep[using the models of][]{bara98} has a color of zero magnitudes for the
$K_S$ band and all IRAC bandpasses, and assuming that our four-band IRAC
detection completeness (Phase~1) is limited to a typical 8.0~$\mu$m 90\%
differential completeness limit of 12$^{th}$ magnitude, we would be sensitive
to a 0.040~$M_{\odot}$ diskless brown dwarf at the distance of Taurus (140~pc),
but only a 0.50~$M_{\odot}$ at the distance of Mon~R2 (830~pc).  However, our
mass sensitivity to Class~II YSOs is somewhat better (ie. lower) than these
values in practice.  We have isolated only sources with excess infrared
emission in the IRAC wavebands in this work, making our sources brighter for a
given mass than the adopted model would predict.  In contrast, the mass
sensitivity limits for deeply embedded sources (e.g. protostars) are likely
higher than the estimates presented, but the mass limits for protostars are
highly dependent on many uncharacterized physical parameters (e.g. accretion
rate, inclination, etc.) and cannot be robustly estimated from these data.  

The mass sensitivity of our Phase~2 classification scheme, which demands at
least $H$, $K$, 3.6, and 4.5~$\mu$m photometry, is typically limited by the
$K_S$-band sensitivity of 2MASS ($K_S$~14).  Under the same model assumptions,
that would improve the Taurus completeness to well below 0.025~$M_{\odot}$ and
the Mon~R2 completeness to 0.11~$M_\odot$, at the cost of increased limitations
at higher extinctions and younger YSO evolutionary states (e.g. protostars).

\subsection{Source Confusion}

Bright infrared sources, bright structured nebulosity, and high source
densities are all quite common at the centers of young clusters.  Given the
angular resolution of 2MASS and {\it Spitzer}, we expect there to be varying
degrees of reduced source completeness as a function of the fluxes of neighbor
sources, the flux and structure of any nearby nebulosity, and the angular
proximity of the source to any other sources.  This bias will have a
considerable effect on the reported peak YSO surface densities as well as the
higher density portions of the nearest neighbor surface density maps.  We
expect only a modest effect on mean YSO surface densities and overall member
counts.  Treating these effects would demand careful modeling of the
completeness by band and position.  While such treatment is outside the scope
of this paper, some method development has already begun, with promising
results \citep{gt05}.

\subsection{Unidentified Class~III (Diskless) Members}

There is some uncharacterized number of diskless YSOs in each region that we
have not attempted to identify nor include in the spatial distribution and
surface density analyses presented here.  While the fraction of such sources is
expected to be small and relatively spatially uniform in the youngest ($<$1~Myr
old) clusters \citep[e.g.][]{gute08}, the membership of the more evolved
regions in the survey, such as IC~348, may be composed of over half diskless
sources \citep{muen07}.  Furthermore, we cannot rule out the possibility that
these sources may be distributed somewhat differently than the YSOs with IR
excess.  Neglecting these sources affects the reported mean surface densities
and membership counts by a factor of each region's disk fraction and may have
some impact on the reported peak surface densities and surface density maps.
Without a complete X-ray survey that is equivalent in area and sensitivity to
{\it Spitzer} observations \citep[as noted by][]{muen07} or a combined
photometric and spectroscopic survey at visible wavelengths for young members
in and around all the surveyed star-forming regions \citep[e.g.][but note much
higher extinction in many of the regions presented here]{hern07}, we can only
infer the presense of these diskless sources statistically from their
overdensity above the field star density.  While such efforts are being
pursued, there is considerable effort still needed to obtain and analyze such
data. 

\subsection{YSO Class Confusion}

Recent radiative transfer modeling work has been used to argue that most flat
spectrum YSOs can be explained as inclined pre-main sequence stars with disks
as opposed to bona fide low envelope density protostars
\citep[e.g.][]{robi07,crap08}.  As a specific example, \citet{crap08} argued
that of an ensemble of pre-main sequence stars with disks, $>39\%$ would be
confused as flat spectrum sources if classified by the slope between their
fluxes at 2 and 24~$\mu$m.  Since we group flat spectrum sources with more
extreme protostars in this work, such severe YSO class ambiguity would yield a
large systematic uncertainty in the Class~II to Class~I source counts and the
ratios of those counts, used here to measure qualitative evolutionary states of
the clusters presented here.  However, with our data, we can put some rather
strong limits on the severity of this issue.

One can use the high YSO counts and ancillary data of IC~348 and IC~5146 to
argue that the likelihood of confusion from inclined disks is small.  Due to
their low mean local extinction and association with visible wavelength
reflection and mid-infrared PAH feature nebulosities, we can consider IC~348
Core-1 and the IC~5146 Core to be regions where local star formation has
recently ceased \citep[This assumption is supported in IC~348 by the detailed
star formation history analysis presented in][]{muen07}.  From this assumption,
we can determine an upper limit to the YSO class confusion effect by assuming
that all protostars identified in these two regions are instead inclined stars
with disks and thus mistaken classifications.  This is an upper limit only,
however, since these sources may indeed be bona fide protostars.  That caveat
aside, our analysis shows that IC~348 Core-1 and the IC~5146 Core have $3.6\%
\pm 2.6\%$ ( 2 / 56 ) and $3.1\% \pm 1.8\%$ ( 3 / 96 ) protostars respectively,
an order of magnitude lower than the $>$39\% suggested by \citet{crap08}. 

A potential weakness in the above argument is that the regions chosen could be
more evolved than is typical in our survey.  Disk evolution might be a
significant effect in IC~348 for example, reducing disk flaring as dust has
time to settle to the mid-plane in the majority of disks \citep{lada06}.  If
the severity of the YSO classification confusion is correlated with disk
flaring, then the effect may be much more significant in young, more active
regions with typically more flared disks.  \citet{lada06} argued that mean
$\alpha_{IRAC}$ is a useful indicator of disk flaring, thus we looked for a
correlation between the ratio of protostars and class~II sources as we have
classified them and the median $\alpha_{IRAC}$ of those sources we classified
as Class~II for the largest ten cluster cores ($N>60$).  We found no such
correlation.  Furthermore, the Class~II sources of the IC~5146 Core have a
median $\alpha_{IRAC}$ that is typical of the Class~II sources in
still-embedded clusters like IRAS~20050+2720, suggesting that its disks are no
more evolved than those in the youngest clusters.  Therefore, we argue that the
mean of the IC~348 Core-1 and IC~5146 Core results ($3.3\% \pm 1.5\%$ or 5 /
152 ) is a reasonable upper limit to edge-on Class~II classification confusion
for the YSO classification system used here.  This corresponds to a Class II /
Class I ratio of 30, whereas the median value for that ratio in this survey is
3.7.







\section{Summary\label{summ}}

We have presented a survey of 36 regions thought to contain nearby young
star-forming clusters and groups of YSOs.  In addition to mid-infrared images
and near-infrared extinction maps, we have identified and classified numerous
YSOs, analyzed their spacings, and performed basic spatial
distribution measurements and analyses.  We summarize the work presented here
as follows:

1. We identified and classified 2548 YSOs with no more than 50 expected to be residual contaminators such as broad-line AGN.  Of those, 2076 are classified as Class~II pre-main sequence stars with disks and 472 are classified as Class~I protostars, including flat spectrum sources in the latter group.

2. We presented a purely algorithmic method to isolate local density enhancements in point source distributions from a more diffuse, poorly sampled, and potentially varying density background that uses no smoothing.  This method was applied to our survey, extracting 39 cluster cores of 10 or more YSO members.  Of the 2548 YSOs identified, 1573 (62$\%$) are members of one of these cluster cores.  

3. We made several basic structural measurements of these cores, finding that the median cluster core is 0.4~pc in size and somewhat elongated (aspect ratio 1.8), relatively low density (60~pc$^{-2}$), small (26 members), young (Class~II / Class~I = 3.7), and partially embedded (mean $A_K$ = 0.8 mag).

4. We demonstrated that protostars are found in regions of marginally higher surface densities than the more evolved pre-main sequence stars with disks (KS similarity probability 0.2\%).  

5. We demonstrated that members of larger-$N$ ($>60$) clusters are typically found in regions of higher YSO surface densities than members of smaller-$N$ ($<60$) groupings (KS similarity probability $1.7 \times 10^{-9}$).  We did not attempt to rule out the possibility that this is a result of projection effects in the larger clusters.

6. We have reported concentrations of sources with similar projected nearest neighbor spacings in many of the cluster cores.  The mean of the median spacings of YSOs (regardless of class) in the 39 cluster cores isolated here is 0.072~$\pm$~0.006~pc.  Our large sample size lends statistical weight to the suggestion by other investigations \citep[e.g.][]{teix06} that Jeans fragmentation is a worthwhile starting point for understanding primordial structure in star-forming regions.  In that light, we infer the natal cloud properties from the mean YSO spacing and an assumed temperature of 20K, arriving at properties that are similar to those reported as probable cluster-forming clumps in infrared dark clouds by \citet{rjs06}.




\appendix

\section{Updating the \citet{gute08} Classification Scheme\label{app1}}

We have used a slightly modified version of the \citet{gute08} multiphase
source classification scheme to classify all sources detected in the survey.
For the convenience of the reader, we present the scheme here and note the
changes made over the original method.

\subsection{Phase 1: IRAC Four-Band Source Characterization\label{iracclass}}

We apply Phase~1 to all sources that have photometric uncertainties
$\sigma<0.2$~mag detections in all four IRAC bands.  We begin by separating out
IR-excess contaminants such as star-forming galaxies and broad-line active
galactic nuclei (AGN) and then isolate YSOs with IR-excess from those without
and finally split the Class~I from the more prominent Class~II YSOs.  

In order to obtain a clean sample of confident YSOs, several contaminating
source types can be isolated and removed by application of constriants in
various color spaces where these sources are considerably more prominent than
YSOs.  

First, we eliminate the active star-forming galaxies, as their strong
PAH-feature emission yields very red 5.8 and 8.0~$\mu$m colors \citep{ster05}.
We utilize customized cuts in the $[4.5]-[5.8]$~vs.~$[5.8]-[8.0]$ and $[3.6]-[5.8]$~vs.~$[4.5]-[8.0]$ color-color spaces to identify and remove these contaminants from our sample.

Any sources are considered PAH galaxies if they follow all of the following constraints:
\begin{displaymath}
[4.5]-[5.8] < \frac{1.05}{1.2}\times([5.8]-[8.0]-1)
\end{displaymath}
\begin{displaymath}
[4.5]-[5.8] < 1.05
\end{displaymath}
\begin{displaymath}
[5.8]-[8.0] > 1
\end{displaymath}
\begin{displaymath}
[4.5]>11.5
\end{displaymath}

In addition, we also consider sources that obey all of the following constraints to be PAH galaxies:
\begin{displaymath}
[3.6]-[5.8] < \frac{1.5}{2}\times([4.5]-[8.0]-1)
\end{displaymath}
\begin{displaymath}
[3.6]-[5.8] < 1.5
\end{displaymath}
\begin{displaymath}
[4.5]-[8.0] > 1
\end{displaymath}
\begin{displaymath}
[4.5]>11.5
\end{displaymath}

\begin{figure}
\epsscale{.80}
\plotone{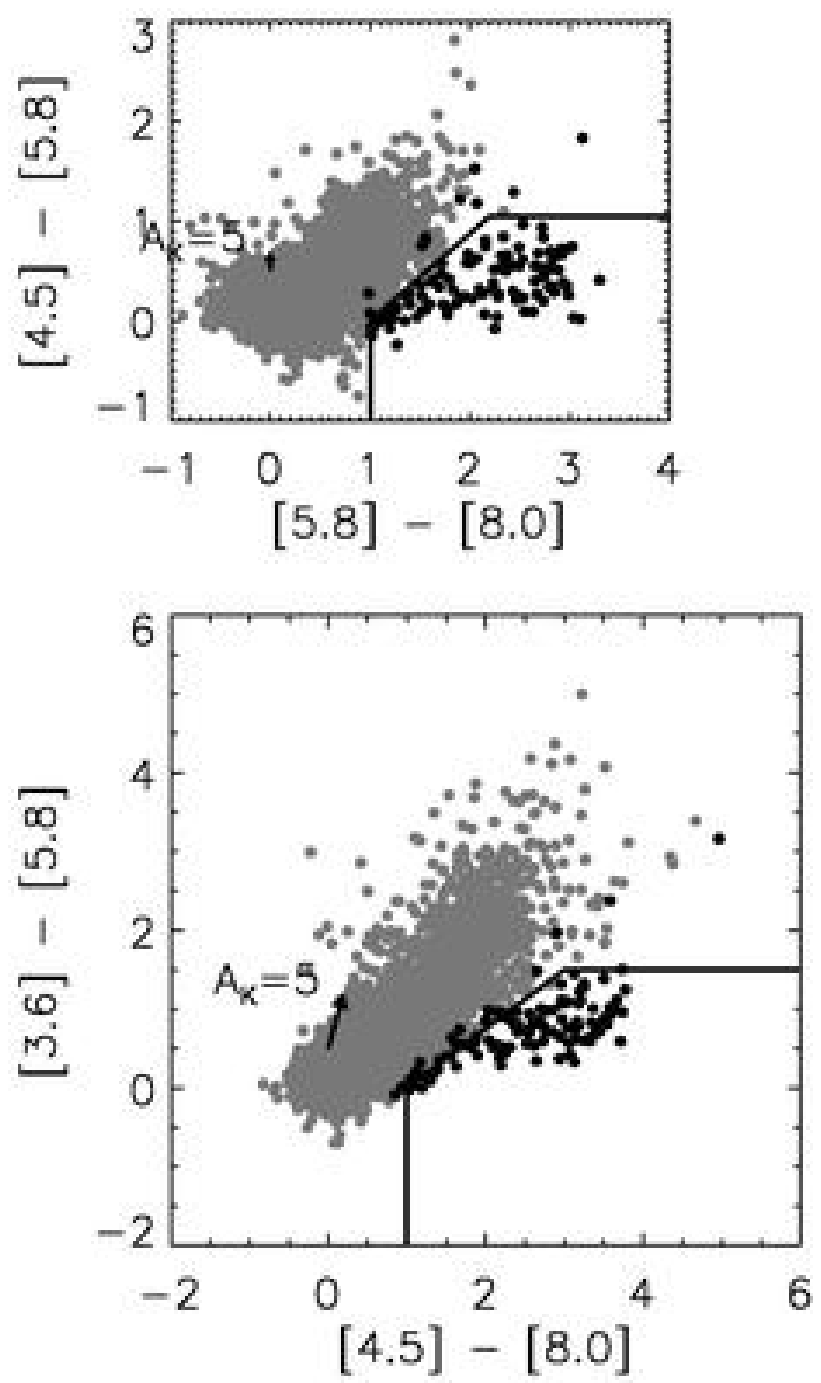}
\caption{Color-color diagrams used for the isolation of unresolved star-forming galaxies (filled black circles).\label{pahgal}}
\end{figure}

See Figure~\ref{pahgal} for displays of these constraints applied to the data.
Once a source has been flagged as a PAH galaxy, it is removed from further
consideration in the classification scheme.  Since these objects have distnct
color differences relative to YSOs, we expect there to be negligible residual
contamination from these objects in our YSO lists.

\begin{figure}
\epsscale{.80}
\plotone{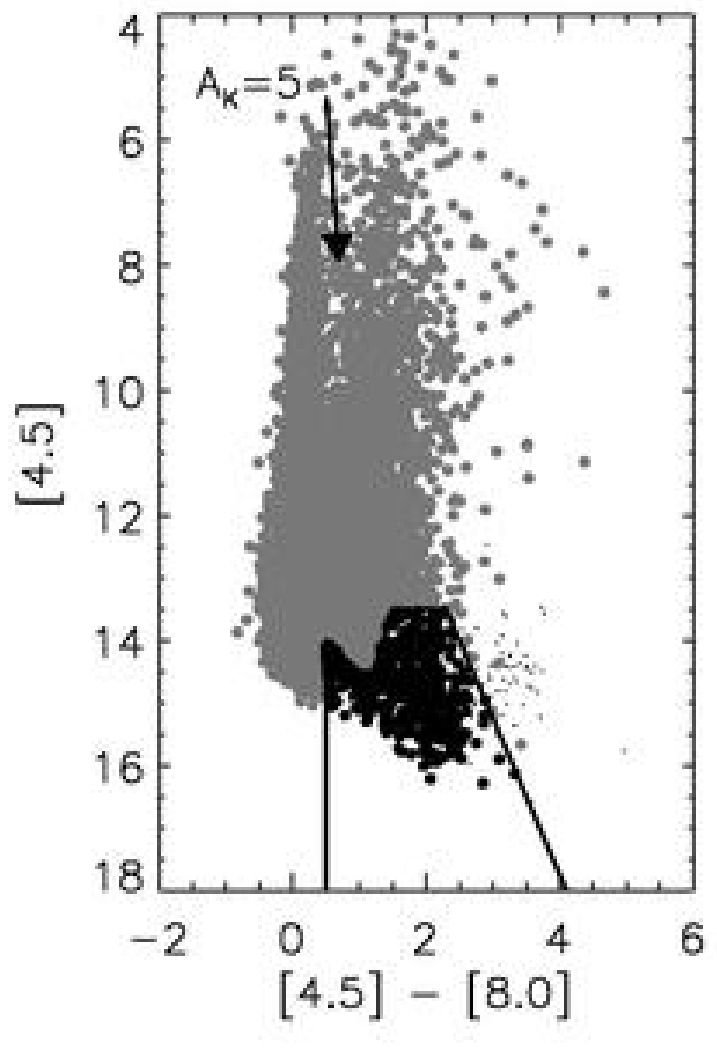}
\caption{Color-magnitude diagram used for the isolation of broad-line active galactic nuclei (filled black circles) from other unclassified sources (filled gray circles).  Previously classified sources are excluded (small black dots).\label{agncmd}}
\end{figure}

Broad-line AGN have mid-IR colors that are largely consistent with YSOs
\citep{ster05}.  We utilize the $[4.5]$~vs.~$[4.5]-[8.0]$ color-magnitude
diagram (Fig.~\ref{agncmd}) to flag as likely AGN all sources that follow all
of these three conditions:

\begin{displaymath}
[4.5]-[8.0] > 0.5
\end{displaymath}
\begin{displaymath}
[4.5] > 13.5+([4.5]-[8.0]-2.3)/0.4
\end{displaymath}
\begin{displaymath}
[4.5]>13.5
\end{displaymath}

Additionally, a source flagged as a likely AGN must follow any one of the following three conditions:

\begin{displaymath}
[4.5] > 14+([4.5]-[8.0]-0.5)
\end{displaymath}
\begin{displaymath}
[4.5] > 14.5-([4.5]-[8.0]-1.2)/0.3
\end{displaymath}
\begin{displaymath}
[4.5] > 14.5
\end{displaymath}

See Figure~\ref{agncmd} for displays of these constraints applied to the data.
Even with these cuts applied, a small number of AGN per square degree are
expected to be brighter than the magnitude-based cutoff, resulting in
contamination of our YSO list.  \citet{gute08} reduced $\sim$7.7 square degrees
of the Bootes Shallow Survey field with the Cluster Grinder data treatment in
order to estimate the residual contamination of the mitigation and YSO
classification effort presented there.  With the updated scheme presented here,
we have improved on the residual contamination estimates presented there.  The
reclassified Bootes field data yields a mean expected contamination of $4.8 \pm
0.8$ Class~I YSOs and $3.1 \pm 0.6$ Class~II YSOs per square degree, for a
total contamination of $7.9 \pm 1.0$ YSOs per square degree.

Unresolved knots of shock emission are often detected in all IRAC bands,
yielding yet another source of contaimation in our YSO samples.  Based on our
findings, all sources with photometry that obeys all of the following
constraints are likely dominated by shock emission and thus are removed:

\begin{displaymath}
[3.6]-[4.5] > \frac{1.2}{0.55} \times (([4.5]-[5.8])-0.3) + 0.8
\end{displaymath}
\begin{displaymath}
[4.5]-[5.8] \le 0.85
\end{displaymath}
\begin{displaymath}
[3.6]-[4.5] > 1.05
\end{displaymath}

\begin{figure}
\epsscale{.60}
\plotone{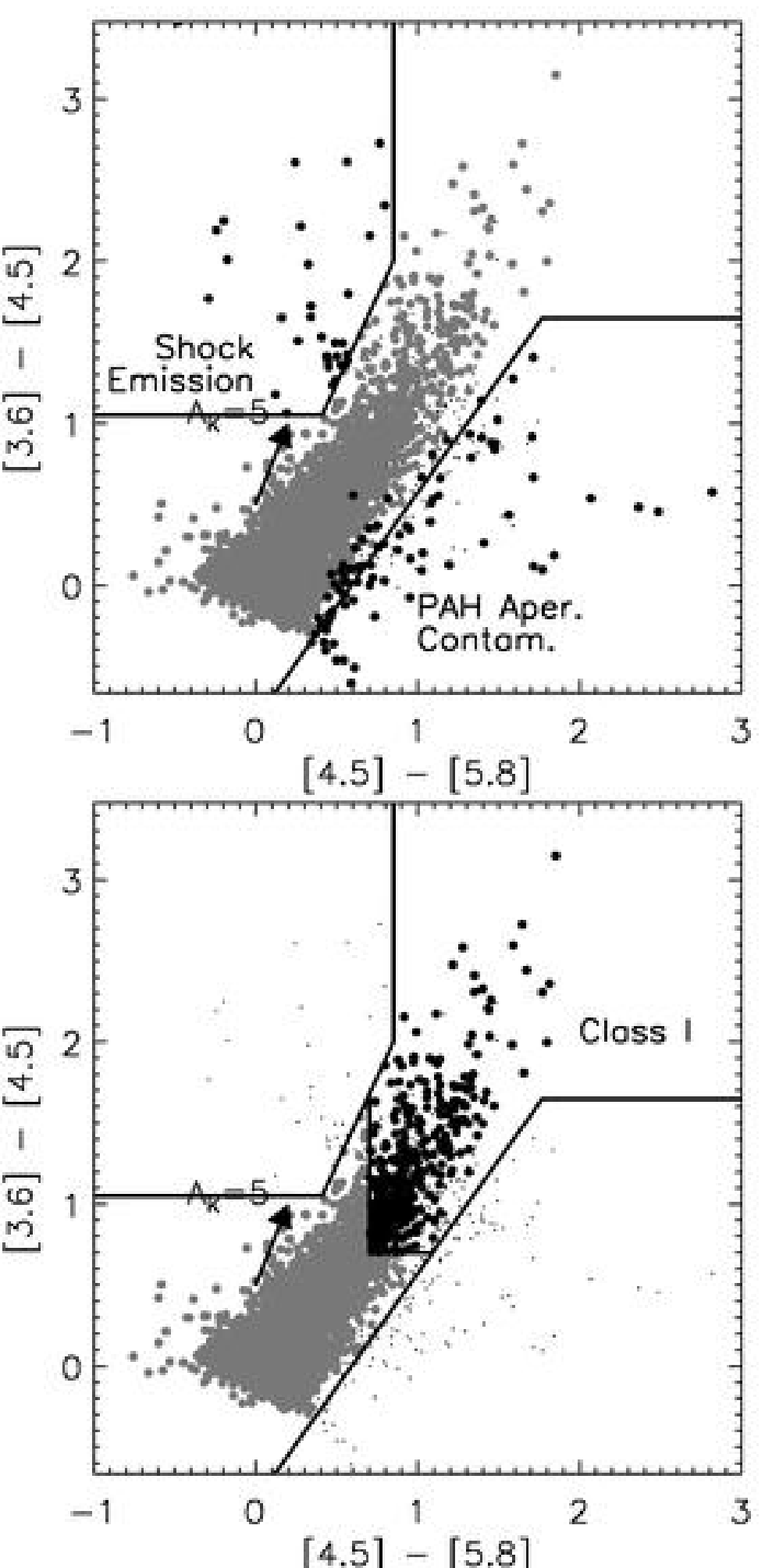}
\caption{Color-color diagram used for the isolation of unresolved shock emission knots, objects that suffer from structured PAH aperture contamination (filled black circles in top plot), and finally Class~I YSOs (filled black circles in bottom plot) from previously unclassified sources (filled gray circles).  Previously classified sources are excluded (small black dots).\label{hpccd}}
\end{figure}

See Figure~\ref{hpccd} for a display of these constraints applied to the data.  
An additional set of contaminant sources are those where resolved structured
PAH-emission has contaminated the photometric apertures of some dim field
stars, leading to spurious excess emission in the 5.8 and 8.0~$\mu$m
bandpasses.  All sources that obey all of the following constraints are
consistent with sources that have PAH-contaminated apertures:

\begin{displaymath}
\sigma_4 =  \sigma\{[[4.5]-[5.8]]\}
\end{displaymath}
\begin{displaymath}
\sigma_5 =  \sigma\{[[3.6]-[4.5]]\}
\end{displaymath}

\begin{displaymath}
[3.6]-[4.5]-\sigma_5 \le 1.4 \times (([4.5]-[5.8])+\sigma_4-0.7) + 0.15
\end{displaymath}
\begin{displaymath}
[3.6]-[4.5]-\sigma_5 \le 1.65
\end{displaymath}

See Figure~\ref{hpccd} for a display of these constraints applied to the data.  

Of the sources that remain, we identify those that obey the following
constraints as Class I YSOs, due to their particularly red discriminant colors:

\begin{displaymath}
[4.5]-[5.8] > 0.7
\end{displaymath}
\begin{displaymath}
[3.6]-[4.5] > 0.7
\end{displaymath}

See Figure~\ref{hpccd} for a display of these constraints applied to the data.  
Note that this is a somewhat simplified version of the protostar criteria of
\citet{gute08}.  It is worth noting that in rare cases, a highly reddened
Class~II source could have the colors of a Class~I as defined above.  For those
objects with 24~$\mu$m detections, we can extract these interlopers and
reclassify them (see Section~\ref{mipsphase} below). 

With those objects extracted, we finally can extract the
Class~II YSOs from the remaining field stars by identifying those sources that
follow these constraints:

\begin{displaymath}
[4.5]-[8.0]-\sigma > 0.5
\end{displaymath}
\begin{displaymath}
[3.6]-[5.8]-\sigma > 0.35
\end{displaymath}
\begin{displaymath}
[3.6]-[5.8]+\sigma \le \frac{0.14}{0.04} \times (([4.5]-[8.0]-\sigma)-0.5) + 0.5
\end{displaymath}
\begin{displaymath}
[3.6]-[4.5]-\sigma > 0.15
\end{displaymath}

\begin{figure}
\epsscale{.80}
\plotone{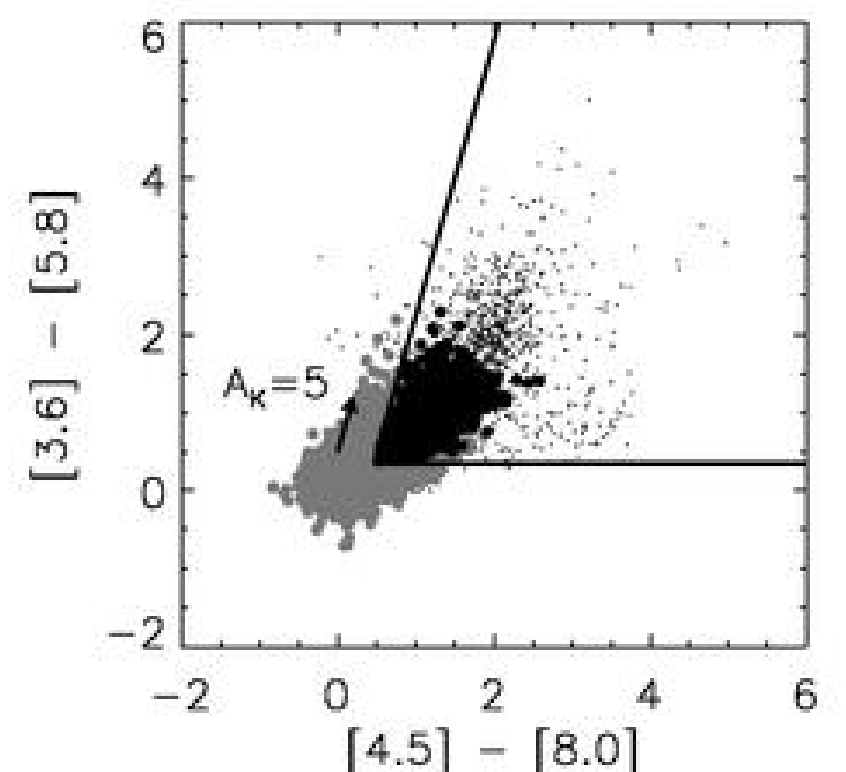}
\caption{Color-color diagram used for the isolation of Class~II YSOs (filled black circles) from previously unclassified sources (filled gray circles). Previously classified sources are excluded (small black dots).\label{c2ccd}}
\end{figure}

See Figure~\ref{c2ccd} for a display of these constraints applied to the data.

\subsection{Phase 2: Additional YSOs Identified via a $JHK_S[3.6][4.5]$ YSO Classification Scheme\label{hk12phase}}

We apply Phase~2 to those sources that lack detections at either 5.8 or
8.0~$\mu$m, but have high quality ($\sigma<0.1$~mag) 2MASS near-infrared
detections in bands ($H$ and $K_S$ are required at minimum, $J$ is used where
present) sufficient to enable distinction between sources with IR-excess and
those that are simply reddened by dust along the line of sight \citep{gt05}.  

First we measure the line of sight extinction to each source as parameterized
by the $E_{H-K}$ color excess from line of sight extinction from dust.  For
objects where we have $J$ photometry in addition to $H$ and $K_S$, we utilize
the $\frac{E_{J-H}}{E_{H-K}}$ color excess ratio with baseline colors based on
the Classical T~Tauri Star (CTTS) locus of \citet{mch97} and standard dwarf
star colors \citep{bb88}.  To accomplish the latter, we force $[J-H]_0 \ge
0.6$, a simplifying approximation for the intrinsic colors of low mass dwarfs.

\begin{figure}
\epsscale{.80}
\plotone{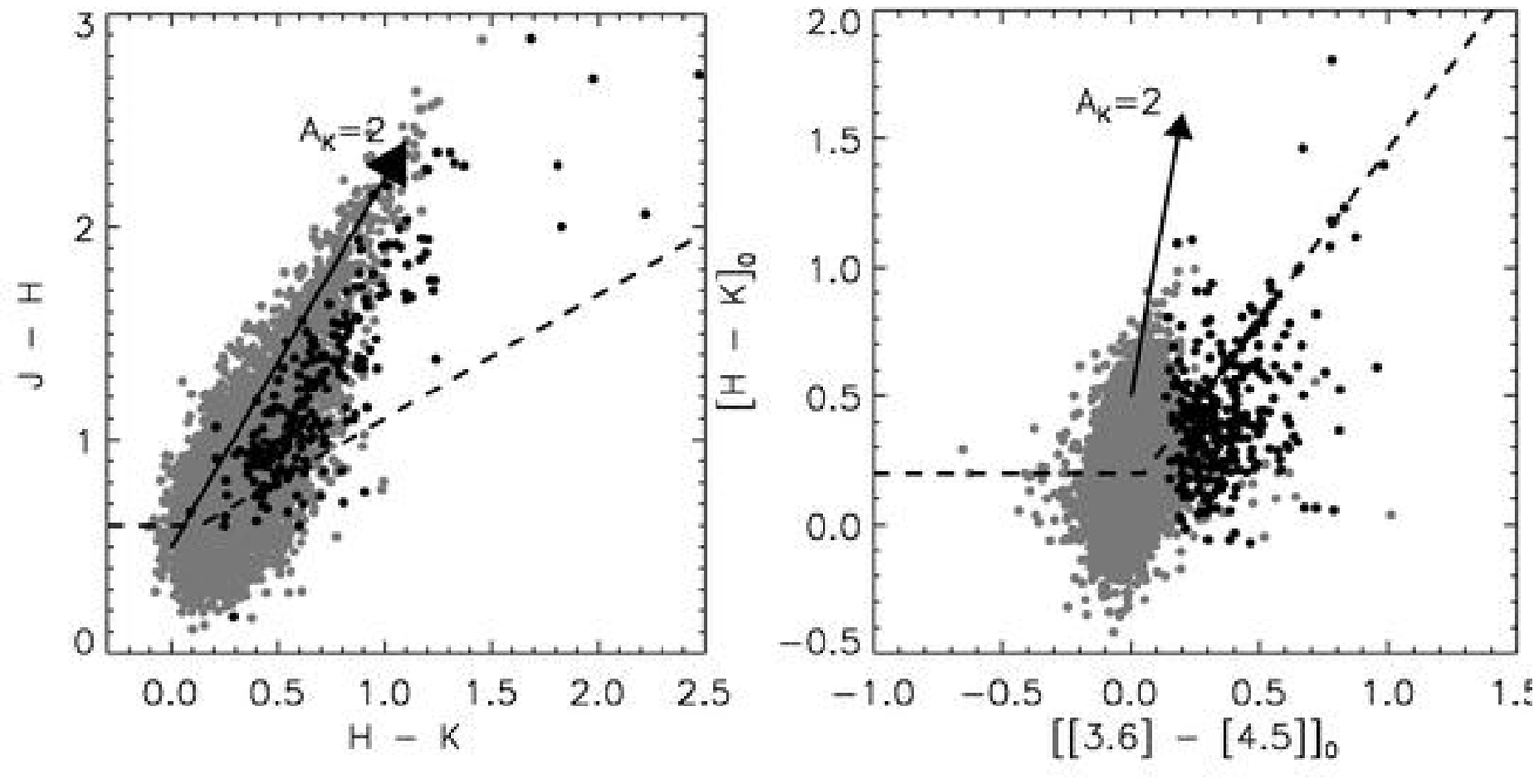}
\caption{Color-color diagrams used for measuring extinction for subsequent dereddening efforts.  YSOs isolated by the Phase~2 classification (see Section~\ref{hk12phase} and Fig.~\ref{k12c}) are plotted as filled black circles.  The reference loci adopted for the extinction measurements are plotted as dashed black lines. \label{k12dr}}
\end{figure}

Here are the equations used to derive the adopted intrinsic colors from the photometry we have measured (see left panel of Fig.~\ref{k12dr}):
\begin{displaymath}
[J-H]_0 = 0.58 \times [H-K]_{0} + 0.52 ; for [H-K]_0 > 0.14
\end{displaymath}
\begin{displaymath}
[J-H]_0 = 0.6 ; for [H-K]_{0} \le 0.14
\end{displaymath}
\begin{displaymath}
[H-K]_0 = [H-K]_{meas} - ([J-H]_{meas} - [J-H]_{0}) \times \frac{E_{H-K}}{E_{J-H}}
\end{displaymath}
\begin{displaymath}
[H-K]_{0} = \frac{[J-H]_{meas} - \frac{E_{J-H}}{E_{H-K}} \times [H-K]_{meas} - 0.52}{0.58-\frac{E_{J-H}}{E_{H-K}}}
\end{displaymath}

For objects that lack $J$ photometry, we use the
$\frac{E_{[3.6]-[4.5]}}{E_{H-K}}$ color excess ratio derived from the color
excess ratios reported in \citet{flah07}, with baseline colors based on the
$H-K vs [3.6]-[4.5]$ color-color space YSO locus, as measured by \citet{gt05},
and standard dwarf star colors \citep{bb88}. To accomplish the latter, we force
$[H-K]_0 \ge 0.2$, a simplifying approximation for the intrinsic colors of low
mass dwarfs (see right panel of Fig.~\ref{k12dr}).

\begin{displaymath}
[H-K]_0 = 1.33 \times [[3.6]-[4.5]]_{0} + 0.133\ ;\ for\ [[3.6]-[4.5]]_0 > 0.06
\end{displaymath}
\begin{displaymath}
[H-K]_0 = 0.2\ ;\ for\ [[3.6]-[4.5]]_{0} \le 0.06
\end{displaymath}

\begin{displaymath}
[H-K]_0 = [H-K]_{meas} - ([[3.6]-[4.5]]_{meas} - [[3.6]-[4.5]]_{0}) \times \frac{E_{H-K}}{E_{[3.6]-[4.5]}}
\end{displaymath}
\begin{displaymath}
[H-K]_{0} = \frac{1.33 \times (C [H-K]_{meas}-[[3.6]-[4.5]]_{meas})-0.133}{1.33 C - 1} ; where\ C = \frac{E_{[3.6]-[4.5]}}{E_{H-K}}
\end{displaymath}

Once we have measured the component of the $H-K$ color excess that is caused by reddening, we compute the dereddened $K-[3.6]$ and $[3.6]-[4.5]$ colors using the color excess ratios presented in \citet{flah07}, specifically $\frac{E_{J-H}}{E_{H-K}}=1.73$, $\frac{E_{H-K}}{E_{K-[3.6]}}=1.49$, and $\frac{E_{H-K}}{E_{K-[4.5]}}=1.17$:

\begin{displaymath}
[K-[3.6]]_0 = [K-[3.6]]_{meas} - ([H-K]_{meas} - [H-K]_0) \times \frac{E_{K-[3.6]}}{E_{H-K}}
\end{displaymath}
\begin{displaymath}
[[3.6]-[4.5]]_0 = [[3.6]-[4.5]]_{meas} - ([H-K]_{meas} - [H-K]_0) \times \frac{E_{[3.6]-[4.5]}}{E_{H-K}}
\end{displaymath}
\begin{displaymath}
\frac{E_{[3.6]-[4.5]}}{E_{H-K}} = \left\{ \left[ \frac{E_{H-K}}{E_{K-[4.5]}} \right]^{-1} - \left[ \frac{E_{H-K}}{E_{K-[3.6]}} \right]^{-1} \right\}^{-1}
\end{displaymath}

\begin{figure}
\epsscale{.80}
\plotone{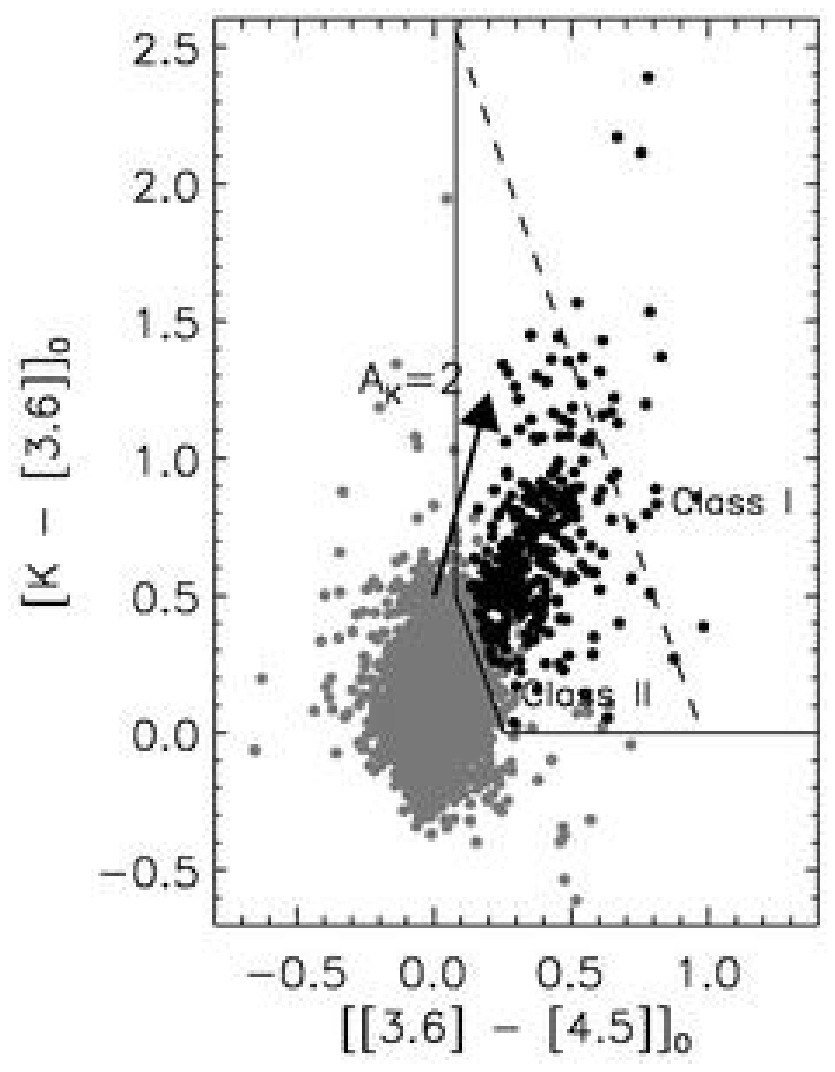}
\caption{Color-color diagram used for isolation of Class~I and Class~II YSOs (filled black circles) in the Phase~2 classification.  The dashed black line denotes the separation between the Class~I and Class~II YSOs.\label{k12c}}
\end{figure}

We now identify additional YSOs as those sources with dereddened colors that obey all of the following constraints (see Fig.~\ref{k12c}):

\begin{displaymath}
\sigma_1 =  \sigma\{[[3.6]-[4.5]]_{meas}\}; \sigma_2 =  \sigma\{[[K]-[3.6]]_{meas}\}
\end{displaymath}
\begin{displaymath}
[[3.6]-[4.5]]_0 - \sigma_1 > 0.101
\end{displaymath}
\begin{displaymath}
[K-[3.6]]_0 - \sigma_2 > 0
\end{displaymath}
\begin{displaymath}
[K-[3.6]]_0 - \sigma_2 > -2.85714 \times ([[3.6]-[4.5]]_0 - \sigma_1 - 0.101) + 0.5
\end{displaymath}

Among these YSOs, sources that also follow the following selection are considered likely protostars, the rest are considered Class~II:
\begin{displaymath}
[K-[3.6]]_0 - \sigma_2 > -2.85714 \times ([[3.6]-[4.5]]_0 - \sigma_1 - 0.401) + 1.7
\end{displaymath}

To minimize the inclusion of dim extragalactic contaminants, we apply a simple brightness limit using the dereddened 3.6~$\mu$m photometry.
All sources classified as Class~II with this method must have $[3.6]_0 < 14.5$ and all protostars must have $[3.6]_0 < 15$.

\subsection{Phase 3: Adding and Checking YSOs with MIPS 24~$\mu$m Photometry\label{mipsphase}}


We apply Phase~3 to all objects with MIPS 24~$\mu$m detections with $\sigma <
0.2$~mag.  Phase~3 is primarily a MIPS 24~$\mu$m reexamination of the entire
catalog, including transition disk sources (Class~III sources as classified by
shorter wavelengths, but with large 24~$\mu$m excess) and deeply embedded YSOs
with insufficient IRAC and 2MASS bandpass coverage to have been included by the
previous two phases but with bright ($[24]<7$) MIPS detections.  Phase~3 also
includes a search for heavily reddened Class~II sources that were misclassified
as Class~I in Phase~1, identified by a lack of adequate 24~$\mu$m excess.  

\begin{figure}
\epsscale{.80}
\plotone{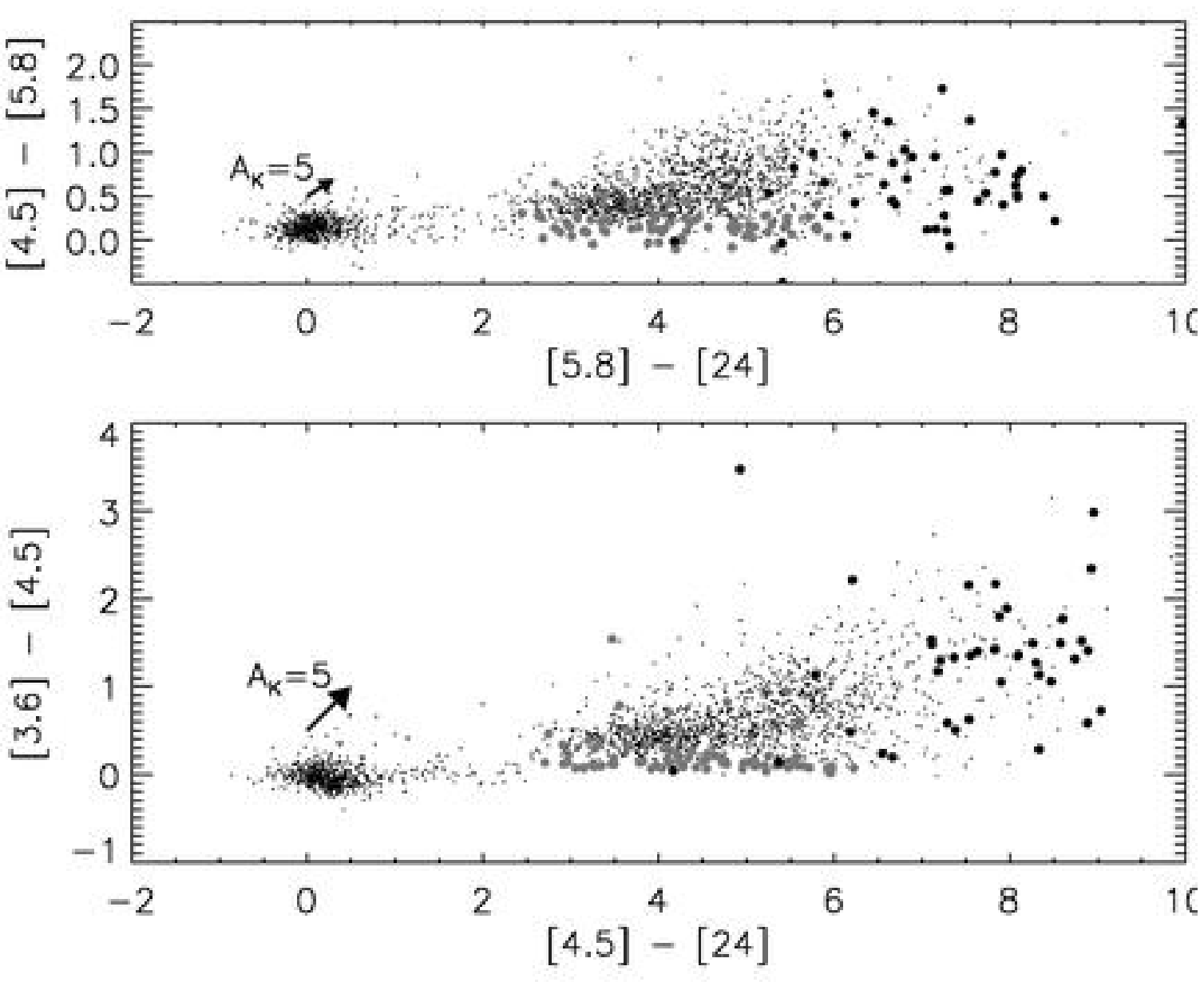}
\caption{Color-color diagrams used for the isolation of transition disk objects (filled gray circles) and deeply embedded protostars (filled black circles).\label{mipsccd}}
\end{figure}

First, sources that were considered photospheric in the previous two phases but
have excess emission at 24~$\mu$m ($[5.8] - [24] > 2.5$ or $[4.5] - [24] > 2.5$) are very interesting (filled gray circles in Fig.~\ref{mipsccd}).
These sources are typically ``Transition Disks'', Class~II sources with
significant dust clearing within their inner disks.  We have required that
valid sources in this group also have $[3.6] < 14$ in an effort to minimize
contamination from extragalactic contaminants.  These objects are marked as
class ``II'' with an asterisk to note their special nature in the tables.

Second, 
many young sources are so deeply embedded that we cannot extract usable
photometry in one or more of the four IRAC bands.  As such, we note any source
as a likely deeply embedded protostar\footnote{Classifying sources as bona fide
Class~0 protostars requires a detailed treatment including submillimeter flux
measurements which are beyond the scope of this paper. As such, these sources
are still labelled Class~I in Table~\ref{ysotable2}, but with the addition of
an asterisk to note their anomalous nature.} if it lacks detections in some
IRAC bands yet has bright MIPS 24~$\mu$m photometry ($[24] < 7$ and $[X]-[24] >
4.5$ mags, where $[X]$ is the photometry for the longest wavelength IRAC
detection that we possess).  The adopted magnitude limit at 24~$\mu$m is needed
to mitigate extragalactic source contamination.  The specific value that we
adopt here is slightly more conservative than that adopted by c2d
\citep{harv06}.  Furthermore, with the addition of the 24~$\micron$ data, we
have more information to filter lower luminosity YSOs from our AGN candidates
and shock emission dominated sources flagged in Section~\ref{iracclass}.  We
reinclude flagged sources as likely protostars if they have both bright MIPS
24~$\mu$m photometry ($[24] < 7$, as before) and convincingly red IRAC/MIPS
colors ($[3.6] - [5.8] > 0.5$ and $[4.5]-[24] > 4.5$ and $[8.0]-[24]  > 4$).  The filled black circles in Fig.~\ref{mipsccd} show the full distribution of these objects in the IRAC and MIPS 24~$\mu$m color planes.

Finally, all previously identified protostars that have 24~$\mu$m detections
are checked to ensure that their SEDs do indeed continue to rise from IRAC to
MIPS wavelengths.  All protostars that have MIPS detections must have $[5.8] -
[24] > 4$ if they possess 5.8~$\mu$m photometry, otherwise they must have
$[4.5] - [24] > 4$.  If a source does not meet this color requirement, it is
likely a highly reddened Class~II and reclassified as such.

\section{Circles, Ellipses, Convex Hulls, and Aspect Ratios\label{app2}}

The choice of method for determining the circular radius of a source grouping,
$R_{circ}$, is an important one, as not only is this an overall cluster size
measurement, it is also a key component of the measurement of aspect ratios of
non-circular distributions.  In the case of an approximately elliptical
distribution, $R_{circ}$ is the semimajor axis of the elliptical area.  As
mentioned in Section~\ref{demcore}, we have departed from the more standard
definition of the circular radius of a cluster \citep[the radial distance from
the mean of the source positions to the furthest source; e.g.][]{cw04}.
Instead, we chose to use the radial distance to the most distant source from use the mean position of those sources that are the vertices of the convex hull. This method is much
less susceptible to bias from the high uncertainty of the mean position of
small-$N$ groupings, and is therefore less likely to overestimate the cluster
area, such as was illustrated by \citet{sk06} for the standard cicular radius
definition.  Note that for large-$N$ uniform or centrally concentrated
groupings, the two methods produce nearly identical results.  

\citet{sk06} defined the ``elongation'' of a cluster, $\xi$, as $R_{circ} /
R_{hull}$ and then performed numerous Monte Carlo realizations of uniform
surface density $N=350$ member clusters distributed in elliptical areas of
varying aspect ratio to demonstrate the efficacy of $\xi$ as a functional
tracer of the aspect ratio of the elliptical area.  This result can be
demonstrated trivially in analytic form if we assume $N$ is sufficiently large
(350 is large), as the resulting convex hull of an elliptical distribution of
sources can be approximated by a true ellipse.  The assumption of large $N$
also ensures that the mean position of sources has very low uncertainty, that
statistical non-uniformity in the overall source distribution is kept to a
relatively low level, and that the convex hull area adjustment for the fraction
of sources making up the vertices of the hull is approximately unity.  As such,
the area of the convex hull is $\pi a b$, where $a$ and $b$ are the semimajor
and semiminor axes of the ellipse, respectively.  It follows then that
$R_{hull} = \sqrt(ab)$ given its definition in Section~\ref{demcore}.  As
mentioned above, $R_{circ}$ in this case is just the semimajor axis, $a$, and
thus the `elongation' is simply $\sqrt(a/b)$, the square root of the aspect
ratio of the ellipse, a result that is born out in the Monte Carlo simulations
of \citet{sk06}.  For conceptual simplicity, we have chosen to ignore the
``elongation'' value, $\xi$, and instead report its square, the elliptical
equivalent aspect ratio.

As can be noted in Tables~\ref{ysotable3}~and~\ref{ysotable3f}, there are
several regions with reported aspect ratios below 1, contrary to the definition
of that quantity.  These typically arise in small-$N$ groups that are
effectively circular, but the granularity of the convex hull area adjustment
for small-$N$ groups results in relatively high uncertainty in the measured
value of $R_{hull}$.  It is important to note the uncertainty in the convex
hull area associated with placing one object more or less on the convex hull
increases rapidly as $N$ decreases and as the fraction of members that are
convex hull vertices increases.

\acknowledgments

Gutermuth would like to thank Ralf Klessen and Nate Bastian for helpful discussions on the utility of the MST and its analysis.
This publication makes use of data products from the Two Micron All Sky Survey,
which is a joint project of the University of Massachusetts and the Infrared
Processing and Analysis Center/California Institute of Technology, funded by
the National Aeronautics and Space Administration and the National Science
Foundation.  This research has made use of the SIMBAD database, operated at
CDS, Strasbourg, France.  This work is based in part on observations made with
the Spitzer Space Telescope, which is operated by the Jet Propulsion
Laboratory, California Institute of Technology under a contract 1407 with NASA.
Support for the IRAC instrument was provided by NASA through contract 960541 issued by JPL.

{\it Facilities:} \facility{Spitzer}.

\input{imfigs.tex}

\clearpage

\clearpage

\input{tab1.tex}

\input{tab2.tex}
\input{tab3.tex}

\input{srctabs.tex}
\input{tab40.tex}
\input{tab41.tex}

\input{tab42.tex}
\input{tab43.tex}

\input{tab44.tex}

\end{document}

%% file: imfigs.tex
\clearpage
 
\begin{figure}
\epsscale{1}
\plotone{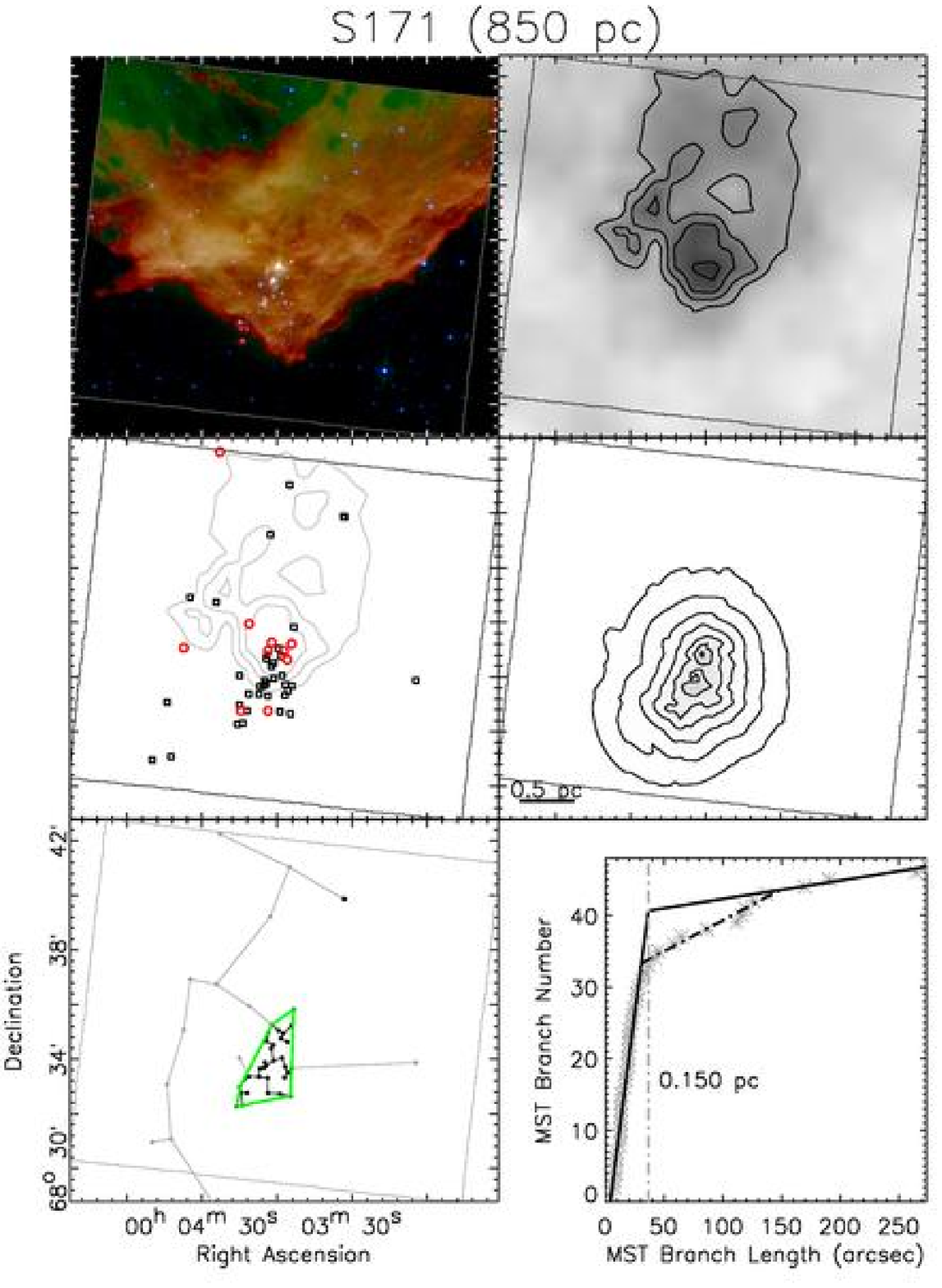}
\caption{S171 at an assumed distance of 850 pc.\label{first}}
\end{figure}
 
\begin{figure}
\epsscale{1}
\plotone{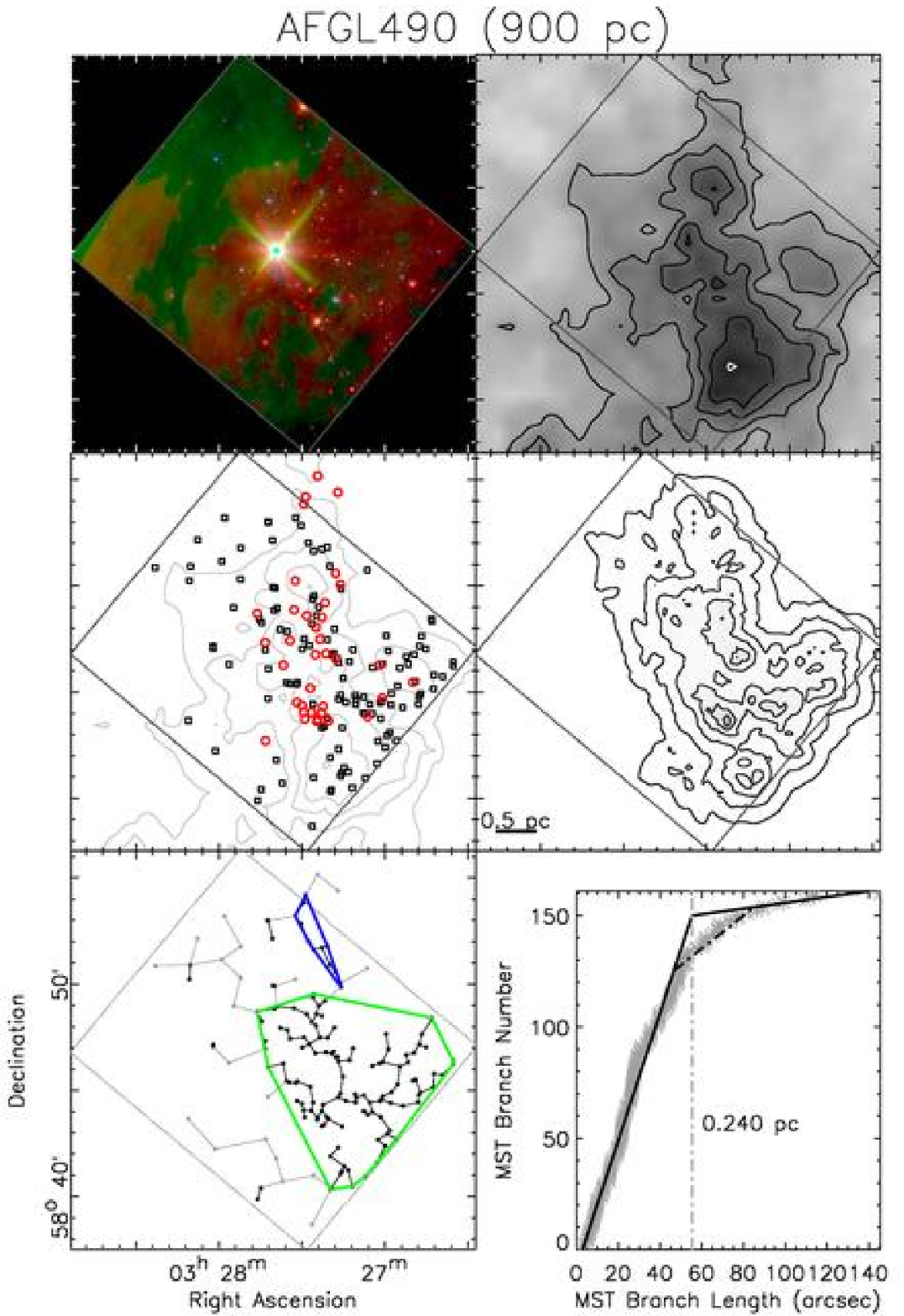}
\caption{AFGL490 at an assumed distance of 900 pc.}
\end{figure}
 
\begin{figure}
\epsscale{1}
\plotone{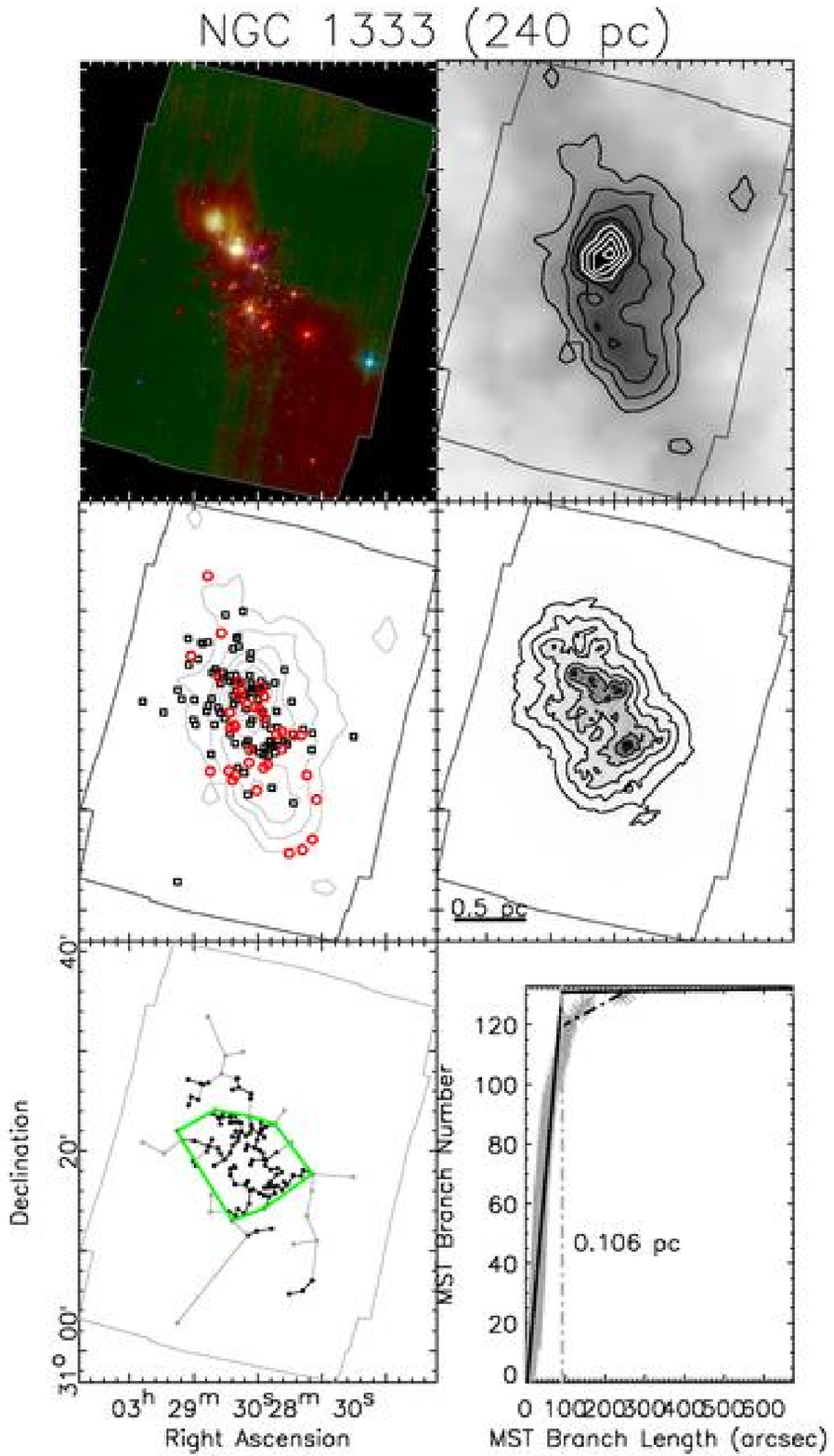}
\caption{NGC 1333 at an assumed distance of 240 pc.}
\end{figure}
 
\begin{figure}
\epsscale{1}
\plotone{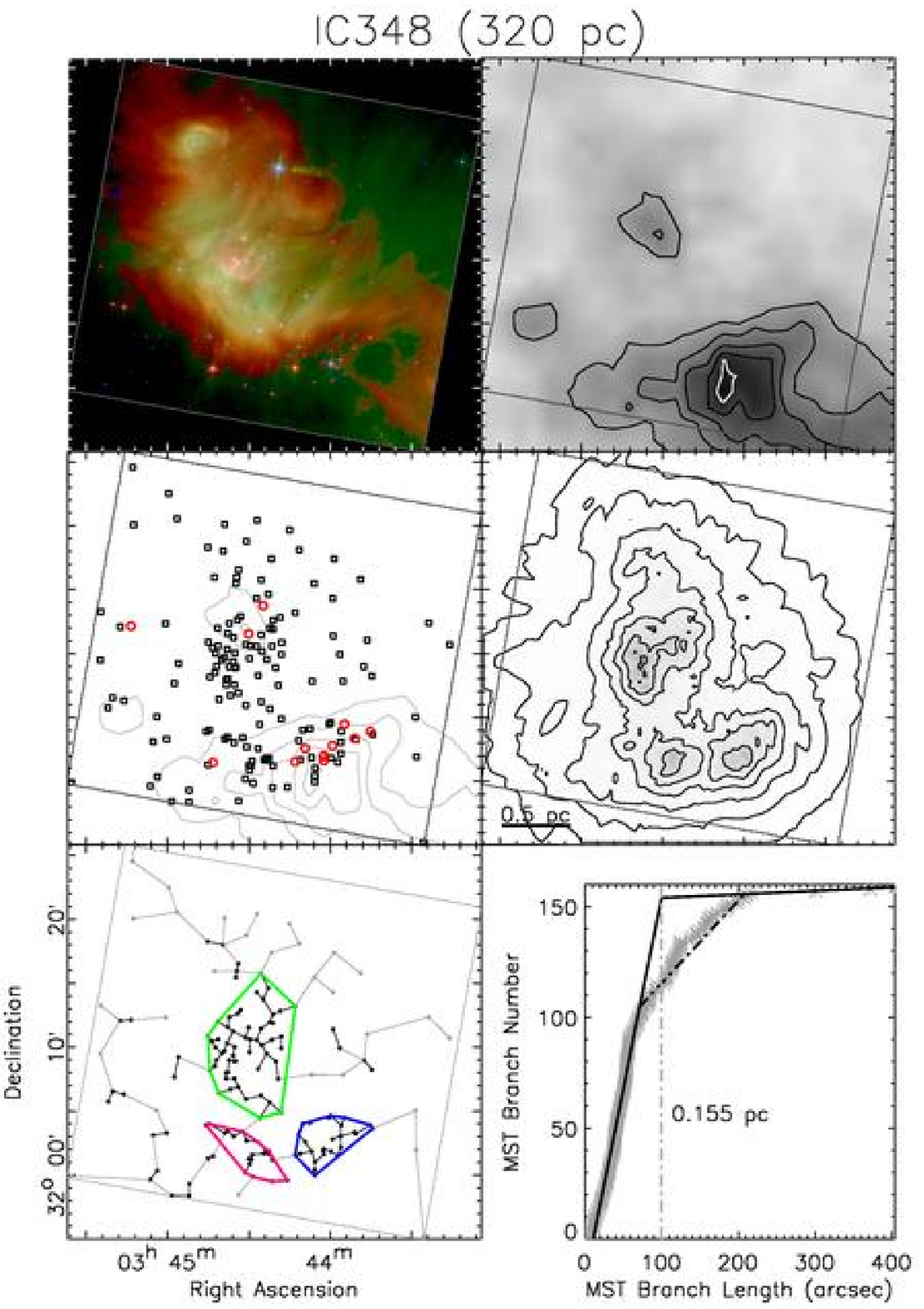}
\caption{IC348 at an assumed distance of 320 pc.}
\end{figure}
 
\begin{figure}
\epsscale{1}
\plotone{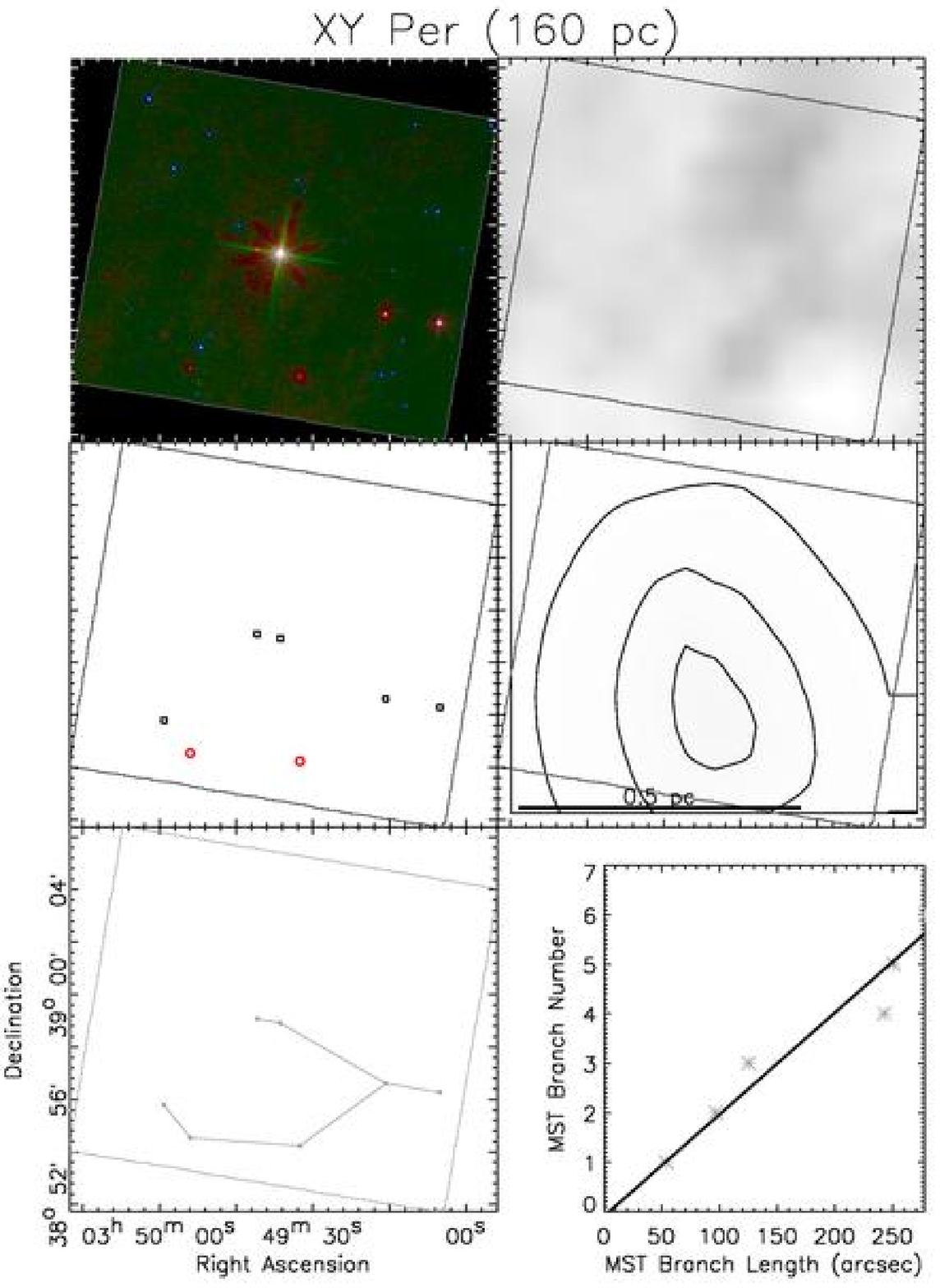}
\caption{XY Per at an assumed distance of 160 pc.}
\end{figure}
 
\clearpage
 
\begin{figure}
\epsscale{1}
\plotone{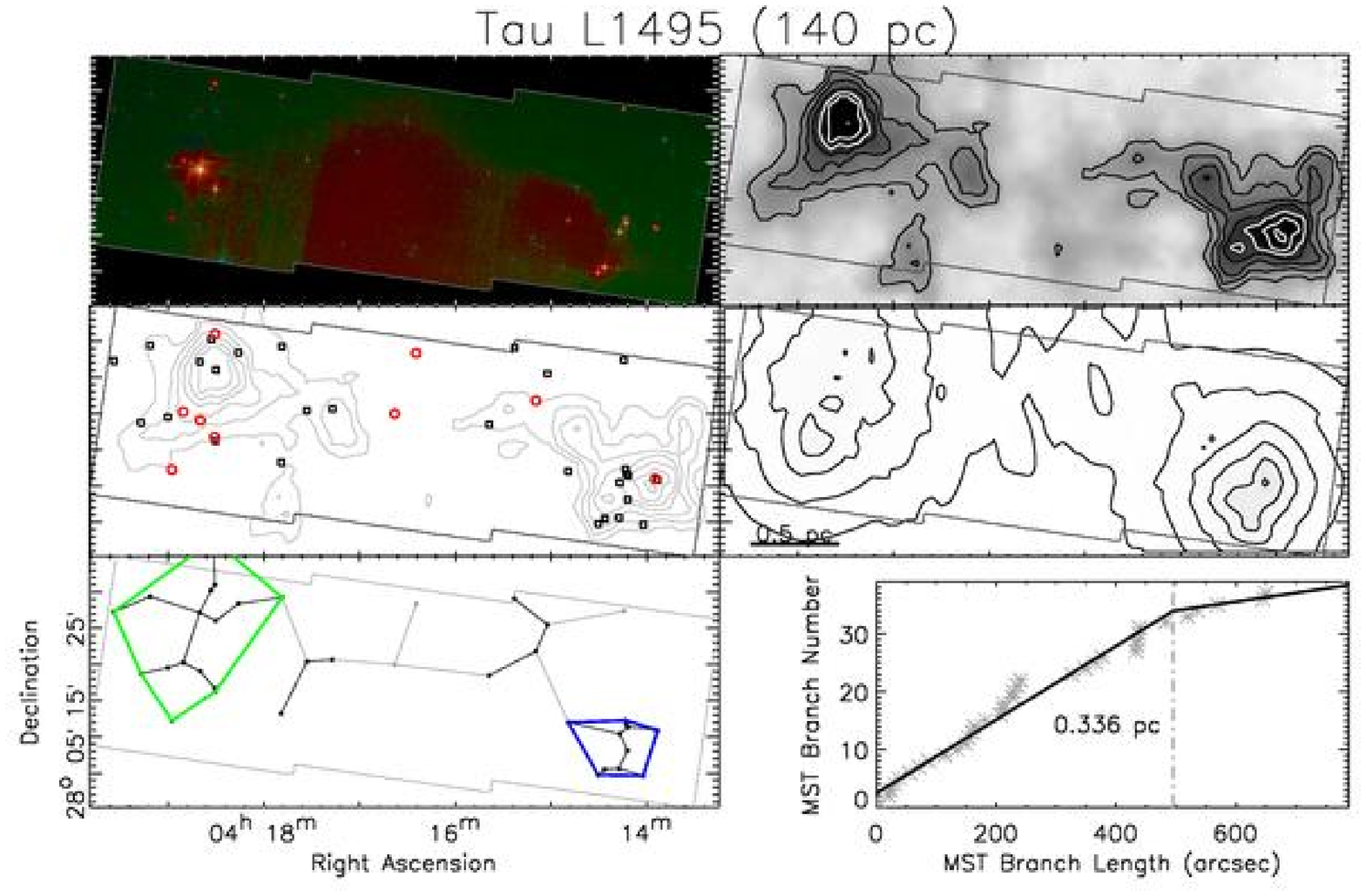}
\caption{Tau L1495 at an assumed distance of 140 pc.}
\end{figure}
 
\begin{figure}
\epsscale{1}
\plotone{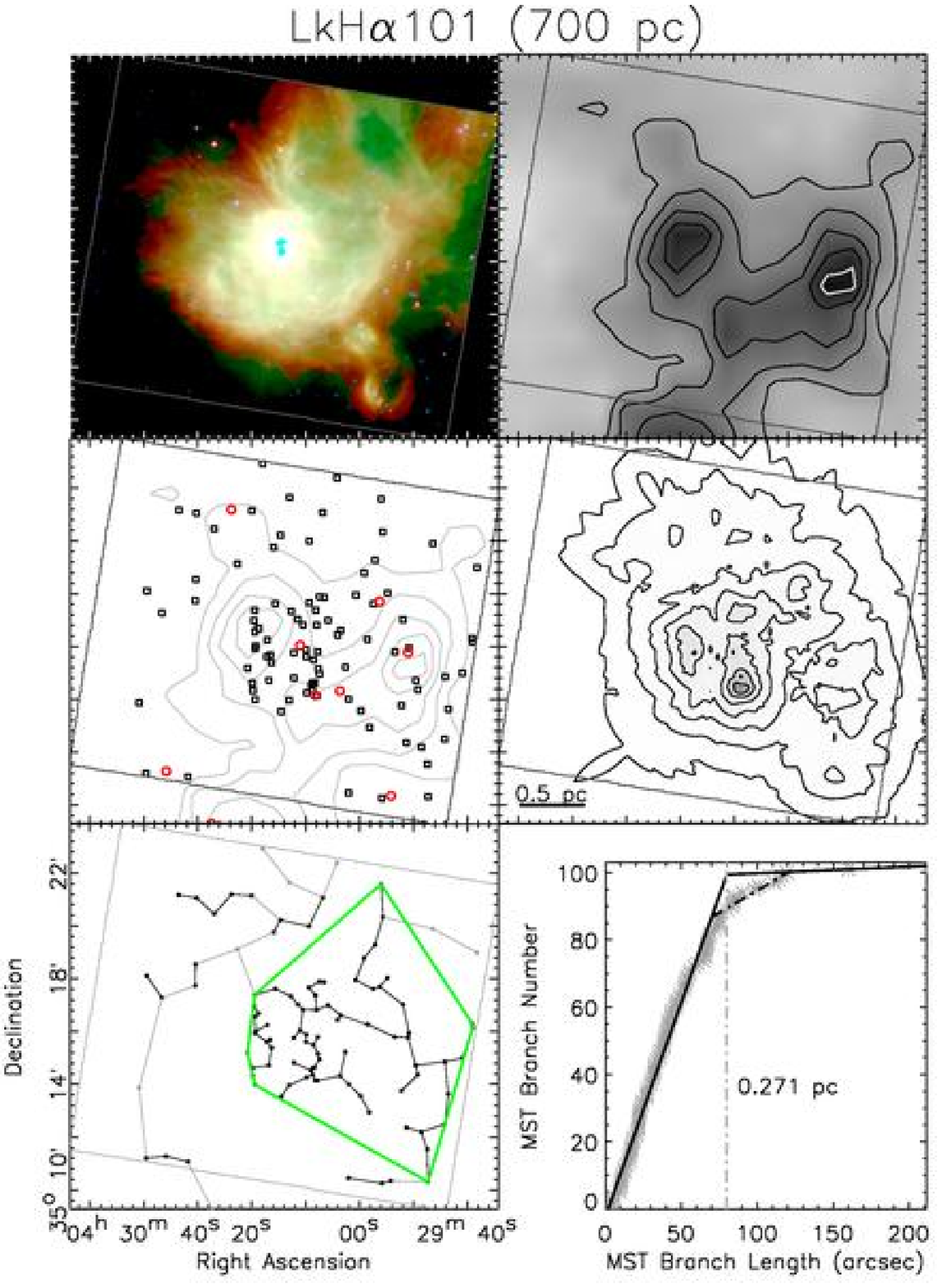}
\caption{LkH$\alpha$101 at an assumed distance of 700 pc.}
\end{figure}
 
\begin{figure}
\epsscale{1}
\plotone{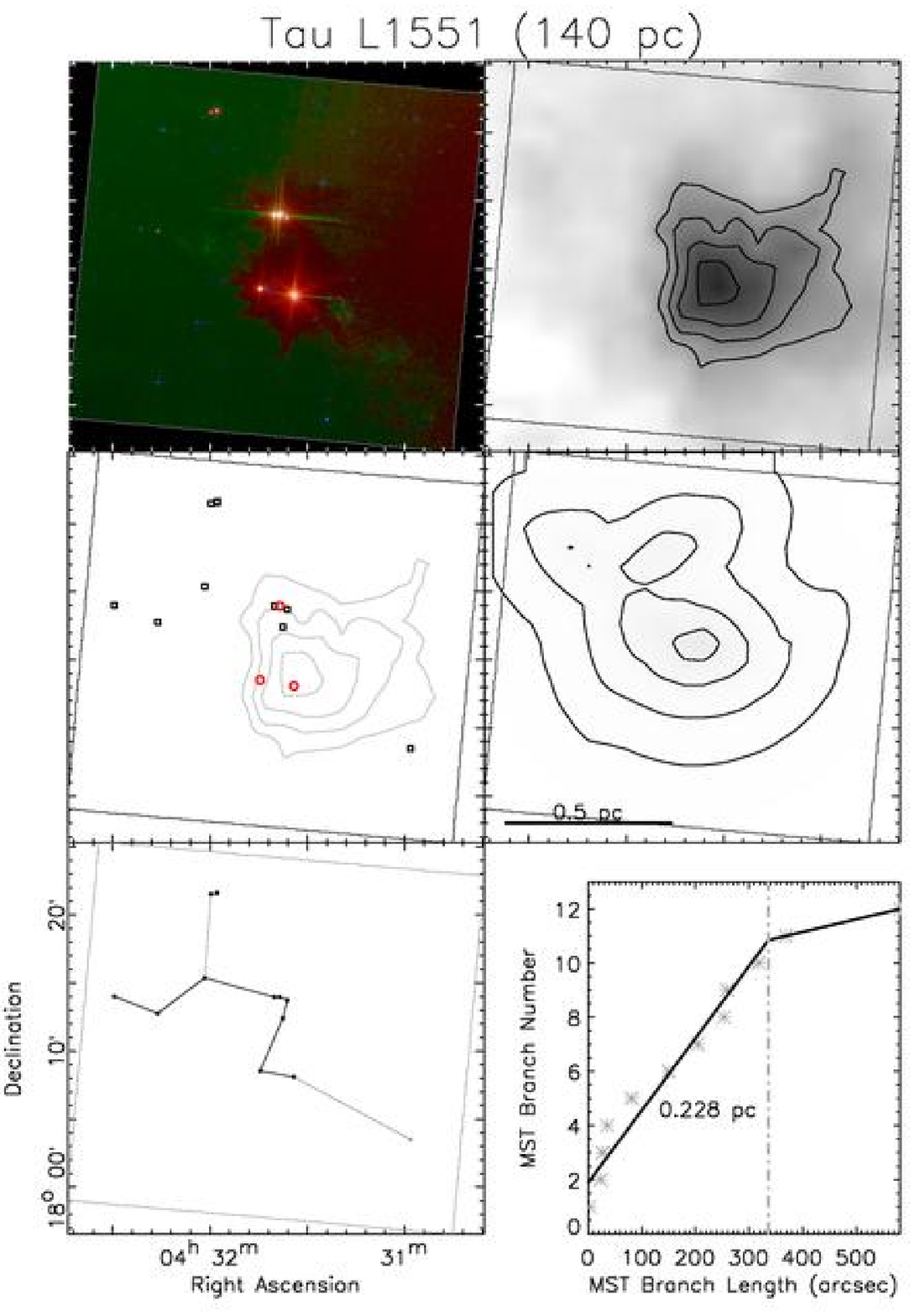}
\caption{Tau L1551 at an assumed distance of 140 pc.}
\end{figure}
 
\begin{figure}
\epsscale{1}
\plotone{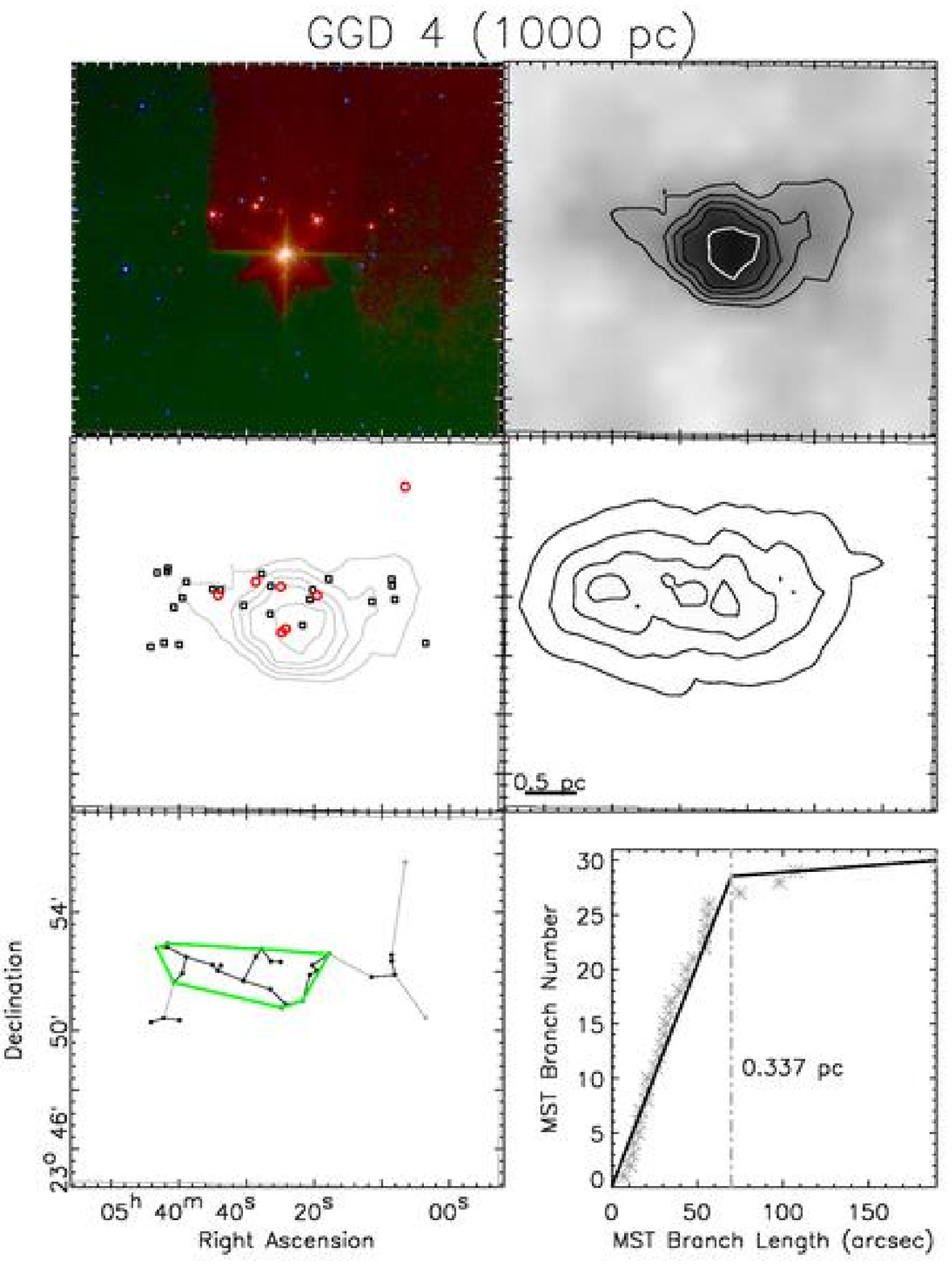}
\caption{GGD 4 at an assumed distance of 1000 pc.}
\end{figure}
 
\begin{figure}
\epsscale{1}
\plotone{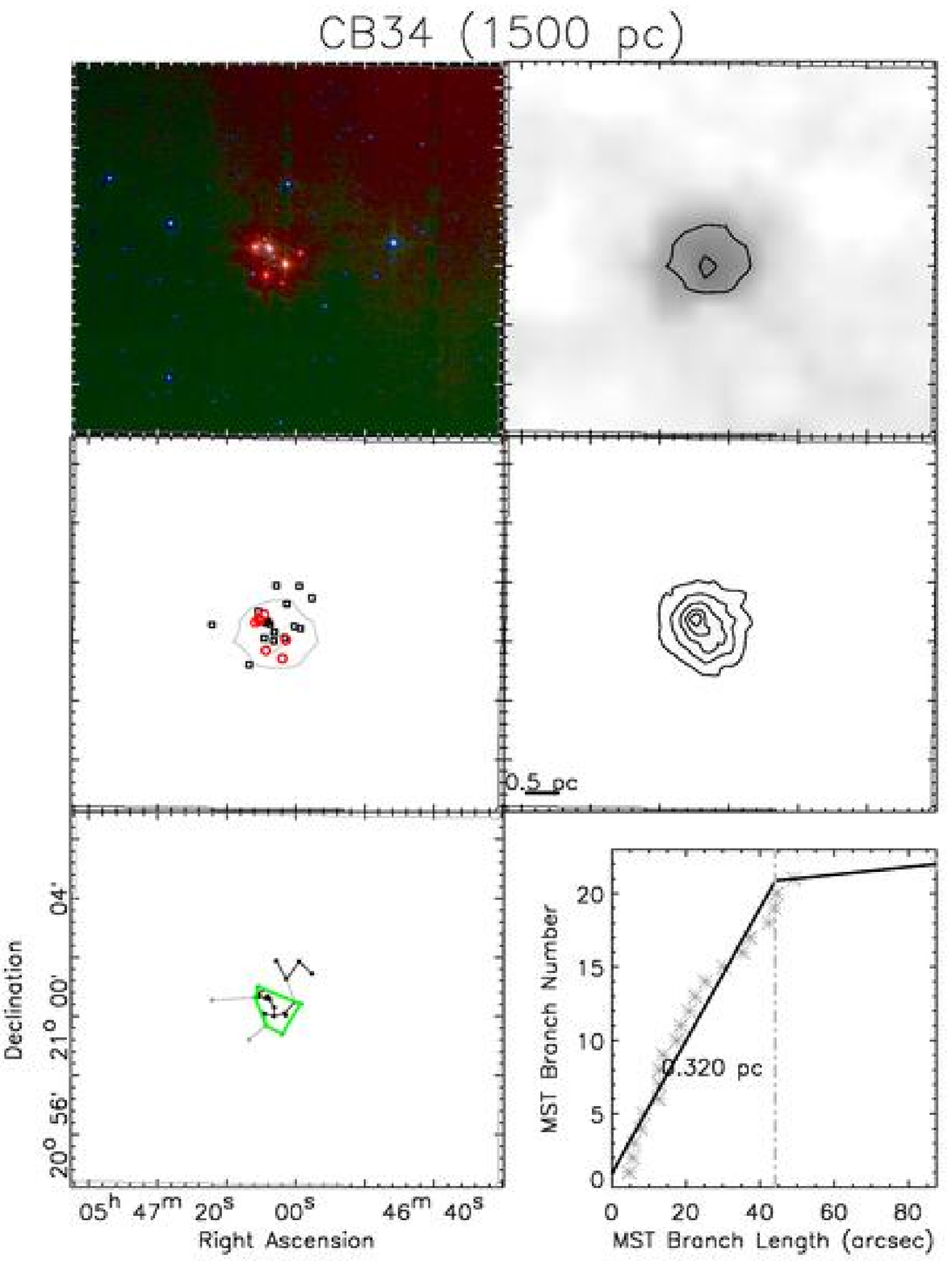}
\caption{CB34 at an assumed distance of 1500 pc.}
\end{figure}
 
\clearpage
 
\begin{figure}
\epsscale{1}
\plotone{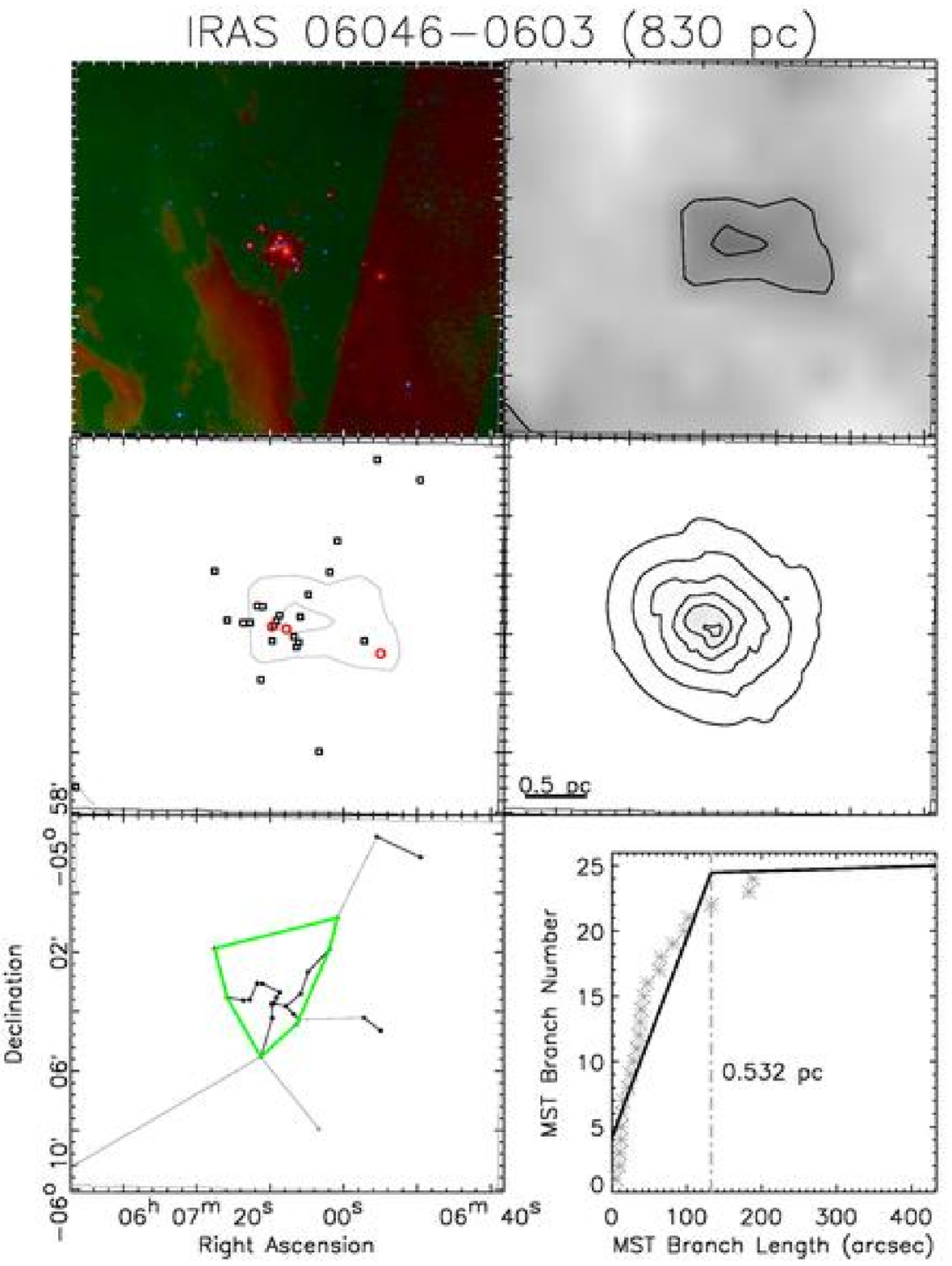}
\caption{IRAS 06046-0603 at an assumed distance of 830 pc.}
\end{figure}
 
\begin{figure}
\epsscale{1}
\plotone{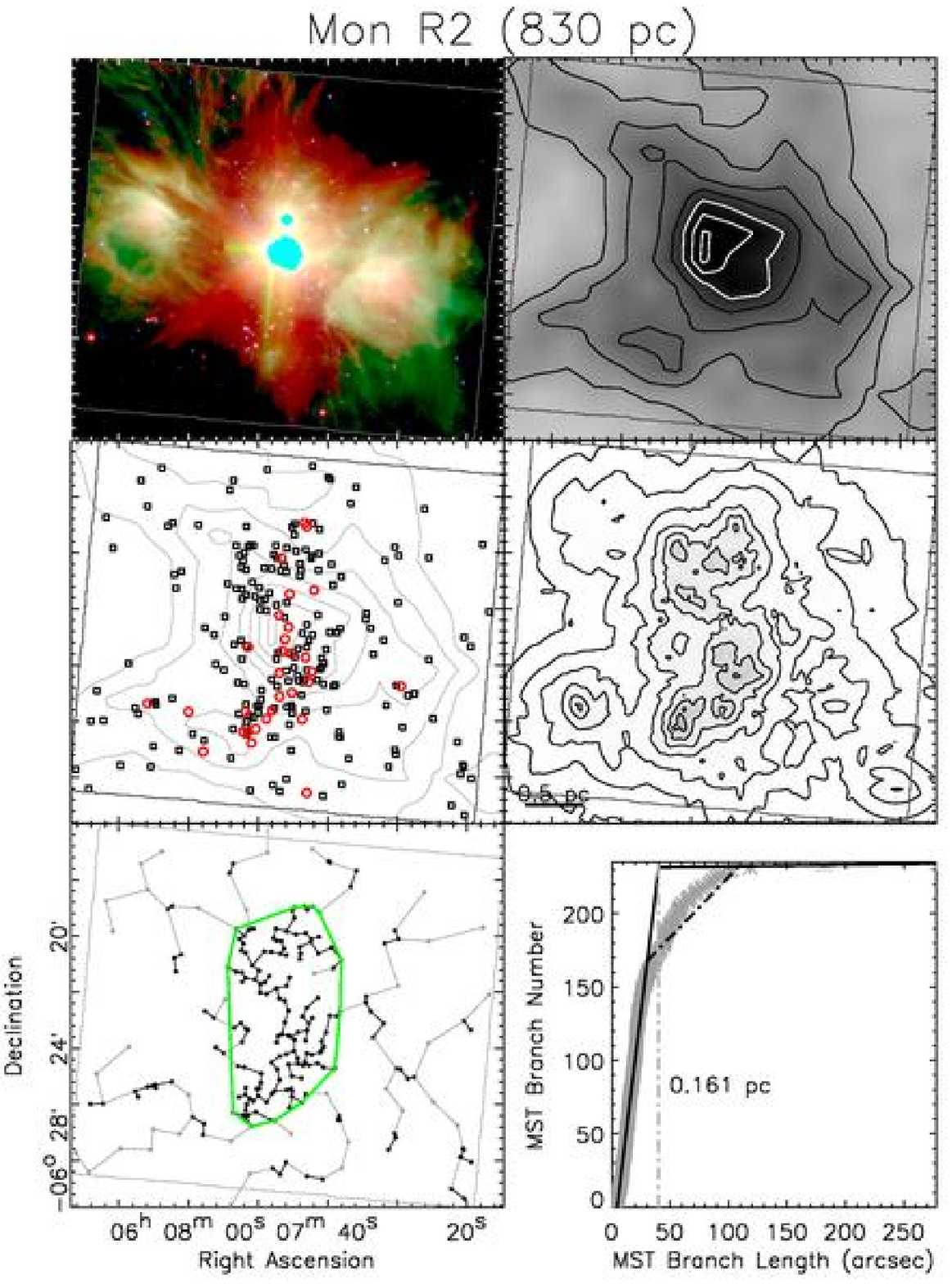}
\caption{Mon R2 at an assumed distance of 830 pc.}
\end{figure}
 
\begin{figure}
\epsscale{1}
\plotone{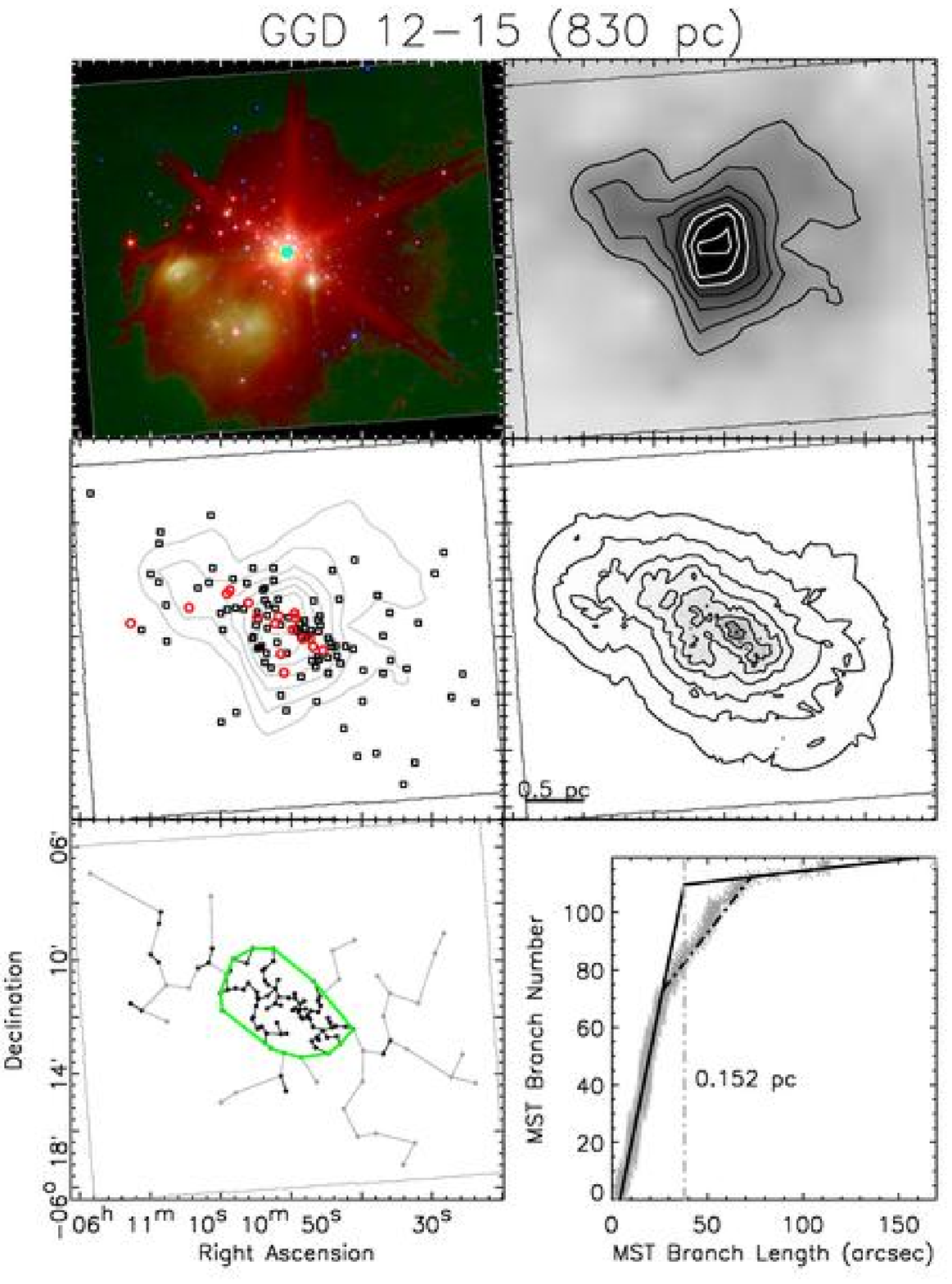}
\caption{GGD 12-15 at an assumed distance of 830 pc.}
\end{figure}
 
\begin{figure}
\epsscale{1}
\plotone{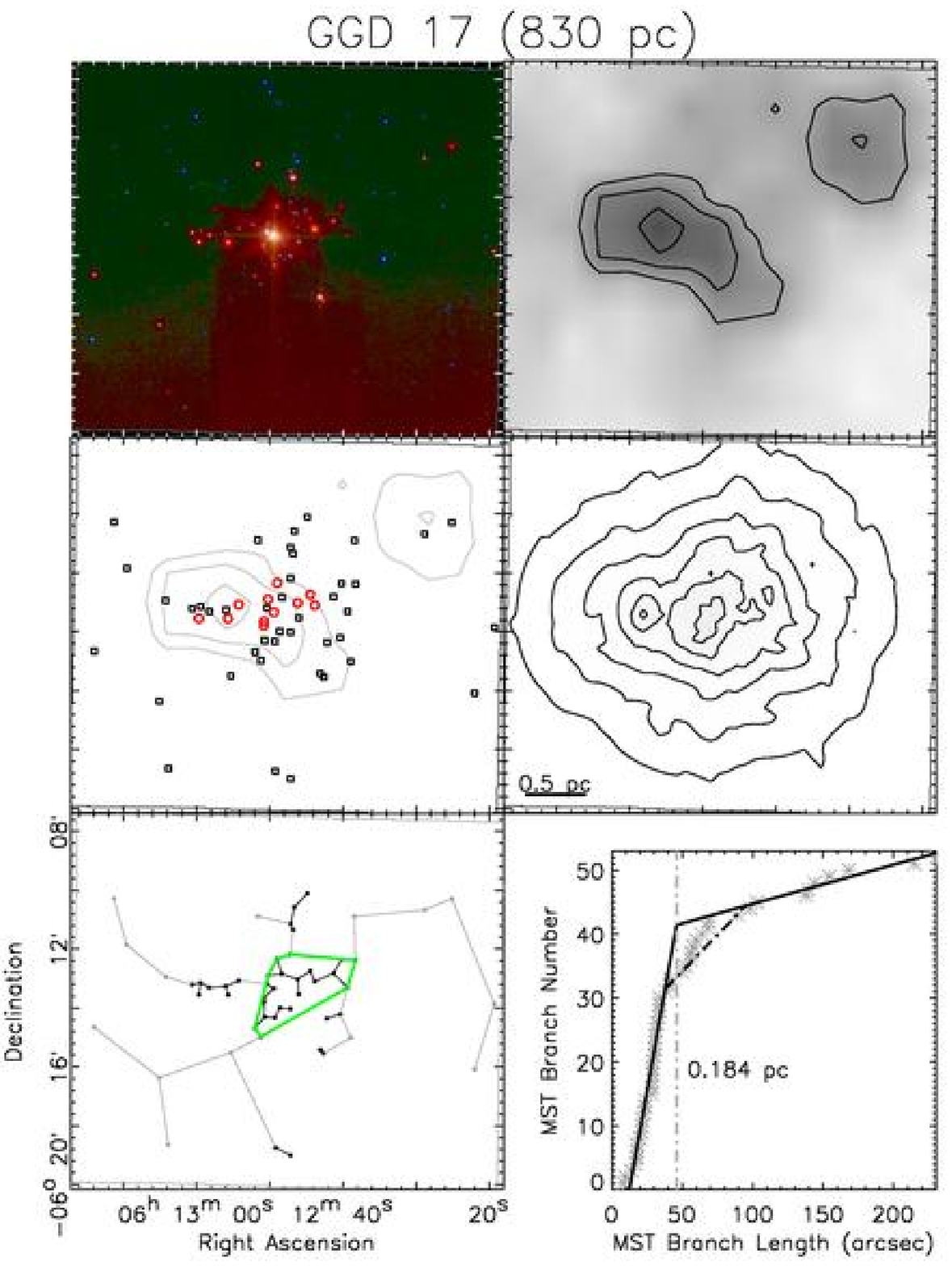}
\caption{GGD 17 at an assumed distance of 830 pc.}
\end{figure}
 
\begin{figure}
\epsscale{1}
\plotone{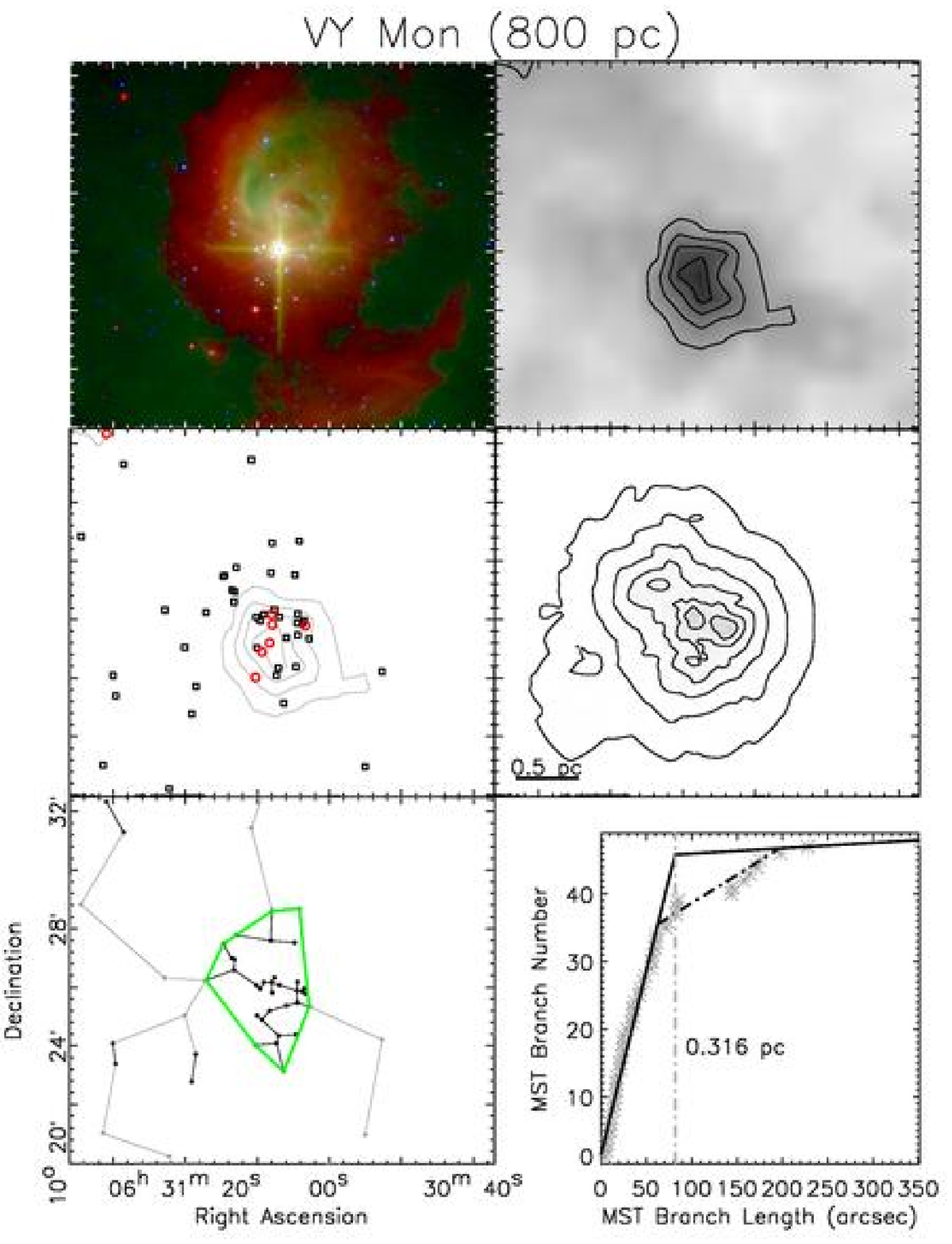}
\caption{VY Mon at an assumed distance of 800 pc.}
\end{figure}
 
\clearpage
 
\begin{figure}
\epsscale{1}
\plotone{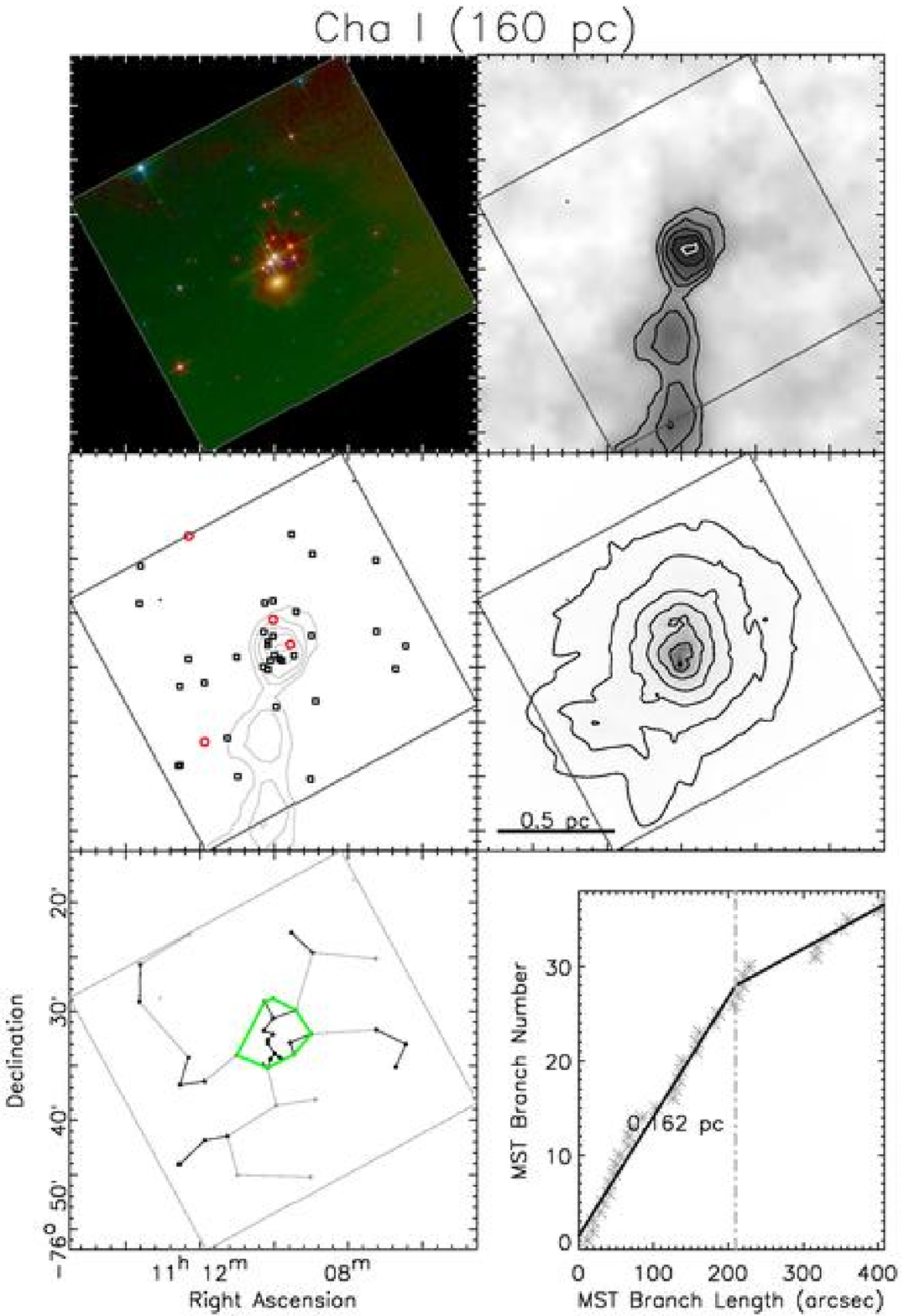}
\caption{Cha I at an assumed distance of 160 pc.}
\end{figure}
 
\begin{figure}
\epsscale{1}
\plotone{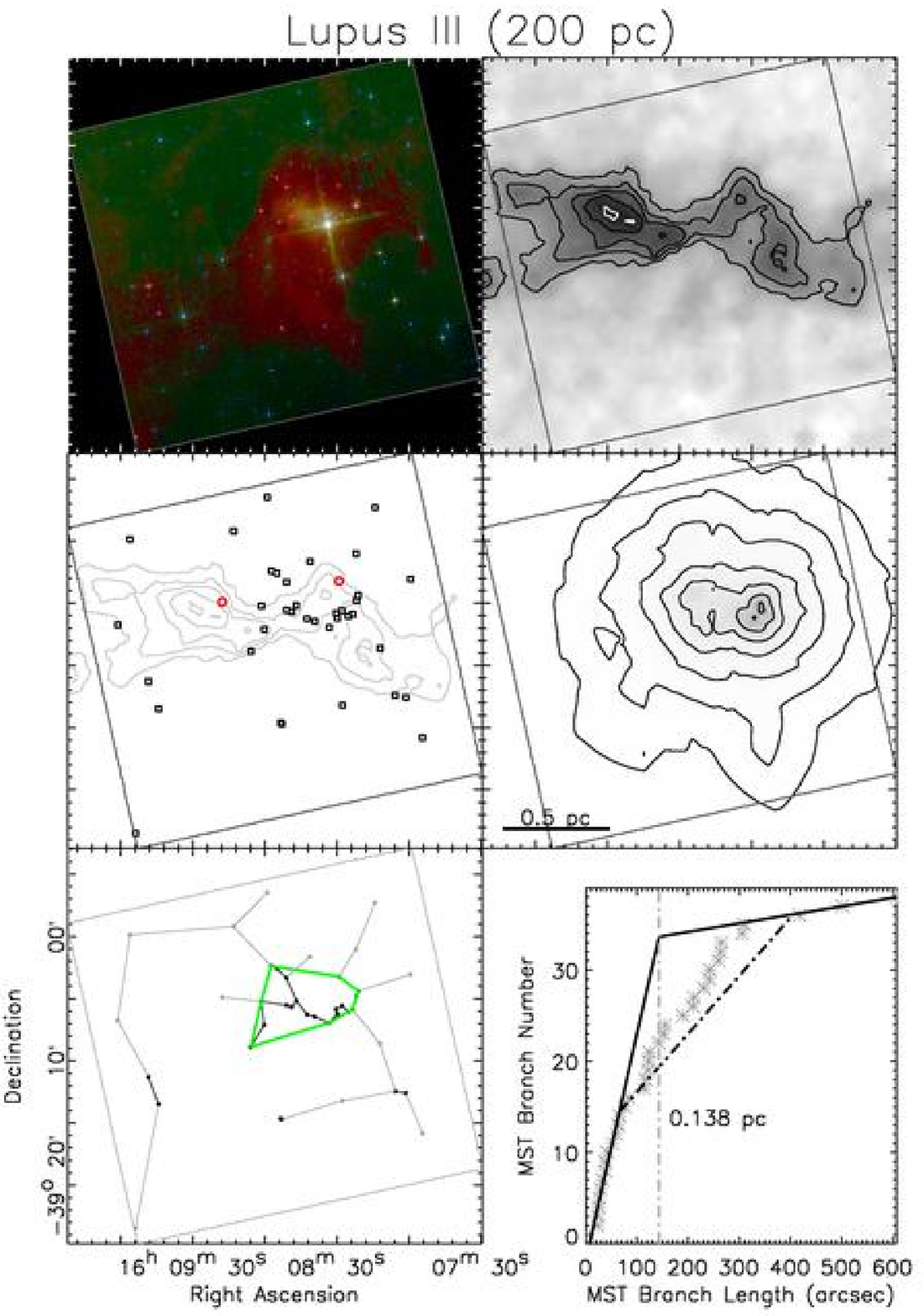}
\caption{Lupus III at an assumed distance of 200 pc.}
\end{figure}
 
\begin{figure}
\epsscale{1}
\plotone{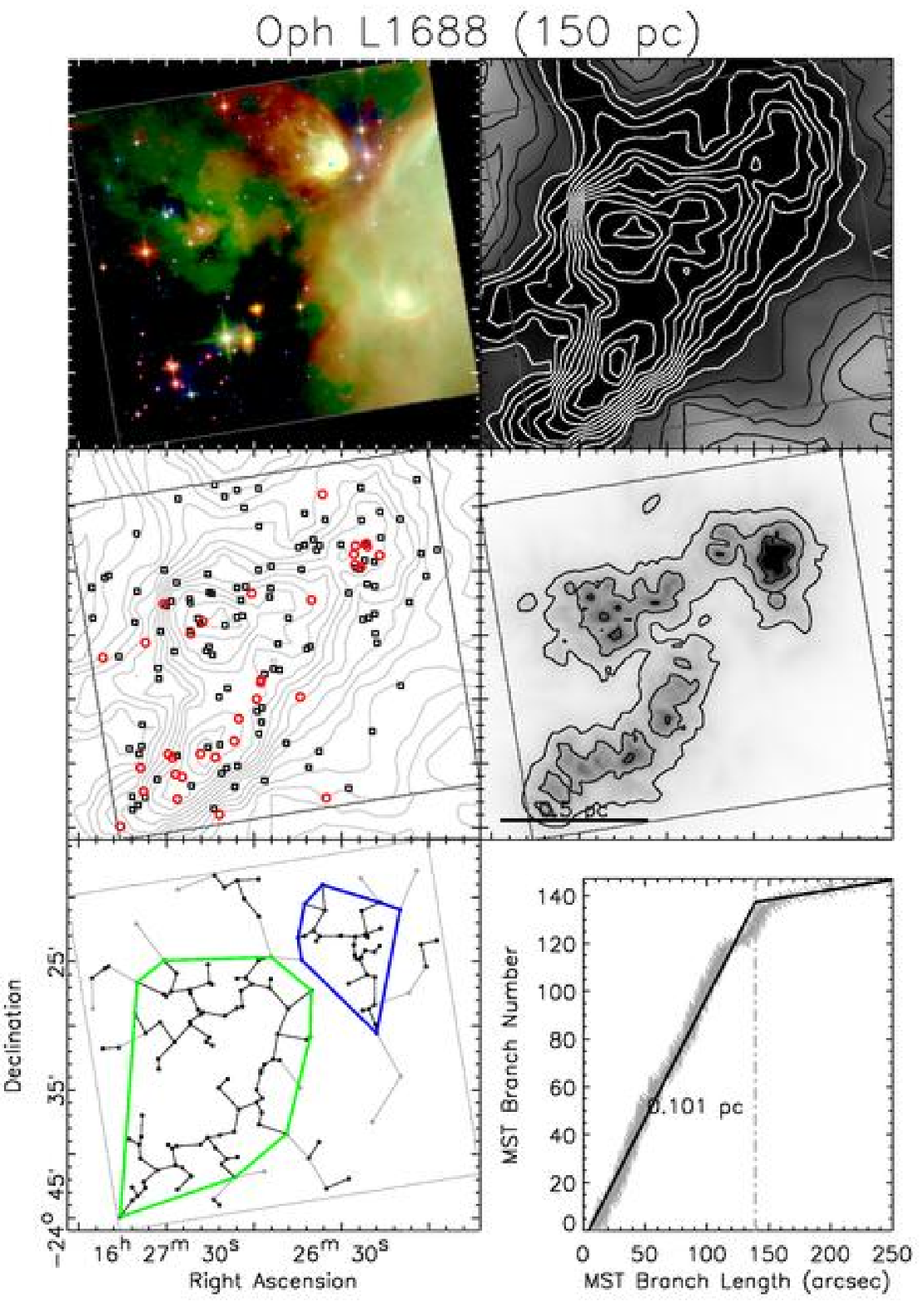}
\caption{Oph L1688 at an assumed distance of 150 pc.}
\end{figure}
 
\begin{figure}
\epsscale{1}
\plotone{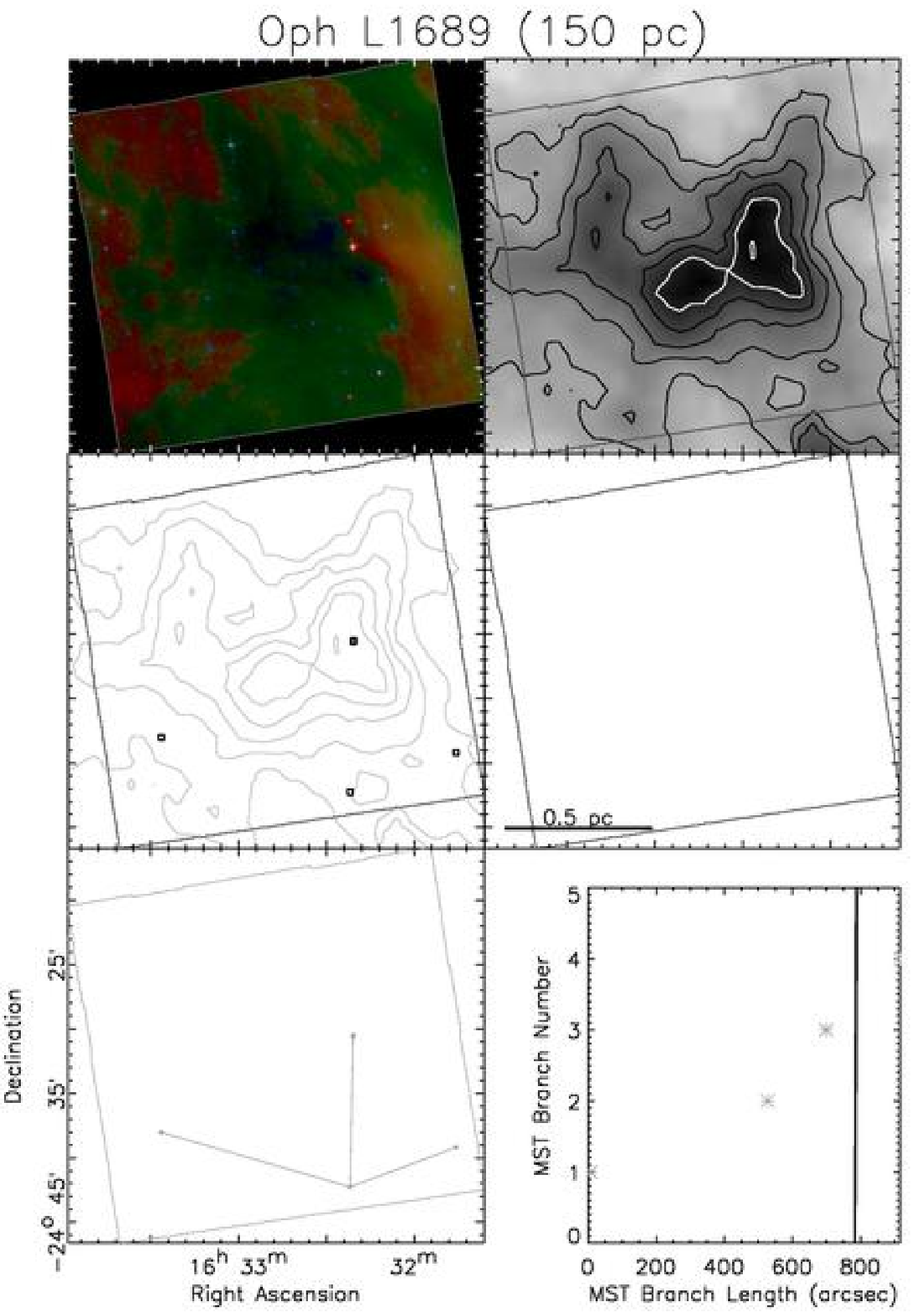}
\caption{Oph L1689 at an assumed distance of 150 pc.}
\end{figure}
 
\begin{figure}
\epsscale{1}
\plotone{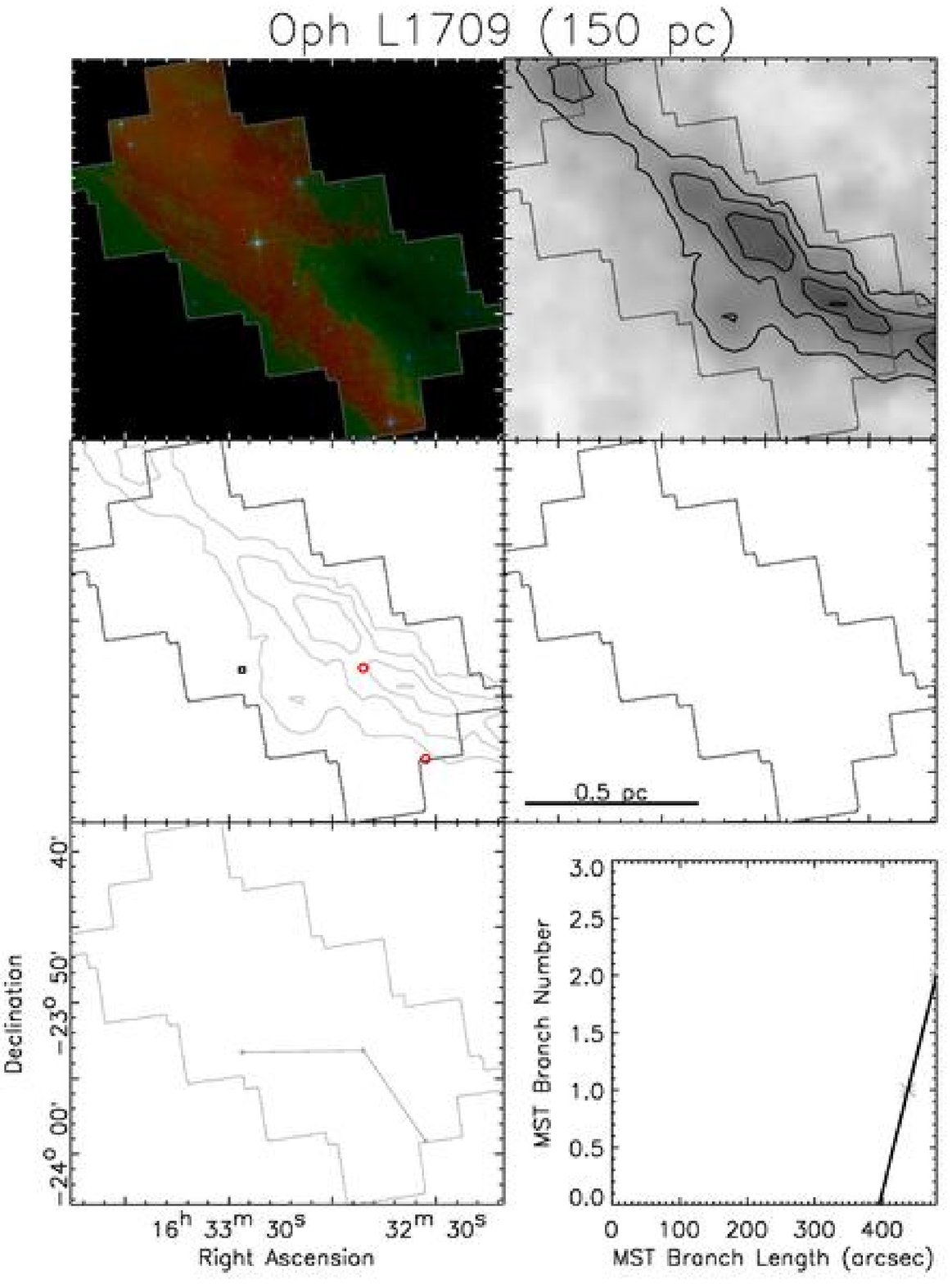}
\caption{Oph L1709 at an assumed distance of 150 pc.}
\end{figure}
 
\clearpage
 
\begin{figure}
\epsscale{1}
\plotone{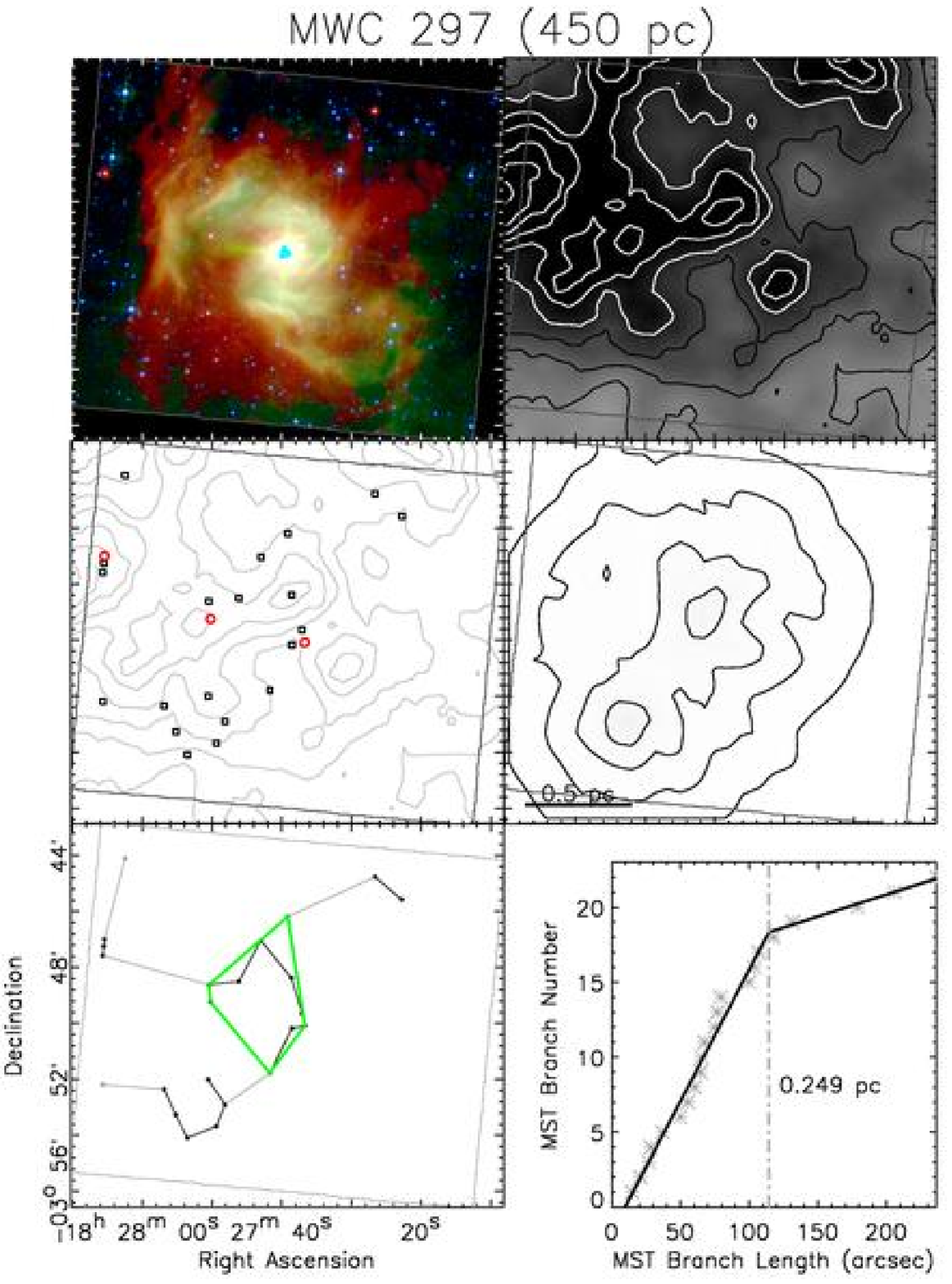}
\caption{MWC 297 at an assumed distance of 450 pc.}
\end{figure}
 
\begin{figure}
\epsscale{1}
\plotone{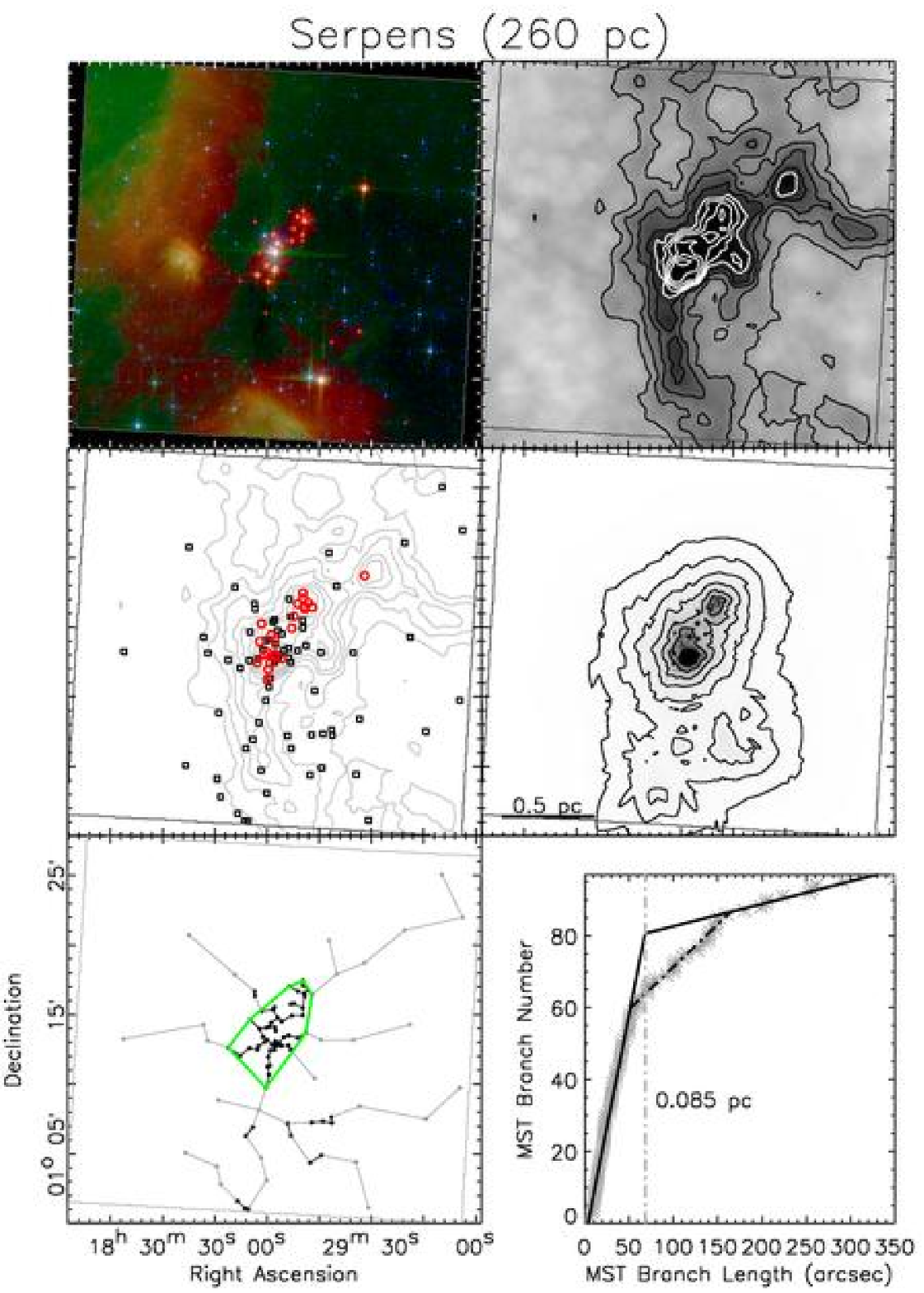}
\caption{Serpens at an assumed distance of 260 pc.}
\end{figure}
 
\begin{figure}
\epsscale{1}
\plotone{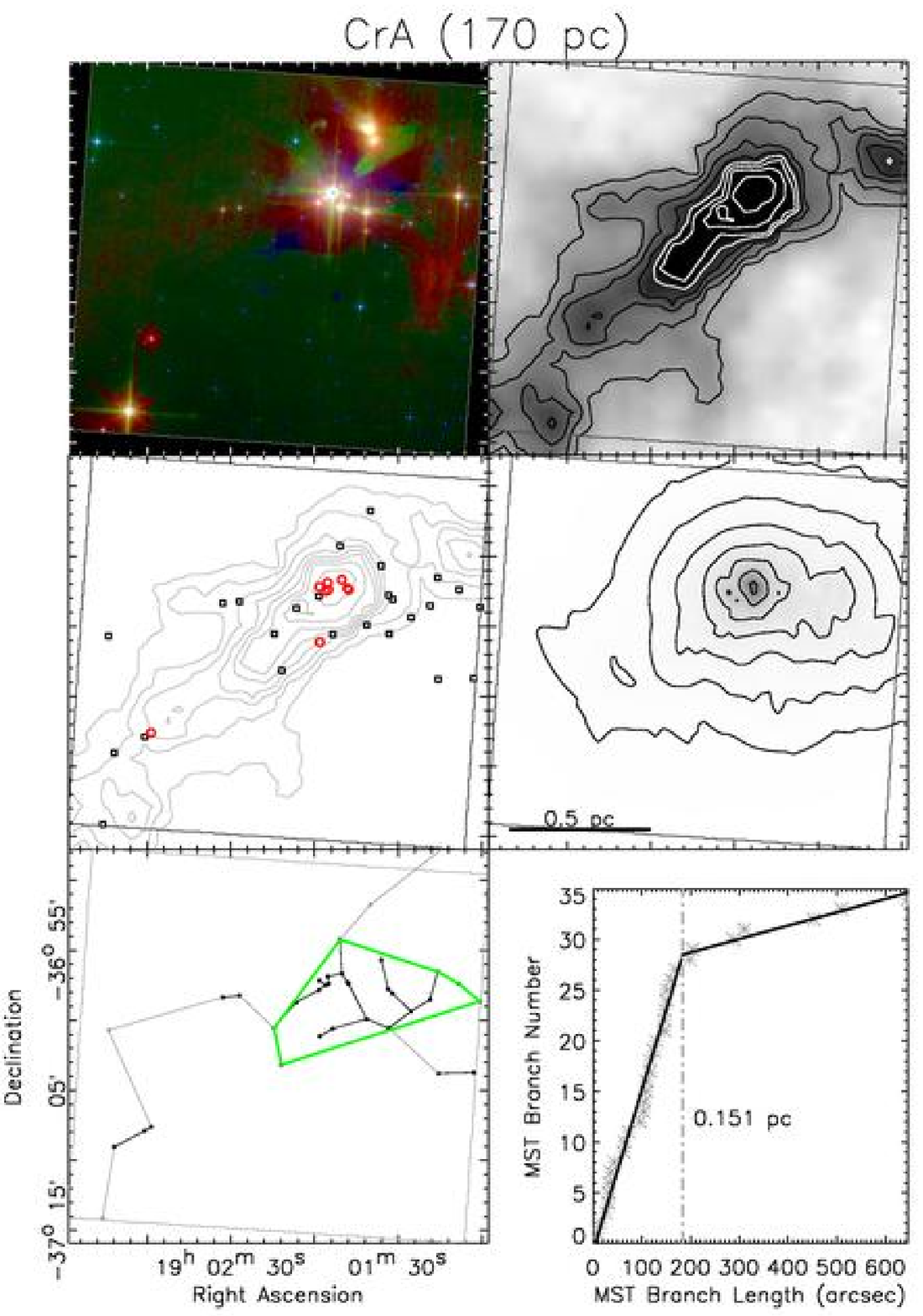}
\caption{CrA at an assumed distance of 170 pc.}
\end{figure}
 
\begin{figure}
\epsscale{1}
\plotone{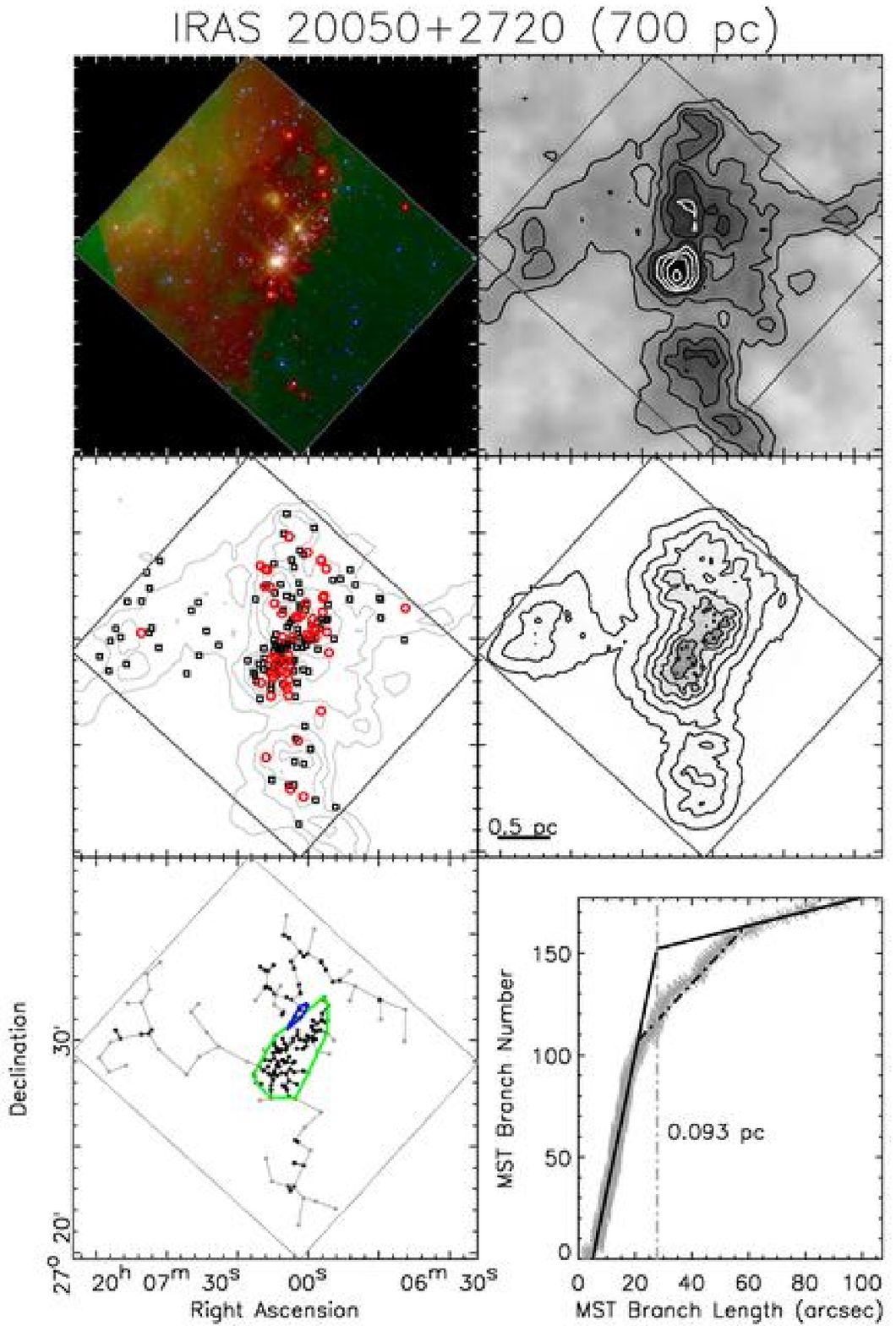}
\caption{IRAS 20050+2720 at an assumed distance of 700 pc.}
\end{figure}
 
\begin{figure}
\epsscale{1}
\plotone{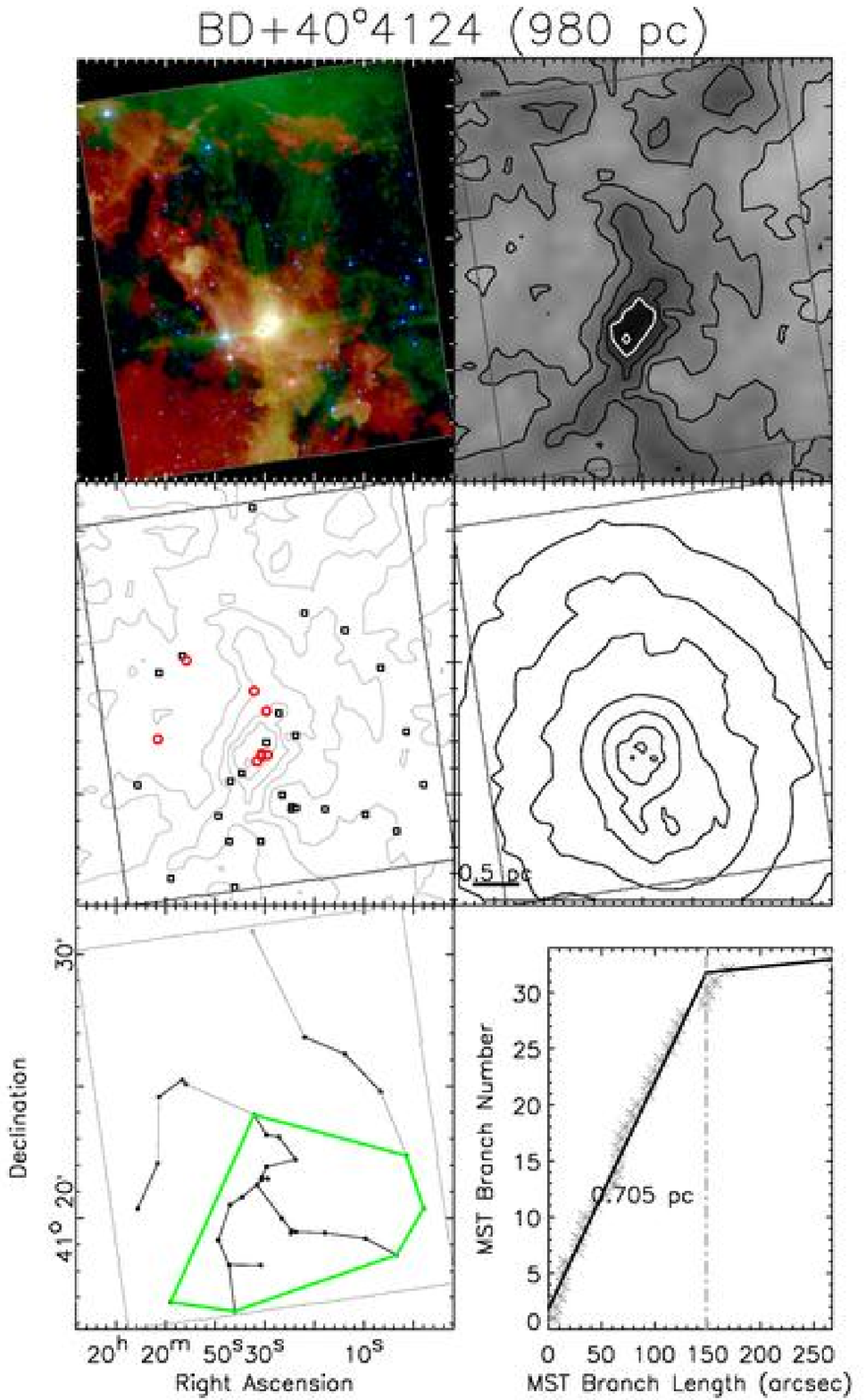}
\caption{BD+40$^{\circ}$4124 at an assumed distance of 980 pc.}
\end{figure}
 
\clearpage
 
\begin{figure}
\epsscale{1}
\plotone{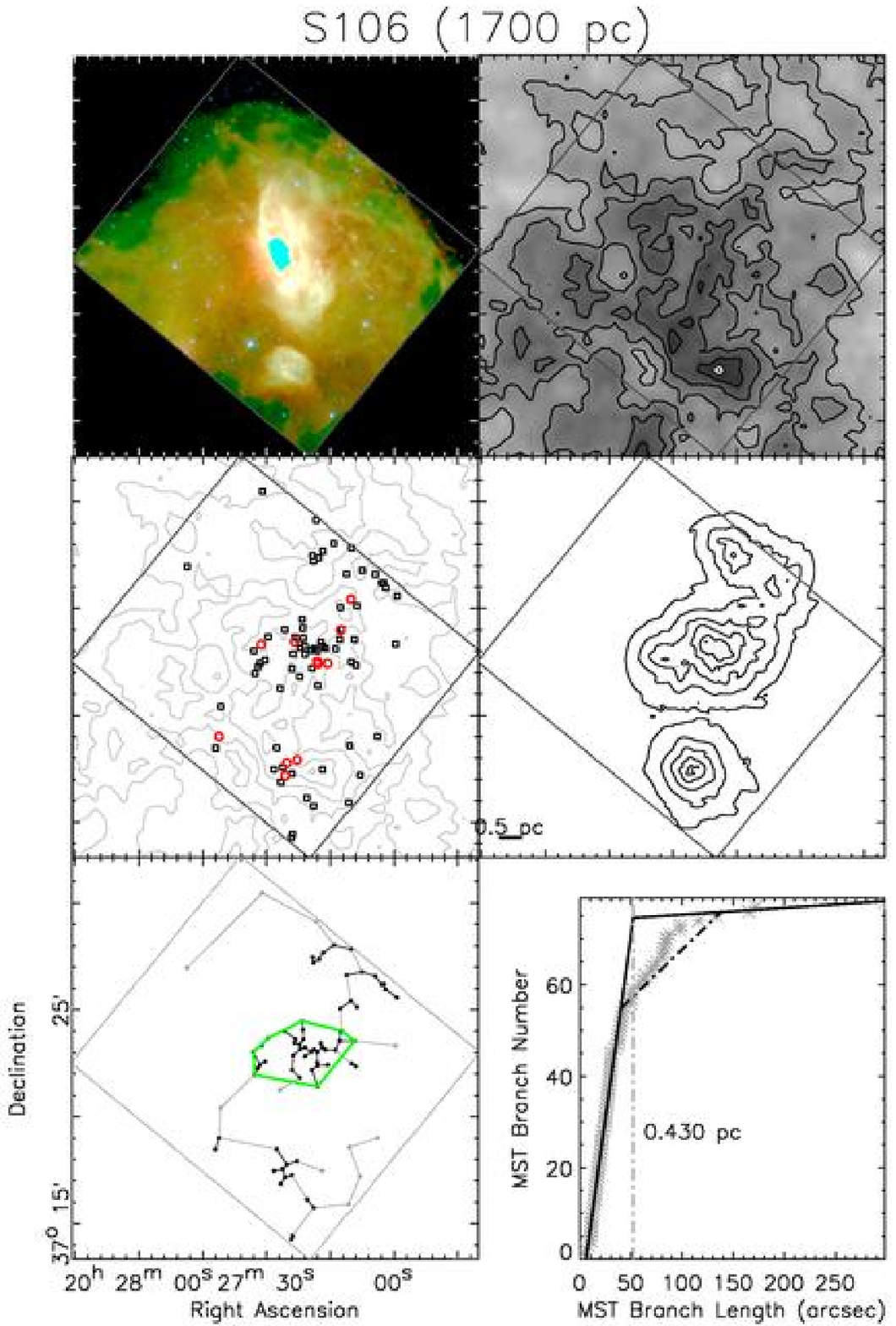}
\caption{S106 at an assumed distance of 1700 pc.}
\end{figure}
 
\begin{figure}
\epsscale{1}
\plotone{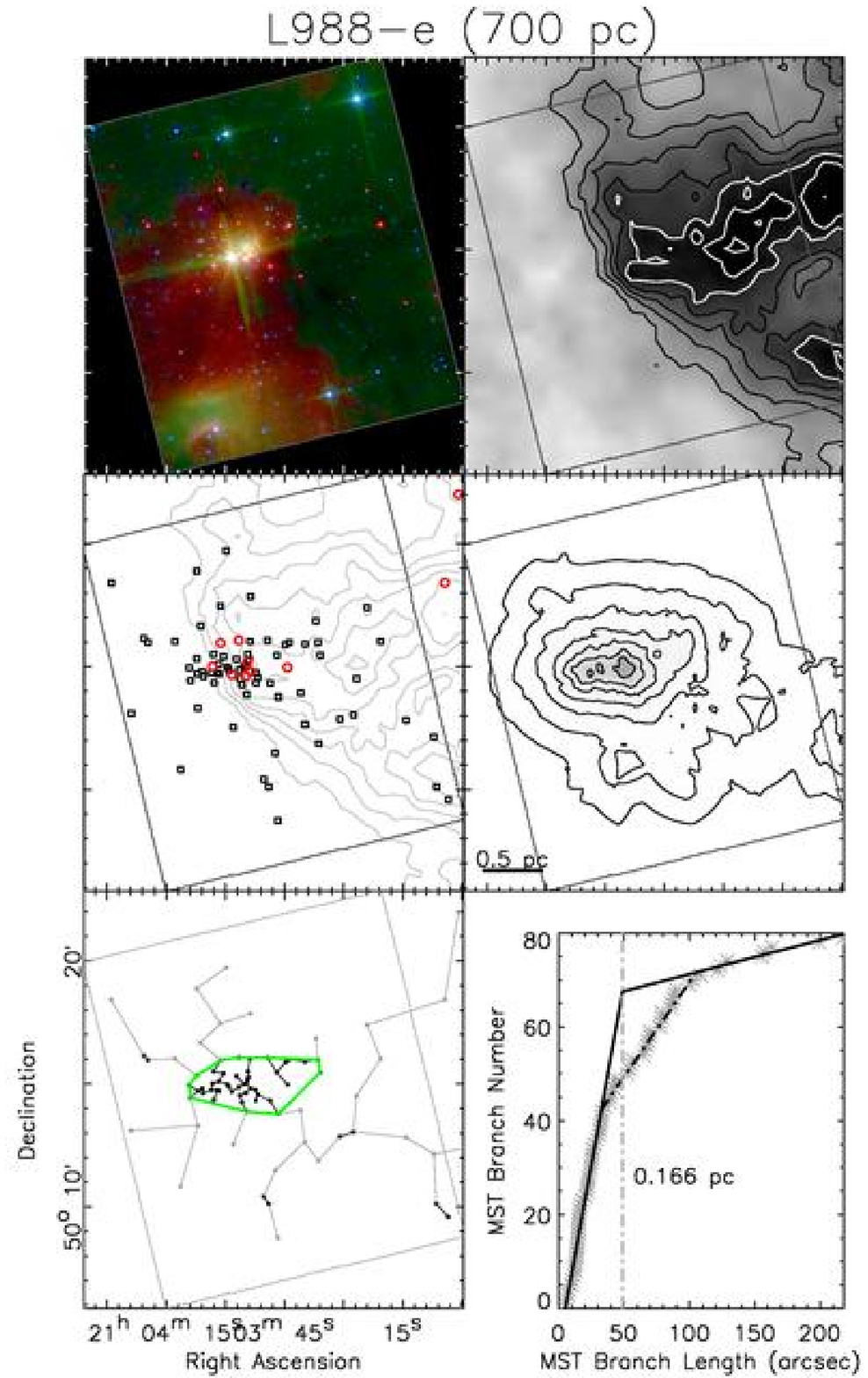}
\caption{L988-e at an assumed distance of 700 pc.}
\end{figure}
 
\begin{figure}
\epsscale{1}
\plotone{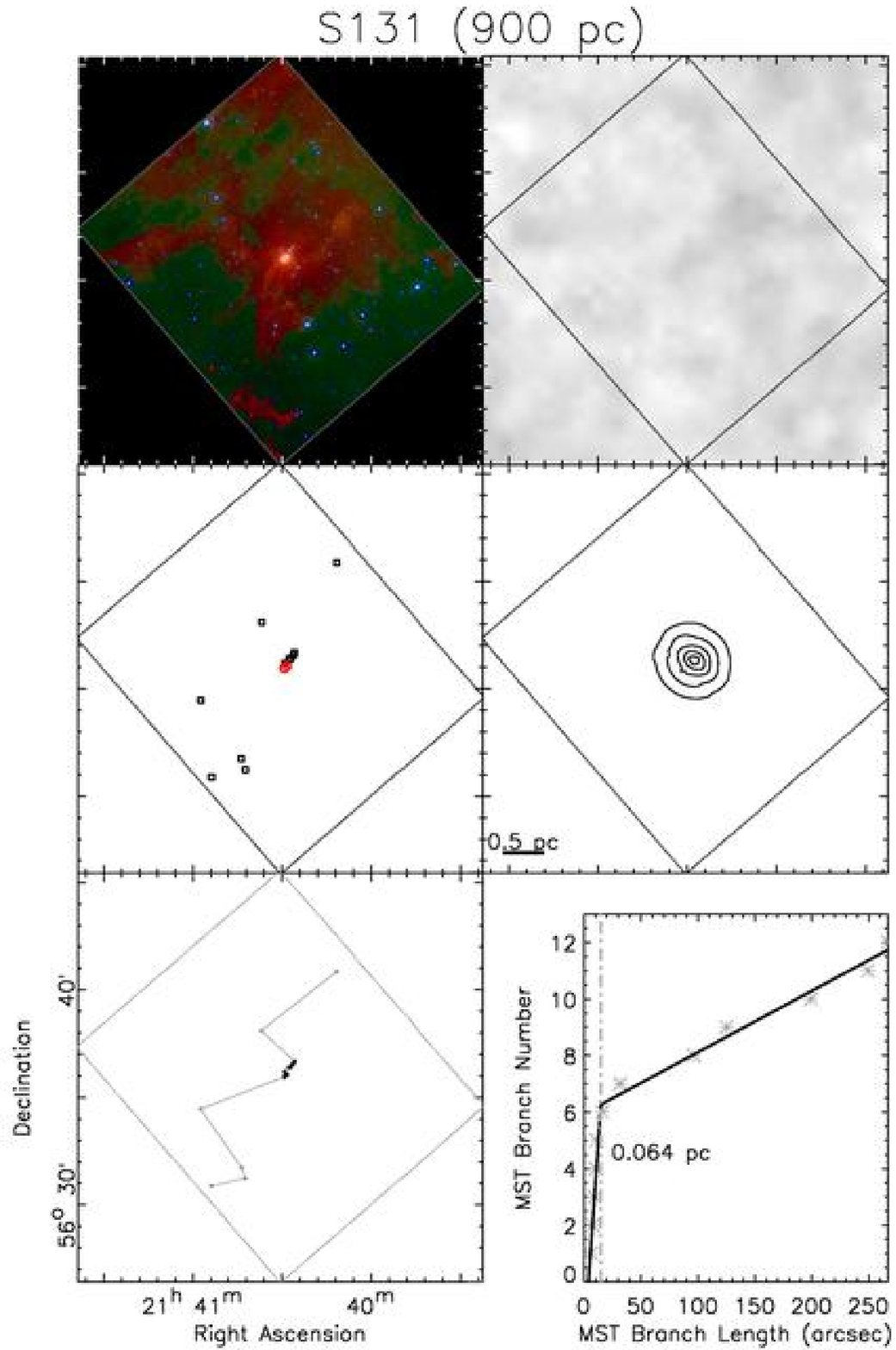}
\caption{S131 at an assumed distance of 900 pc.}
\end{figure}
 
\begin{figure}
\epsscale{1}
\plotone{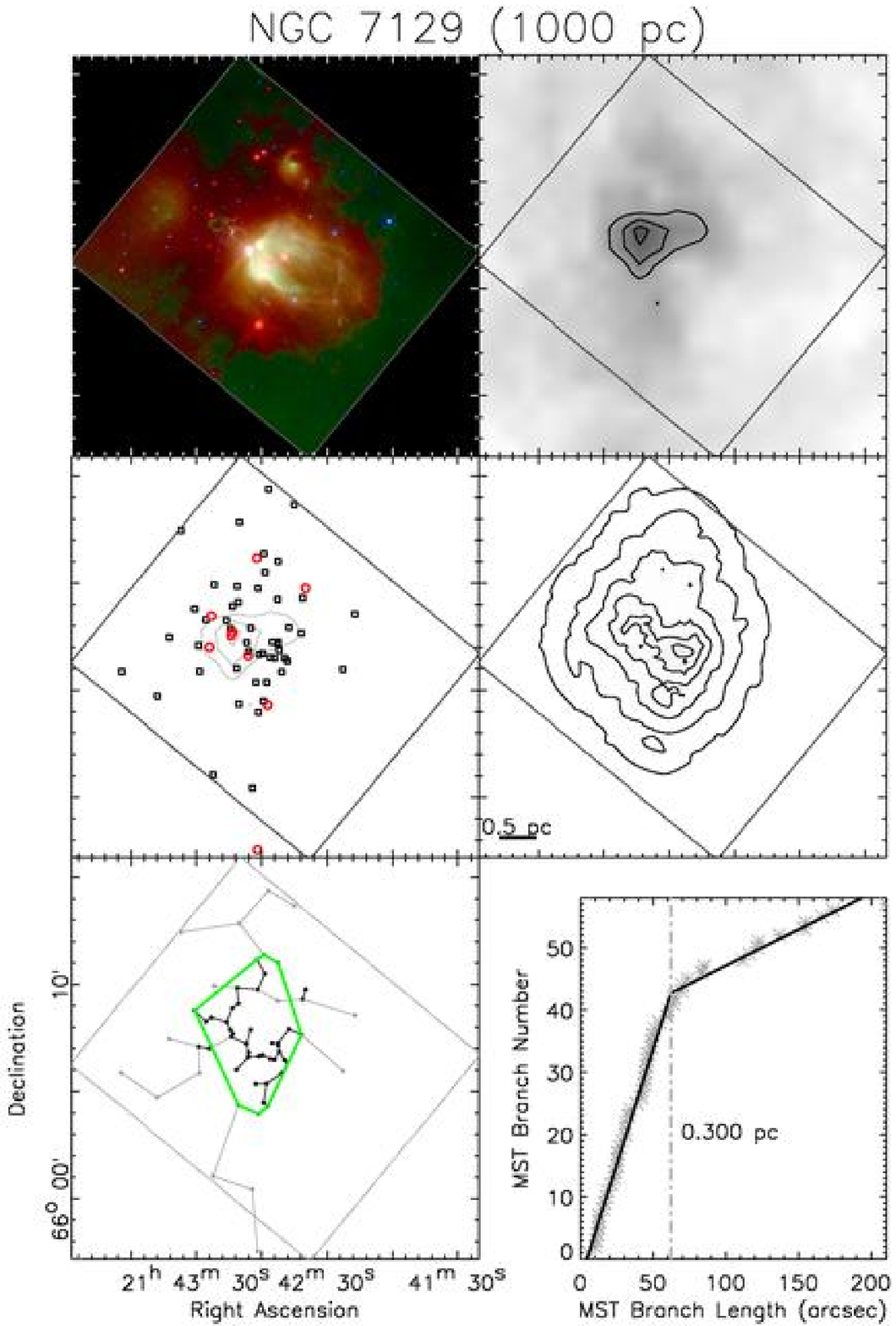}
\caption{NGC 7129 at an assumed distance of 1000 pc.}
\end{figure}
 
\begin{figure}
\epsscale{1}
\plotone{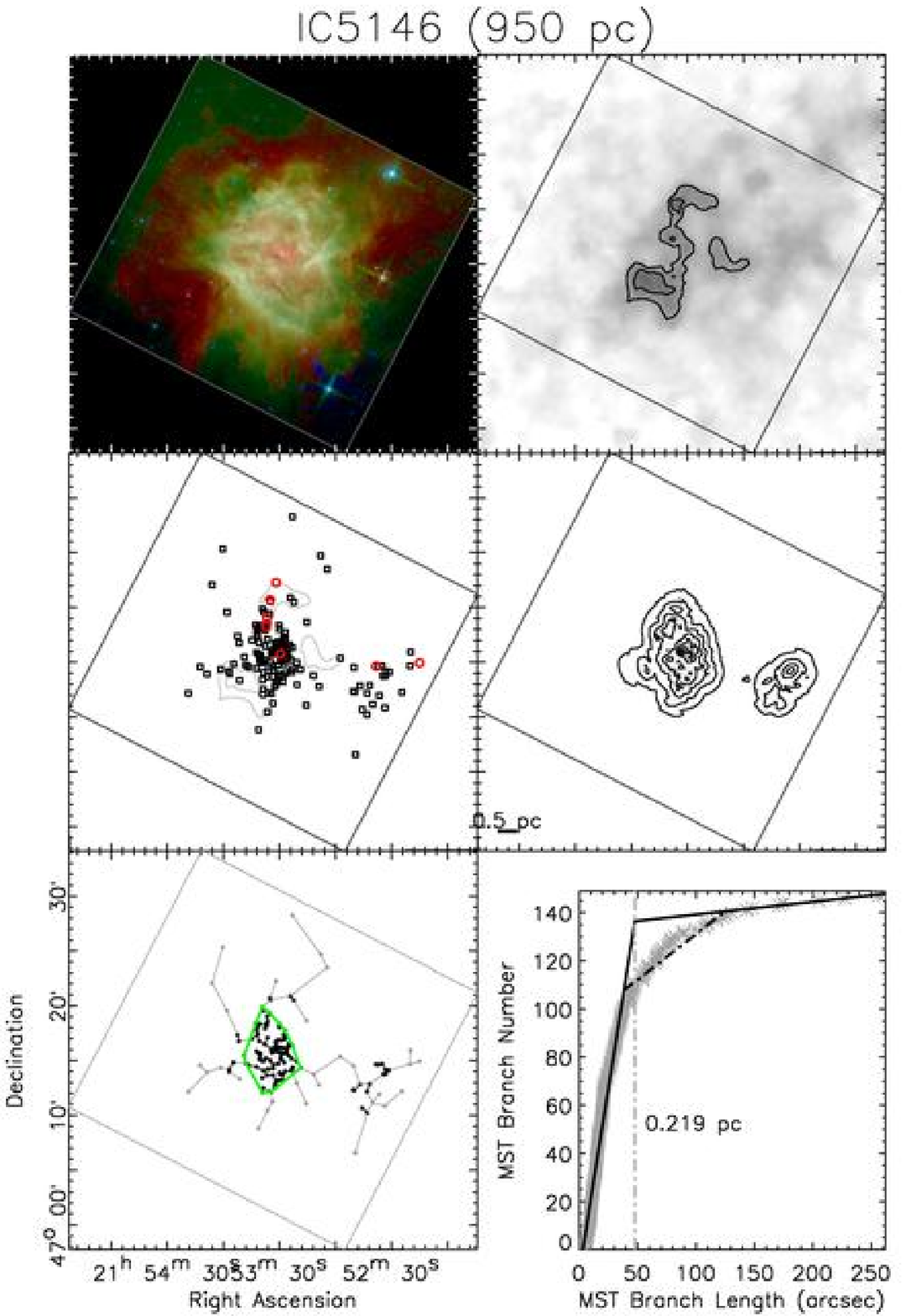}
\caption{IC5146 at an assumed distance of 950 pc.}
\end{figure}
 
\clearpage
 
\begin{figure}
\epsscale{1}
\plotone{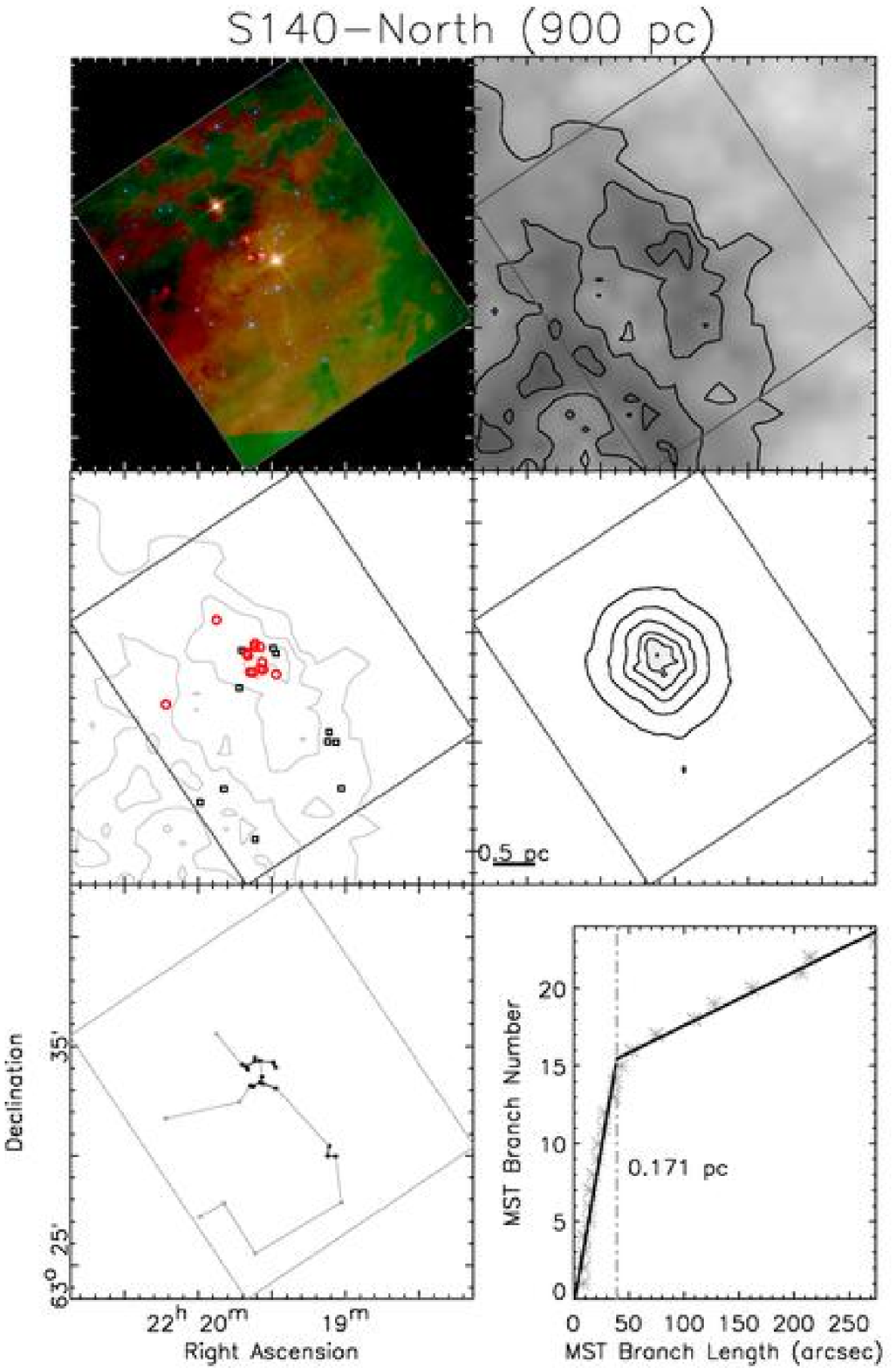}
\caption{S140-North at an assumed distance of 900 pc.}
\end{figure}
 
\begin{figure}
\epsscale{1}
\plotone{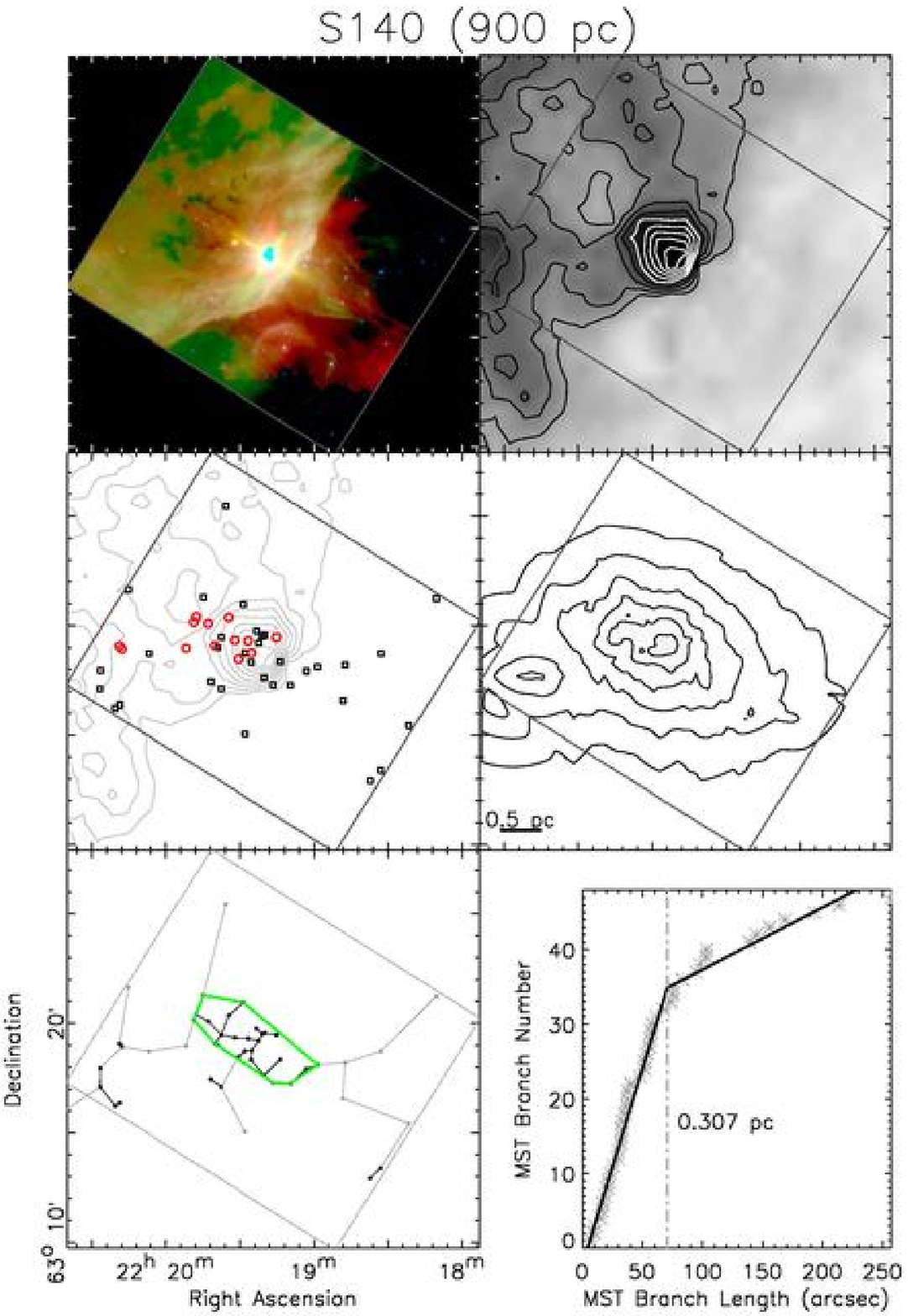}
\caption{S140 at an assumed distance of 900 pc.}
\end{figure}
 
\begin{figure}
\epsscale{1}
\plotone{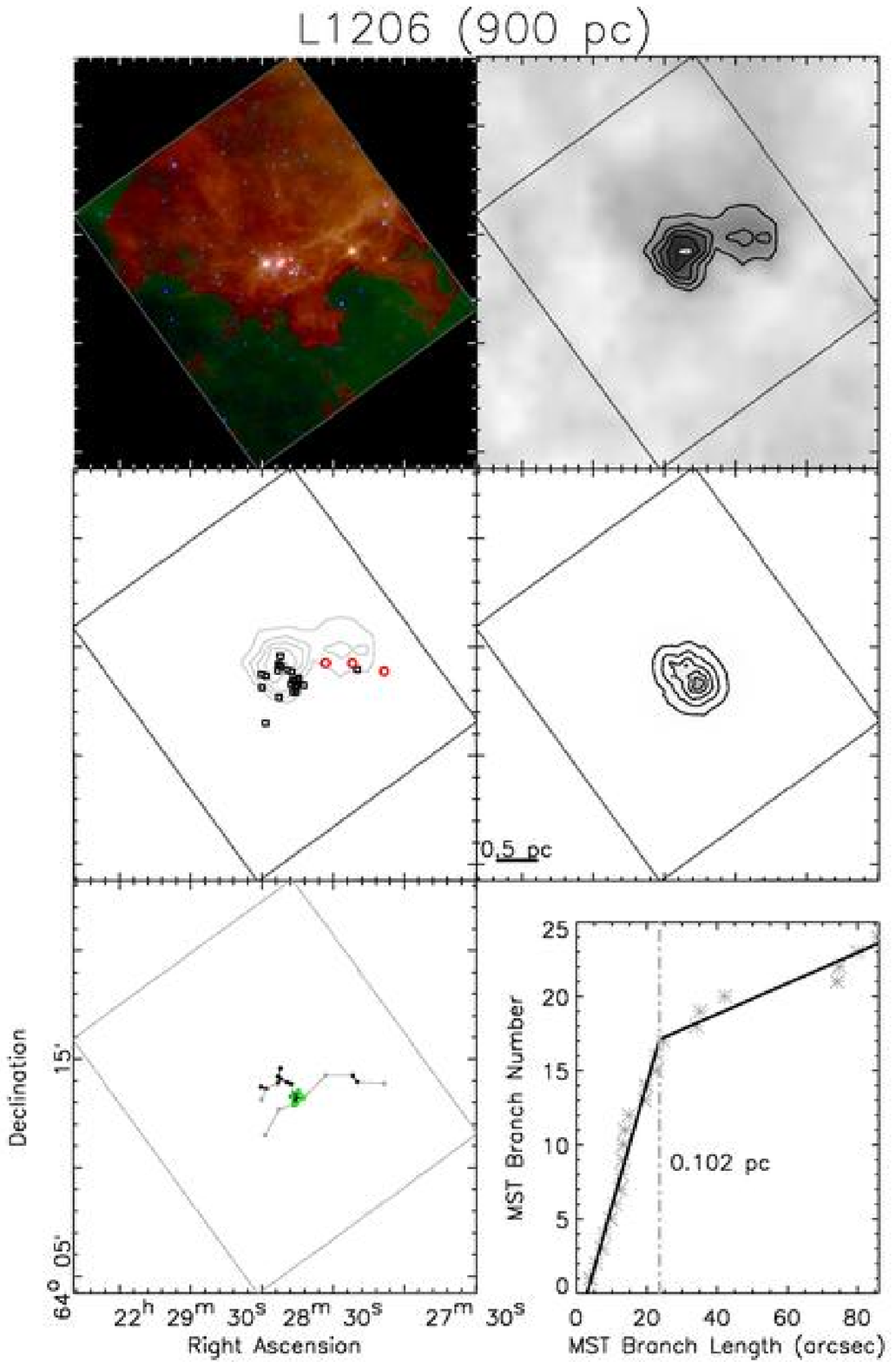}
\caption{L1206 at an assumed distance of 900 pc.}
\end{figure}
 
\begin{figure}
\epsscale{1}
\plotone{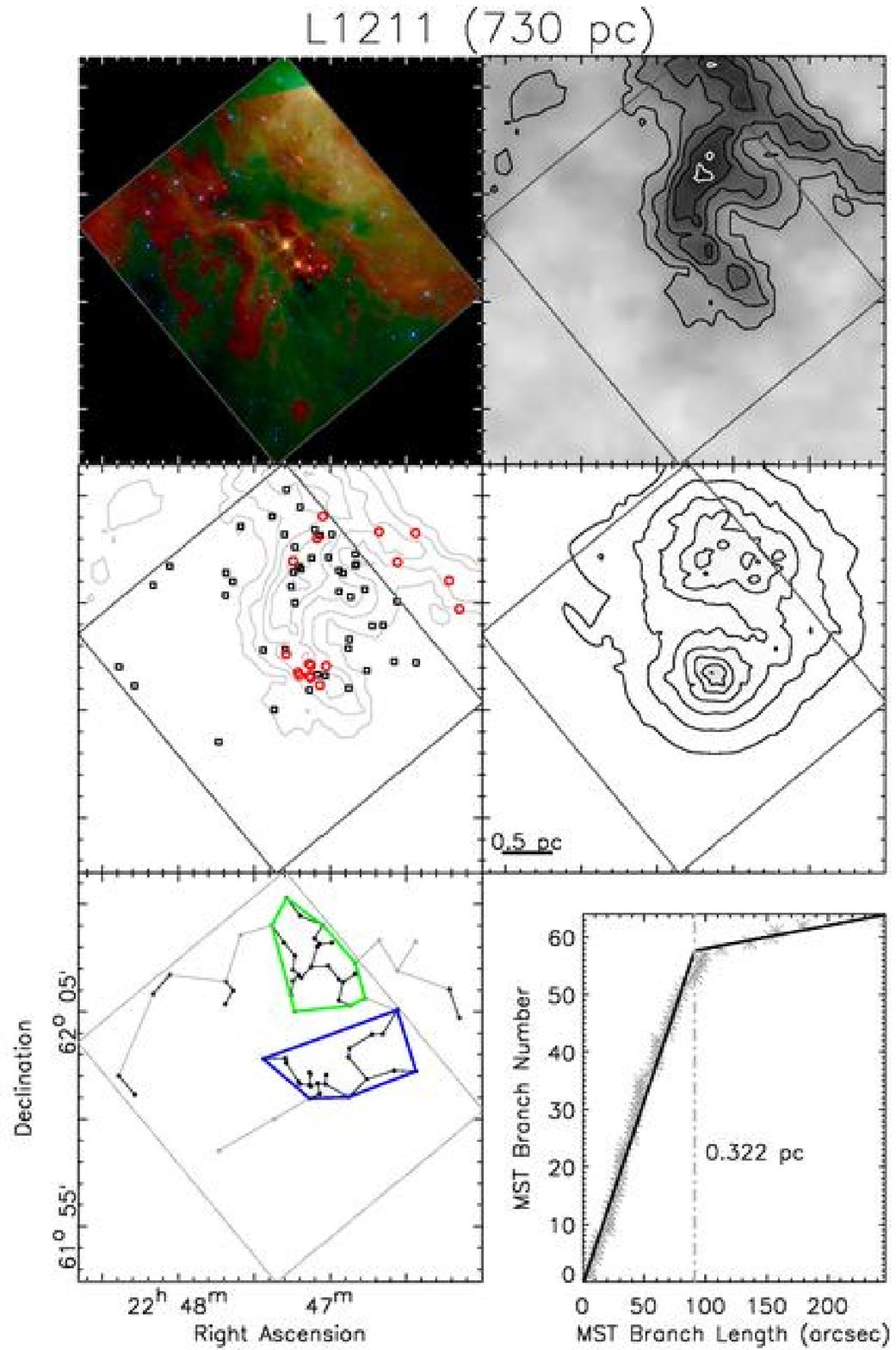}
\caption{L1211 at an assumed distance of 730 pc.}
\end{figure}
 
\begin{figure}
\epsscale{1}
\plotone{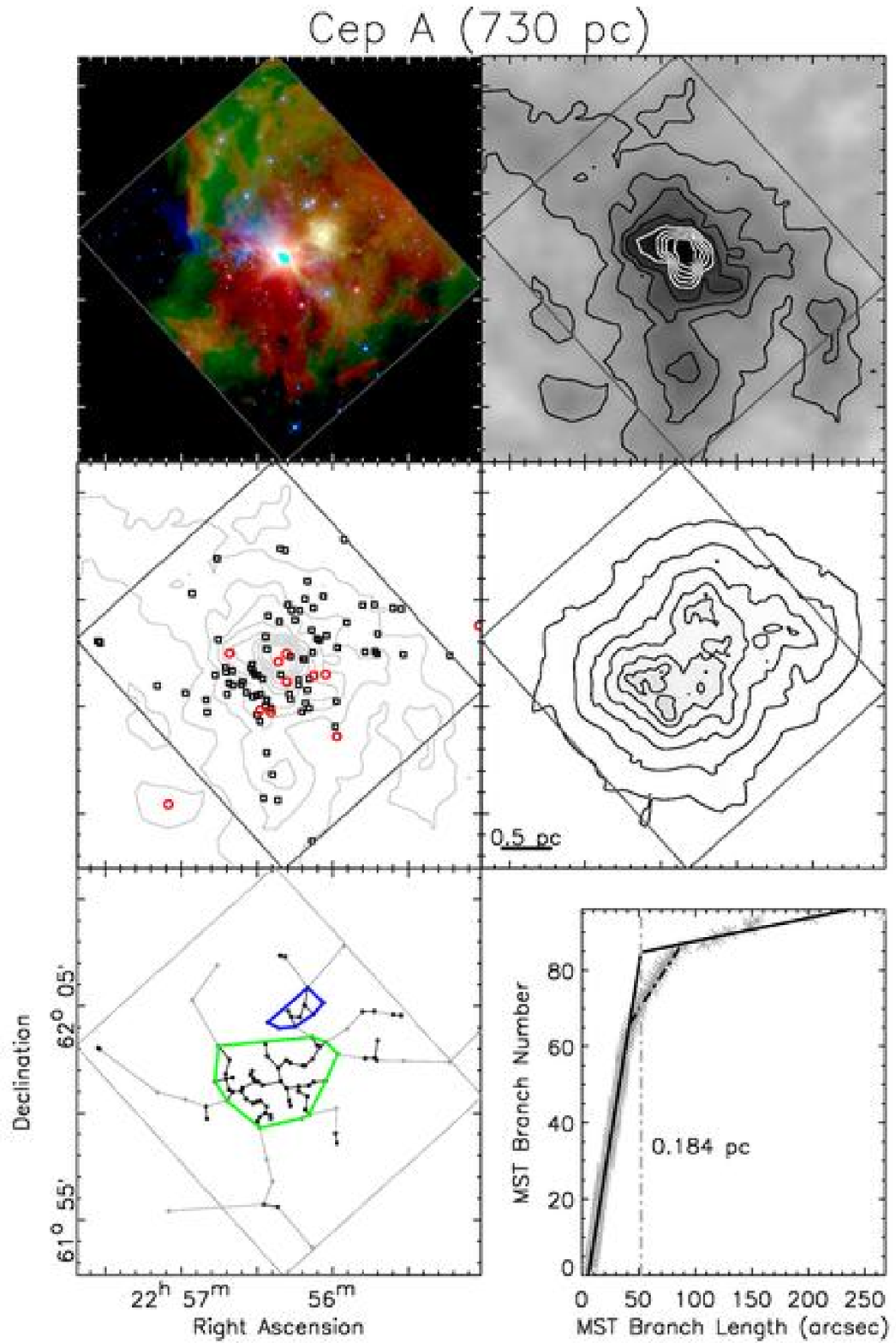}
\caption{Cep A at an assumed distance of 730 pc.}
\end{figure}
 
\clearpage
 
\begin{figure}
\epsscale{1}
\plotone{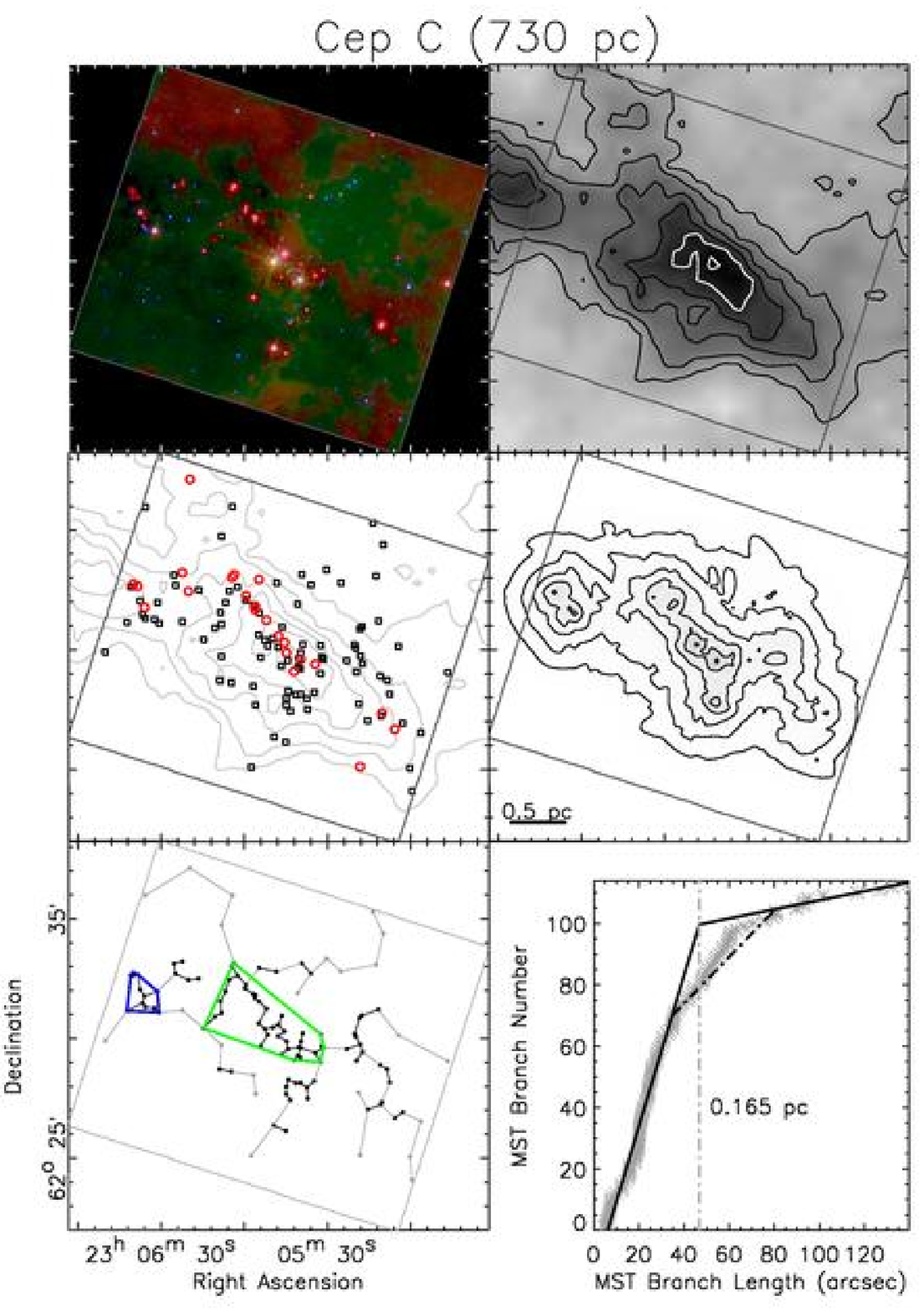}
\caption{Cep C at an assumed distance of 730 pc.\label{last}}
\end{figure}

%% file: srctabs.tex
\input{tab4.tex}
\input{tab5.tex}
\input{tab6.tex}
\input{tab7.tex}
\input{tab8.tex}
\input{tab9.tex}
\input{tab10.tex}
\input{tab11.tex}
\input{tab12.tex}
\input{tab13.tex}
\input{tab14.tex}
\input{tab15.tex}
\input{tab16.tex}
\input{tab17.tex}
\input{tab18.tex}
\input{tab19.tex}
\input{tab20.tex}
\input{tab21.tex}
\input{tab22.tex}
\input{tab23.tex}
\input{tab24.tex}
\input{tab25.tex}
\input{tab26.tex}
\input{tab27.tex}
\input{tab28.tex}
\input{tab29.tex}
\input{tab30.tex}
\input{tab31.tex}
\input{tab32.tex}
\input{tab33.tex}
\input{tab34.tex}
\input{tab35.tex}
\input{tab36.tex}
\input{tab37.tex}
\input{tab38.tex}
\input{tab39.tex}